\documentclass[letter,journal]{IEEEtran}

\usepackage{amsmath} 
\usepackage{amssymb}  
\usepackage{hyperref} 
\usepackage{amsthm}
\usepackage{enumerate,color,float}
\usepackage[justification=centering]{caption}
\usepackage{amsfonts}
\usepackage{mathtools}
\usepackage{arydshln}
\usepackage{stmaryrd}
\usepackage{cases}
\usepackage{multicol}
\usepackage{algorithm,algorithmic}
\usepackage{tikz,pgfplots,rotating}
\usepackage{graphicx}
\usepackage{bm}
\usepackage{cite}

\pgfplotsset{compat = 1.3}
\usetikzlibrary{shapes, positioning, arrows, decorations.markings}
\tikzset{
    -*/.style={
        shorten >=#1,
        decoration={
            markings,
            mark={
                at position 1
                with {
                    \draw[fill] circle [radius=#1];
                }
            }
        },
        postaction=decorate
    },
    -*/.default=1.5pt
}
\tikzset{
    *-/.style={
        shorten <=#1,
        decoration={
            markings,
            mark={
                at position 0
                with {
                    \draw[fill] circle [radius=#1];
                }
            }
        },
        postaction=decorate
    },
    *-/.default=1.5pt
}
\tikzset{
    *-*/.style={
        shorten <=#1,
        shorten >=#1,
        decoration={
            markings,
            mark={
                at position 0
                with {
                    \draw[fill] circle [radius=#1];
                }
            },
            mark={
                at position 1
                with {
                    \draw[fill] circle [radius=#1];
                }
            },
        },
        postaction=decorate
    },
    *-*/.default=1.5pt
}
\newtheoremstyle{remboldstyle}
  {}{}{}{}{\bfseries}{.}{.5em}{{\thmname{#1 }}{\thmnumber{#2}}}

\theoremstyle{plain}
\newtheorem{theorem}{Theorem}

\theoremstyle{definition}
\newtheorem{definition}[theorem]{Definition}

\newtheorem{example}[theorem]{Example}

\theoremstyle{remboldstyle}
\newtheorem{remark}[theorem]{Remark}
\theoremstyle{corollary}
\newtheorem{corollary}[theorem]{Corollary}
\theoremstyle{lemma}
\newtheorem{lemma}[theorem]{Lemma}
\newcommand{\mb}[1]{\mathbf{#1}}
\newcommand{\tmb}[1]{\tilde{\mathbf{#1}}}
\newcommand{\mbb}[1]{\mathbb{#1}}
\newcommand{\GF}[1]{\mbb{F}_{#1}}
\newcommand{\vect}[2] {\mathbf{#1}_{#2}}

\newcommand{\Rm}[1]{\mathrm{#1}}
\newcommand{\Cal}[1]{\mathcal{#1}}
\newcommand{\codeB}{\mathcal{B}}
\newcommand{\codeBe}{\mathcal{B}_{\mathrm{e}}}

\newcommand{\om}{\omega}               
\newcommand{\ob}{\overline{\om}} 

\newcommand{\uO}{\underline{0}}
\newcommand{\uI}{\underline{1}}
\newcommand{\defeq}{\triangleq}
\newcommand{\msf}[1]{\mathsf{#1}}
\newcommand{\fI}{\makebox[2ex][c]{$I$}}
\newcommand{\fX}{\makebox[2ex][c]{$X$}}
\newcommand{\fY}{\makebox[2ex][c]{$Y$}}
\newcommand{\fZ}{\makebox[2ex][c]{$Z$}}
\newcommand{\fo}[1]{\makebox[1ex][c]{$0$}}
\newcommand{\fl}[1]{\makebox[1ex][c]{$1$}}
\newcommand{\fw}[1]{\makebox[1ex][c]{$\om$}}
\newcommand{\fwb}[1]{\makebox[1ex][c]{$\bar{\om}$}}
\newcommand{\matr}[1]{\mathbf{#1}}
\newcommand{\vg}{\mathbf{g}}
\newcommand{\bvg}{\overline{\mathbf{g}}}
\newcommand{\vm}{\mathbf{m}}

\newcommand{\vt}{\mathbf{t}}
\newcommand{\set}[1]{\mathcal{#1}}
\newcommand{\tset}[1]{\tilde{\mathcal{#1}}}
\newcommand{\graph}[1]{\mathcal{#1}}
\newcommand{\code}[1]{\mathcal{#1}}

\newcommand{\tcode}[1]{\tilde{\mathcal{#1}}}

\newcommand{\syp}{\mathsf{S}}
\newcommand{\tr}{\mathsf{T}}
\newcommand{\herp}{\mathsf{H}}
\newcommand{\Ca}{\mathcal{C}_{\mathrm{a}}}
\newcommand{\Cb}{\mathcal{C}_{\mathrm{b}}}

\newcommand\TSpace{\rule{0pt}{2.6ex}}       
\newcommand\BSpace{\rule[-1.2ex]{0pt}{0pt}} 
\newcommand{\qedblack}{\hspace*{\fill}%
  $\blacksquare$\smallskip}

\newcommand{\defend}{\qedblack} 
\newcommand{\exampleend}{\qedblack}

\newcommand{\cB}{\mathcal{B}}
\newcommand{\ctB}{\tilde{\mathcal{B}}}
\newcommand{\cC}{\mathcal{C}}

\newcommand{\cS}{\mathcal{S}}
\newcommand{\cG}{\mathcal{G}}
\newcommand{\cGbar}{\overline{\mathcal{G}}}

\newcommand{\cP}{\mathcal{P}}
\newcommand{\cQ}{\mathcal{Q}}

\newcommand{\ci}{\mathsf{i}} 

\newcommand{\SP}{\mathsf{S}}
\newcommand{\HP}{\mathsf{H}}
\newcommand{\tell}{\tilde{\ell}}

\newcommand{\CASE}[1]{\STATE \textbf{case} #1\textbf{:} \begin{ALC@g}}
\newcommand{\ENDCASE}{\end{ALC@g}}

\newcommand{\DEFAULT}{\STATE \textbf{default:} \begin{ALC@g}}
\newcommand{\ENDDEFAULT}{\end{ALC@g}}
\newcommand{\DEFAULTLINE}[1]{\STATE \textbf{default:} }
\newcommand{\sq}[2][0]{
  \mbox{$\medmuskip=#1mu\displaystyle#2$}%
}
\newlength{\myl}
\newcommand{\vtag}[2]{
  \settowidth{\myl}{{\scriptsize $\sq[.05]{#2}$}}
  \node at ([yshift=7pt]#1) [rotate=90,draw=none,anchor=east] 
	{\parbox{\myl}{\raggedleft\scriptsize $\sq[.05]{#2}$}}}

\title{
Normal Factor Graph Representations \\of Stabilizer Quantum Codes with Applications
}

\author{July X.\ Li
        and 
	Pascal O.\ Vontobel%
\thanks{The authors are with the  
        Department of Information Engineering,
        The Chinese University of Hong Kong,
        Shatin, N.T., Hong Kong,
        (email: \texttt{july.x.li@ieee.org},
                \texttt{pascal.vontobel@ieee.org}).}%
}
\begin{document}

\maketitle

\begin{abstract}

We develop normal factor graph (NFG) representations of stabilizer quantum error-correcting codes and entanglement-assisted quantum error-correcting codes. We introduce two complementary representations. Quantum-code NFGs describe stabilizer and normalizer label codes and provide a graphical framework for establishing self-orthogonality, constructing codes from component codes, and identifying encoding and decoding circuits. Quantum-probability NFGs represent the quantum-mechanical probabilities associated with encoding, noise, and measurement. We also introduce a double-edge formulation that makes the relationship between these two representations explicit. We apply the framework to convolutional and tail-biting convolutional stabilizer codes, as well as to two classes of low-density parity-check stabilizer codes. Finally, we extend the quantum-probability representation to entanglement-assisted codes.
\end{abstract}
\section{Introduction}
\label{sec:introduction:1}
Quantum error correction is essential for reliable quantum information storage, computation, and communication~\cite{lidar2013quantum}. 
Stabilizer quantum error-correcting codes (QECCs) are particularly useful because their algebraic structure can be described using classical codes equipped with suitable symplectic or trace inner products~\cite{calderbank1998quantum}. 
In this description, the stabilizer label code is self-orthogonal, and its dual is the normalizer label code. This connection allows classical coding techniques to be adapted to the construction and decoding of stabilizer QECCs.

Graphical models provide powerful tools for representing classical codes and developing decoding algorithms~\cite{kschischang2001factor, forney2001codes, loeliger2004introduction}. 
Extending these tools to stabilizer QECCs require accounting for both the self-orthogonality constraint and code degeneracy. 
Normal factor graphs (NFGs) offer a natural framework for studying these requirements through local constraints and graph duality.
A central challenge is to develop graphical representations that expose the self-orthogonality constraints and support decoding methods that account for degeneracy.

The first representation, which we call a quantum-code NFG, describes a stabilizer QECC through its stabilizer and normalizer label codes. Building on the graphical framework introduced in~\cite{vontobel2008stabilizer} and developed further in~\cite{LiVontobel2016factor}, we establish duality results that provide sufficient conditions for the self-orthogonality of the represented stabilizer label code. In particular, we provide a complete proof of Theorem~\ref{thm:selforthogonal_thm}. 
Closing boxes in these graphs yields constructions of larger codes from component codes, whereas opening boxes reveals decompositions into elementary components associated with Clifford gates.
With closing-the-box operations, the quantum-code NFG representation provides a method to construct new stabilizer QECCs based on known stabilizer QECCs.
We also study the Clifford group on $n$ qubits and show that it is isomorphic to a group with a newly defined operation, where related discussions can be found in~\cite{dehaene2003clifford, aaronson2004improved, maslov2018shorter}.
Hence by opening-the-box operations, the quantum-code NFG representation of a stabilizer QECC can be decomposed into basic blocks representing additive codes over $\GF{4}$ associated with a single Hadamard, phase, or CNOT gate, respectively.
Moreover, when the stabilizer QECC's stabilizer label code is linear over $\GF{4}$, the counterpart result is that the quantum-code NFG representation can be decomposed into basic blocks representing linear codes over $\GF{4}$ associated with some quantum circuits constructed using Hadamard, phase, and CNOT gates only.

The second representation, which we call a quantum-probability NFG, describes the quantum-mechanical probabilities associated with stabilizer QECCs, including measurement outcomes. This construction specializes the factor graph framework for quantum probabilities in~\cite{loeliger2017factor} to stabilizer coding systems. 
We further introduce a double-edge representation that groups paired variables and makes the structural relationship with quantum-code NFGs more transparent.

Besides introducing the NFG representations for stabilizer QECCs in the general way, we also discuss several useful classes of stabilizer QECCs in detail.
The first class of code is (tail-biting) convolutional stabilizer QECCs.
The quantum-code NFG representations of a (tail-biting) convolutional stabilizer QECC are almost the same except that the tail-biting convolutional stabilizer QECCs require the initial and final memory states to coincide. 
The key step of obtaining the quantum-code NFG representation is to use our algorithms generating the stabilizer QECCs represented by the small blocks of the quantum-code NFG representation.
We first show how to find the quantum-code NFG representation of a (tail-biting) convolutional stabilizer QECC whose stabilizer label code is additive over $\GF{4}$ with single basic generator and then demonstrate that the steps can be extended to other cases like when there are multiple basic generators or when the stabilizer label code is linear over $\GF{4}$.
We introduce the overall constraint length for a (tail-biting) convolutional stabilizer QECC, which is the number of memory qubits required when the stabilizer label code is additive over $\GF{4}$ and is the upper bound of the number of memory qubits required when the stabilizer label code is linear over $\GF{4}$.
This construction requires fewer qubits for memory compared with earlier constructions, e.g., see~\cite{wilde2009quantum}.
We also introduce an equivalent quantum-code NFG representation of a (tail-biting) convolutional stabilizer QECC, which reveals the structure of the encoding/decoding circuit and provides a simple and elegant proof of the self-orthogonality of the stabilizer label code.

As for LDPC stabilizer QECCs, there are two classes of codes of special interest, which are both Calderbank-Shor-Steane type of codes defined by two binary linear codes $\cC_{X}$ and $\cC_{Z}$, i.e., the stabilizer label code is the product code of $\cC_{X}^{\perp}$ and $\cC_{Z}^{\perp}$. 
For this type of codes, we prove the equivalent condition of the self-orthogonality of stabilizer label codes, i.e., $\cC_{Z}^{\perp} \subseteq \cC_{X}$ or $\cC_{X}^{\perp} \subseteq \cC_{Z}$, with the help of duality results of NFGs and switching between image-presentation type of NFGs and kernel-representation-type of NFGs.
One of two classes of codes was introduced by Tillich and Z\'emor~\cite{tillich2014quantum} with $\cC_{X}$ and $\cC_{Z}$ constructed from two binary linear codes, where the NFGs representing $\cC_{X}^{\perp}$ and $\cC_{Z}^{\perp}$ are taking advantage of the special product code structure of them.
The Toric code by Kitaev~\cite{kitaev2003fault} has minimum distance proportional to $\sqrt{n}$ even though the row weight of the stabilizer label code is a constant, which does not grow with the block length $n$.
The NFGs representing $\cC_{X}^{\perp}$ and $\cC_{Z}^{\perp}$, which lie on a torus surface, can be used to easily show the self-orthogonality of the stabilizer label code since it is a direct consequence due to the special structures of the NFGs. 

Furthermore, we also introduce the quantum-probability NFG representation for general entanglement-assisted QECCs, which can be easily extended to other entanglement-assisted quantum codes, e.g, quantum convolutional with entanglement-assisted codes~\cite{wilde2010convolutional}.
\subsection{Paper structure}
\label{sec:introduction:1a}
This paper is structured as follows. 
In Section II, we introduce some basics of stabilizer QECCs. 
In Section III, we discuss the two types of NFG representations of general stabilizer QECCs including the quantum-code NFG representations and quantum-probability NFG representations.
In Section IV, we give the quantum-code NFG representations of (tail-biting) convolutional stabilizer QECCs. 
In Section V, we demonstrate how to use duality results of graphs to prove the self-orthogonality of stabilizer label codes of two classes of LDPC stabilizer QECCs.
In Section VI, we extend the quantum-probability NFG representations for general entanglement-assisted QECCs.
\subsection{Notation}
\label{sec:introduction:1b}
We use the following notations throughout this paper.
Let $\langle \GF{2}, +, \,\cdot\, \rangle$ and $\langle \GF{4}, +, \,\cdot\, \rangle$ be the finite fields of size $2$ and $4$, respectively, 
where $\GF{2} \defeq \{ 0, 1\}$, $\GF{4} \defeq \{ 0, 1, \om, \ob\}$ with $\om^2 = 1 + \om$ and the conjugate of $a$ is defined as $\overline{a} \defeq a^2\ \forall a \in \GF{4}$.
Let $\mbb{Z}$, $\mbb{Z}_{\geq 0}$, and $\mbb{Z}_{>0}$ be the set of integers, non-negative integers, and positive integers, respectively.
For all $n\in \mbb{Z}_{>0}$, let $\langle \mbb{Z}_n, +,\, \cdot\, \rangle$ with $\mbb{Z}_n \defeq \{ 0, 1, \ldots, n \! - \! 1 \}$ be the ring of integers with addition and multiplication modulo $n$.
For all $n\in\mbb{Z}_{>0}$, define the index sets $\set{I}_{n} \defeq \{ 0, 1, \ldots, n \! - \! 1 \}$ and $\set{I}_{n}^{+} \defeq \set{I}_{n} \backslash \{ 0 \}$.
Let $\mbb{C}$ be the set of complex numbers and $\ci$ be the imaginary unit of $\mbb{C}$.
For $m, n\in \mbb{Z}_{>0}$, let $\mb{I}_n$, $\mb{0}_n$, and $\mb{0}_{m \times n}$ be the $n \times n$ identity matrix, the $n \times n$ all-zero matrix, and the $m \times n$ all-zero matrix, respectively; their indices may be omitted when the dimensions are clear from the context.
For any $n \in \mbb{Z}_{>0}$, let $\mb{\Lambda}_n \in \GF{2}^{2n \times 2n}$ be defined as 
$\mb{\Lambda}_n \defeq \left[\begin{array}{cc} 
\mb{0}_n & \mb{I}_n\\
\mb{I}_n & \mb{0}_n
\end{array}\right]$.
For any matrix $\mb{u}$ over $\GF{2}$, $\GF{4}$ or $\mbb{C}$, let 
$\mb{u}^{\msf{T}}$ be the transpose of $\mb{u}$ and 
$\mb{u}^{\msf{H}} \defeq \overline{\mb{u}}^{\msf{T}}$ be the conjugate transpose of $\mb{u}$.
Let $v_i$ be the $i^{\Rm{th}}$ entry of any vector $\mb{v}$, 
$\mb{m}_i$ be the $i^{\Rm{th}}$ row of any matrix $M$, and 
$m_{i,j}$ be the $(i,j)^{\Rm{th}}$ entry of any matrix $M$,
respectively.
When $S$ is a statement, $[S] \defeq 1$ if $S$ is true and $[S] \defeq 0$ otherwise.

The Hadamard, phase, and controlled-NOT (CNOT) gates are Clifford transformations defined by the unitary matrices 
\begin{equation*}
\Rm{H} \defeq \frac{1}{\sqrt{2}}\left[\begin{array}{cc} 1 & 1 \\ 1 & -1 \end{array}\right],\ 
\Rm{P} \defeq \left[\begin{array}{cc} 1 & 0 \\ 0 & \ci \end{array}\right],\ 
\Rm{CNOT} \defeq \left[\begin{array}{cc} \mb{I}_2 & \mb{0}_2 \\ \mb{0}_2 & X \end{array}\right],
\end{equation*}
respectively. 
\section{Stabilizer QECCs}
\label{sec:stabilizerQECC:2}
In this section, we review the stabilizer formalism.
\subsection{Qubits and Pauli Group}
\label{sec:Qubits:2a}
The basis for a quantum state of one qubit $| \phi \rangle$ is usually denoted as $| 0 \rangle$ and $| 1 \rangle$ and hence $| \phi \rangle = \alpha_{0} | 0 \rangle + \alpha_{1} | 1 \rangle$ for some $\alpha_{0}, \alpha_{1} \in \mbb{C}$ with $|\alpha_{0}|^2+|\alpha_{1}|^2=1$.
The Pauli matrices 
\begin{equation*}
X \defeq 
\left[\begin{array}{cc}
0 & 1\\
1 & 0
\end{array}\right],\ 
Y \defeq 
\left[\begin{array}{cc}
0 & -\ci\\
\ci & 0
\end{array}\right],\ 
Z \defeq
\left[\begin{array}{cc}
1 & 0\\
0 & -1
\end{array}\right],
\end{equation*}
along with $I \defeq \mb{I}_2$
form the single-qubit Pauli group $\cG_1 \defeq  \left\{ c \cP \ \middle|\ c\in\left\{ \pm 1, \pm \ci \right\}, \cP\in \left\{ I, X, Y, Z \right\} \right\}$ under matrix multiplication.
A quantum state of $n$ qubits is described as a vector of a $2^n$-dimensional Hilbert space, which is the tensor product of $n$ $2$-dimensional Hilbert spaces. 
We consider an error model where all the possible errors of an $n$-qubit system are in the Pauli group $\cG_n$,
where $\cG_n$ is the tensor product of $\cG_1$, i.e.,
$\cG_n = \cG_1^{\otimes n}$ and $\langle \cG_n,\,\cdot\, \rangle $ forms a group under matrix multiplication.
For $j \in \left\{ 1,2, \ldots, n\right\}$, denote 
\begin{align*}
X_j &\defeq \underbrace{I \otimes \cdots \otimes I}_{j-1\text{ times}} \otimes X \otimes \underbrace{I \otimes \cdots \otimes I}_{n-j\text{ times}},\\
Z_j &\defeq \underbrace{I \otimes \cdots \otimes I}_{j-1\text{ times}} \otimes Z \otimes \underbrace{I \otimes \cdots \otimes I}_{n-j\text{ times}}.
\end{align*}
The set $\left\{ X_1, \ldots, X_n, Z_1, \ldots, Z_n \right\}$ together with the phase factor $c,\ c\in\left\{ \pm 1, \pm \ci \right\}$, forms a basis for $\cG_n$.
The elements of $\cG_{n}$ are called Pauli operators.
For simplicity, a Pauli operator $c\cP_1 \otimes \cP_2 \otimes \cdots \otimes \cP_n \in \cG_n$ is abbreviated as $c\cP_1 \cP_2 \cdots\cP_n$, 
where $\cP_1, \cP_2,  \ldots, \cP_n \in \cG_1$ and $c\in\{\pm 1, \pm \msf{i}\}$.
\subsection{Stabilizer QECCs}
\label{sec:stabilizerQECC:2b}
An $\llbracket n,k \rrbracket$ stabilizer QECC encodes $k$ logical qubits into $n$ physical qubits and has a code space of dimension $2^k$.
\begin{definition}[Stabilizer QECCs]
Consider the stabilizer $\cS\subseteq \cG_n$ which is a set generated by $n-k$ independent and commuting Pauli operators such that $-I \notin S$.
A stabilizer QECC $\cC$ associated with the stabilizer $\cS$ is defined as the set of quantum states stabilized by all Pauli operators in the stabilizer $\cS$, i.e.,
\begin{equation}
\label{eq:QECC_def1}
\cC \defeq \left\{ |\overline{\psi}\rangle \in \mbb{C}^{2^n} \ \middle|\ g|\overline{\psi}\rangle = |\overline{\psi}\rangle\ \forall g \in \cS\right\}.
\end{equation}
The normalizer $N(\cS)$ is defined as 
\begin{equation}
N(\cS) \defeq \left\{ g\in\cG_n \ \middle|\ gg' = g'g\ \forall g'\in \cS \right\}.
\end{equation}
The minimum distance $d$ is defined to be the minimum weight of Pauli operators in $N(\cS)\backslash \cS$, 
where the weight of a Pauli operator is the number of terms in its tensor product not equal to the identity matrix assuming there is at most one Pauli matrix corresponding to each qubit. 
Then the parameter of the code $\cC$ is $\llbracket n, k, d \rrbracket$.
\defend
\end{definition}
\subsection{Stabilizer Label Codes and Normalizer Label Codes}
\label{sec:SLC_NLC:2c}
Instead of working on stabilizer QECCs directly, equivalently we use NFGs to represent the stabilizer label codes and normalizer label codes of stabilizer QECCs, where results of classical codes can be applied.

Since the phase factors of Pauli operators are irrelevant to the commutativity of Pauli operators,
we consider the set of equivalence classes of Pauli operators from $\cG_n$ which is denoted as $\overline{\cG}_n$, where Pauli operators in the same equivalence class differ by a phase factor.

The stabilizer label code of a stabilizer QECC is obtained using some label mappings.
We use the following label mappings in this paper. 

\begin{table}
\begin{center}
\caption{\sc{\\mappings $\ell$ and $\Rm{L}$ from $\overline{\cG}_1$ to $\GF{2}^2$ and $\GF{4}$.}}
\label{tab:l_L}
\begin{tabular}{|c|cccc|}
\hline
$\cP \in \cGbar_1$ & $I$ & $X$ & $Y$ & $Z$ \TSpace\BSpace \\
\hline
$\ell(\cP)$ & $[0,0]$ & $[1,0]$ & $[1,1]$ & $[0,1]$ \TSpace \\
$\Rm{L}(\cP)$ & $0$ & $\om$ & $1$ & $\ob$ \BSpace \\
\hline
\end{tabular}
\end{center}
\end{table}

\begin{definition}[Label Mapping]
The label mappings $\ell^{(n)}$ and $\Rm{L}^{(n)}$ are isomorphisms from $\overline{\cG}_n$ to $\left(\GF{2}^2\right)^n$ and $\GF{4}^n$, respectively, defined as 
for any $\cP = \cP_1 \cdots \cP_n \in \overline{\cG}_n$ with $\cP_j \in \overline{\cG}_1\ \forall j$,
\begin{align}
\label{eq:ell}
\ell^{(n)}(\cP_1 \cdots \cP_n) &\defeq \big[\ell(\cP_1), \ldots, \ell(\cP_n)\big],\\
\Rm{L}^{(n)}(\cP_1 \cdots \cP_n) &\defeq \big[\Rm{L}(\cP_1), \ldots, \Rm{L}(\cP_n)\big],
\end{align}
where $\ell$ and $\Rm{L}$ are defined in Table~\ref{tab:l_L}.
In fact, $\ell^{(n)}_X(\cP_j)$ and $\ell^{(n)}_Z(\cP_j)$ collect the bits relevant to the bit flip $X$ and the phase flip $Z$, respectively.
Note that $Y$ is considered as a combination of bit flip and phase flip.
By suitable re-ordering and grouping w.r.t. $X$ and $Z$ of~\eqref{eq:ell}, we get another isomorphism $\tell^{(n)}:\ \overline{\cG}_n \rightarrow \left(\GF{2}^n\right)^2$ such that for any $\cP = \cP_1 \cdots \cP_n\in \overline{\cG}_n$ with $\cP_j \in \overline{\cG}_1\ \forall j$,
\begin{equation}
\tell^{(n)}(\cP) \defeq \big[\ell^{(n)}_X(\cP), \ell^{(n)}_Z(\cP)\big],
\end{equation}
where $[\ell_X(\cP_i), \ell_Z(\cP_i)] \defeq \ell(\cP_i)\ \ \forall \cP_i \in \cGbar_1$,
\begin{align*}
\ell^{(n)}_X\left(\cP_1 \cdots \cP_n\right) &\defeq \big[\ell_X(\cP_1), \ldots, \ell_X(\cP_n)\big],\text{ and }\\
\ell^{(n)}_Z\left(\cP_1 \cdots \cP_n\right) &\defeq \big[\ell_Z(\cP_1), \ldots, \ell_Z(\cP_n)\big].
\end{align*}
\defend
\end{definition}
When the number of qubits is clear from the context, we usually omit the upper index ``$(n)$'' in the above mappings.

\begin{definition}[Stabilizer Label Code]
For an $\llbracket n,k \rrbracket$ stabilizer QECC $\cC$ associated with the stabilizer $\cS \subseteq \cG_n$, 
its stabilizer label code is defined to be
\begin{align}
\ctB &\defeq \tell(\cS), \text{ or }
\cB \defeq \Rm{L}(\cS),
\end{align}
where $\ctB$ is a binary linear code of length $2n$ and $\cB$ is an additive code over $\GF{4}$ of length $n$.
\defend
\end{definition}

The connection between the commutativity of Pauli operators in the stabilizer $\cS$ and self-orthogonality of the stabilizer label codes can be made after introducing the normalizer label codes.
First we need to introduce symplectic, trace and Hermitian inner products, which play a key role in the above mentioned connection.

\begin{definition}[Symplectic, Trace and Hermitian Inner Products]
The symplectic inner product $\langle \cdot, \cdot \rangle_{\syp}: \left(\GF{2}^n\right)^2 \times \left(\GF{2}^n\right)^2 \rightarrow \GF{2}$,
the trace inner product\footnote{An equivalent definition of the trace inner product is $\langle \mb{u}, \mb{v} \rangle_{\tr} \defeq \Rm{Tr}(\mb{u} \mb{v}^{\msf{H}})$, where $\Rm{Tr}(a) \defeq a + \overline{a},\ a\in \GF{4}$.}
 $\langle \cdot, \cdot \rangle_{\tr}: \GF{4}^n \times \GF{4}^n \rightarrow \GF{2}$, and 
the Hermitian inner product $\langle \cdot, \cdot \rangle_{\herp}: \GF{4}^n \times \GF{4}^n \rightarrow \GF{2}$ 
are defined as 
\begin{align}
\label{eq:symplectic_inner_product}
\langle \mb{u}, \mb{v} \rangle_{\syp} &\defeq \mb{u} \mb{\Lambda}_n \mb{v}^{\msf{T}} \in \GF{2}\ \forall \mb{u}, \mb{v} \in \left(\GF{2}^n\right)^2,\\
\label{eq:trace_inner_product}
\langle \mb{u}, \mb{v} \rangle_{\tr} &\defeq \mb{u} \mb{v}^{\msf{H}} + \mb{v} \mb{u}^{\msf{H}} \in \GF{2}\ \forall \mb{u}, \mb{v} \in \GF{4}^n,\\
\label{eq:Hermitian_inner_product}
\langle \mb{u}, \mb{v} \rangle_{\herp} &\defeq \mb{u}^{\msf{H}} \mb{v} \in \GF{4}\ \forall \mb{u}, \mb{v} \in \GF{4}^n.
\end{align}
\defend
\end{definition}

It is easy to check that for any $\cP, \cQ \in \cGbar_n$, we have 
\begin{equation}
\label{eq:commutativity_S_inner_prod}
\Cal{PQ} = (-1)^{\langle \tell(\cP), \tell(\cQ) \rangle_{\syp}}\Cal{QP} 
= (-1)^{\langle \Rm{L}(\cP), \Rm{L}(\cQ) \rangle_{\tr}}\Cal{QP}.
\end{equation}
Hence 
$\cP$ and $\cQ$ commute if and only if $\big\langle \tell(\cP), \tell(\cQ) \big\rangle_{\syp} = \big\langle \Rm{L}(\cP), \Rm{L}(\cQ) \big\rangle_{\tr} = 0$, while
$\cP$ and $\cQ$ anti-commute if and only if $\big\langle \tell(\cP), \tell(\cQ) \big\rangle_{\syp} = \big\langle \Rm{L}(\cP), \Rm{L}(\cQ) \big\rangle_{\tr} = 1$.

\begin{definition}[Normalizer Label Code]
For a stabilizer QECC $\cC$ with its stabilizer $\cS$ and stabilizer label code $\ctB \defeq \tell(\cS)$ or $\cB \defeq \Rm{L}(\cS)$, its normalizer label code is defined as the dual code of the stabilizer label code, i.e.,
\begin{align}
\ctB^{\perp_{\SP}} &\defeq \left\{ \mb{u}\in \left(\GF{2}^n\right)^2\ \middle|\ \langle \mb{u}, \mb{v} \rangle_{\SP}=0\ \forall \mb{v} \in \ctB\right\}, \text{ or }\\
\cB^{\perp_{\tr}} &\defeq \Bigl\{ \mb{u}\in \GF{4}^n\ \Bigm|\ \langle \mb{u}, \mb{v} \rangle_{\tr}=0\ \forall \mb{v} \in \cB\Bigr\},
\end{align}
respectively.
When the inner product is clear from the context, we simply omit $\SP$ and $\tr$.
\defend
\end{definition}

The relation of a stabilizer QECC and its stabilizer and normalizer label codes can be summarized as follows according to~\cite{calderbank1998quantum}.
For a stabilizer QECC $\cC$ with its stabilizer $\cS$ and stabilizer label code $\ctB$ or $\cB$, its normalizer label code $\tilde{\codeB}^{\perp_{\msf{S}}}$ or $\codeB^{\perp_{\msf{T}}}$, we have 
$\ctB$ is self-orthogonal under the symplectic inner product and 
$\cB$ is self-orthogonal under the trace inner product, i.e., 
\begin{equation}
\ctB \subseteq \ctB^{\perp_{\msf{S}}} \text{ and } \cB \subseteq \cB^{\perp_{\msf{T}}}.
\end{equation}
If $\cB$ is a linear code over $\GF{4}$, then $\cB$ is self-orthogonal under the Hermitian inner product.

Errors that differ by a stabilizer element have the same action on the code space. Consequently, degenerate decoding seeks the most probable equivalence class of errors modulo the stabilizer, conditioned on the measured syndrome. After choosing a reference error consistent with that syndrome, these equivalence classes can be indexed by the quotient $\cB^\perp/\cB$.
\subsection{Encoding and Decoding Circuits of Stabilizer QECCs}
\label{sec:enc_dec:2d}
In this section, we discuss the Clifford group and its isomorphic vector space with a redefined operation, which is related to finding possible encoding and decoding circuits of stabilizer QECCs and which is also about how to decompose a quantum-code NFG into basic blocks corresponding to some basic quantum gates.

From the encoding perspective, for a stabilizer QECC $\cC$ there is a unitary transformation $\Cal{V}:\ \mbb{C}^{2^n}\rightarrow \mbb{C}^{2^n}$ such that 
\begin{equation}
\label{eq:QECC_def2}
\cC \defeq \left\{ |\overline{\psi} \rangle = \Cal{V}\left( | 0 \rangle^{\otimes n-k} \otimes  | \psi \rangle \right) \ \middle|\ | \psi \rangle \in \mbb{C}^{2^k}\right\}.
\end{equation}
For every stabilizer QECC, an encoding unitary $\Cal{V}$ can be chosen from the Clifford group, whose elements normalize the n-qubit Pauli group:
$\Cal{V}\mathcal{G}_n \Cal{V}^{\mathrm H}
=\{\Cal{V}\Cal{P}\Cal{V}^{\mathrm H}\mid \Cal{P}\in \Cal{G}_n\}
= \Cal{G}_n$.
Throughout this section, Clifford unitaries that differ only by a global phase are identified; thus, $\mathrm{Cl}_n$ denotes the Clifford group modulo global phases.

\begin{definition}[Encoding Matrix]
For each Clifford transformation $\Cal{V}$ on $n$ qubits, its associated encoding matrix $V$ is defined as a $2n \times 2n$ binary matrix such that 
\begin{equation}
\label{eq:encoding_matrix}
\tell(\Cal{V} \cP \Cal{V}^{\msf{H}}) = \tell(\cP) V\ \forall \cP \in \cGbar_n.
\end{equation}
The $j^{\Rm{th}}$ and $(j+n)^{\Rm{th}}$ rows of $V$ are obtained by substituting $\cP = X_j$ and $\cP = Z_j$ into~\eqref{eq:encoding_matrix}, respectively, i.e., 
\begin{align}
\mb{v}_j &= \tell(X_j) V = \tell( \Cal{V} X_j \Cal{V}^{\msf{H}} ),\\
\mb{v}_{j+n} &= \tell(Z_j) V = \tell( \Cal{V} Z_j \Cal{V}^{\msf{H}} ).
\end{align}
\defend
\end{definition}

\begin{theorem}
A matrix $V \in \GF{2}^{2n \times 2n}$ is an encoding matrix of some Clifford transformation $\Cal{V}$ acting on the Pauli group $\cG_n$ if and only if $V\in \Rm{Sp}_{2n}(\GF{2})$, where $\Rm{Sp}_{2n}(\GF{2})$ is the symplectic group, i.e., 
$\Rm{Sp}_{2n}(\GF{2}) \defeq \left\{ V \ \middle|\ V \mb{\Lambda}_n V^{\msf{T}} = \mb{\Lambda}_n \text{ over }\GF{2} \right\}$.
\end{theorem}
\begin{proof}
See~\cite{poulin2009quantum}.
\end{proof}

Only considering the symplectic group is not enough to have an isomorphism of the Clifford group. 
We need to consider the signs of Pauli operators as well.
\begin{definition}[Signed encoding matrix]
The signed encoding matrix of $\Cal{V}$ is defined to be the matrix $\hat{V}\in\GF{2}^{2n\times 2n}\times \mbb{Z}_4^{2n}$ such that 
\begin{equation*}
\hat{\mb{v}}_j \defeq \hat{\ell}\left(\Cal{V} X_j \Cal{V}^{\HP}\right) \text{ and }
\hat{\mb{v}}_{j+n} \defeq \hat{\ell}\left(\Cal{V} Z_j \Cal{V}^{\HP}\right)\ \forall j,
\end{equation*}
where $\hat{\ell}: \cG_n \rightarrow \GF{2}^{2n} \times \mbb{Z}_4$ is the signed label mapping
\begin{equation*}
\hat{\ell}(\cP) \defeq [x_1,\ldots,x_n,z_1,\ldots,z_n\,|\,r] \text{ s.t. }\cP={\msf{i}}^r \prod_{i=1}^{n}X_i^{x_i} Z_i^{z_i}.
\end{equation*}
\defend
\end{definition}

Now we introduce a function $p$ which is used to define the new operation.
Let the function $p: \left(\GF{2}^{2n}\right)^2 \rightarrow \mbb{Z}_{4}$ be defined as
\begin{equation*}
p\left([\mb{x}_1, \mb{z}_1], [\mb{x}_2, \mb{z}_2]\right) \defeq \mb{z}_1\mb{x}_2^{\tr}\in\mbb{Z}_{4}\ \forall [\mb{x}_1, \mb{z}_1], [\mb{x}_2, \mb{z}_2]\in\left(\GF{2}^{n}\right)^2.
\end{equation*}
Note that $p$ is the partial standard symplectic inner product.

\begin{lemma}
\label{lm:addition_binary_rep}
For any $\cP_1, \ldots, \cP_m \in \cG_n$, 
suppose $[\mb{v}_i \,|\, r_i] \defeq \hat{\ell}(\cP_i)\ \forall i$.
Then we have 
\begin{equation}
\label{eq:signed_add}
\hat{\ell}(\cP_1 \cdots \cP_m)  = 
\left[\sum_{i}\mb{v}_i \,\middle|\, \sum_{i}r_i + 2\sum_{j,k:\, j<k} p(\mb{v}_j,\mb{v}_k)\right].
\end{equation}
\end{lemma}
\begin{proof}
We first show that~\eqref{eq:signed_add} holds for $n=1$.
Now suppose $n=1$ and $[x_i,z_i \,|\, r_i] \defeq \hat{\ell}(\cP_i)\ \forall i$.
Then we have 
\begin{align*}
\cP_1 \cdots \cP_m &= (\msf{i}^{r_1} X^{x_1} Z^{z_1}) \cdots (\msf{i}^{r_m} X^{x_m} Z^{z_m})  \\
&= \msf{i}^{\sum_i r_i} (-1)^{\sum_{j,k:\,j<k} z_j x_k} X^{\sum_i x_i} Z^{\sum_i z_i}\\
&= \msf{i}^{\sum_i r_i+2 \sum_{j,k:\,j<k} p(\mb{v}_j, \mb{v}_k)} X^{\sum_i x_i} Z^{\sum_i z_i}.
\end{align*}
Therefore, we have shown that \eqref{eq:signed_add} holds for $n=1$, i.e., 
$\hat{\ell}(\cP_1 \cdots \cP_m)  = \left[\sum_i x_i \,\middle|\, \sum_i z_i, \sum_{i}r_i + 2\sum_{j,k:\, j<k} p(\mb{v}_j,\mb{v}_k)\right]$.
The above proof can be extended to show that~\eqref{eq:signed_add} holds for any positive integer $n$.
\end{proof}

\begin{theorem}
\label{thm:clifford_transformation_property}
The set of signed encoding matrices for Clifford transformations on $n$ qubits is 
\begin{equation*}
\hat{G}_n \defeq \Big\{ [V\,|\, \mb{r}^{\tr}]\in\Rm{Sp}_{2n}(\GF{2}) \times \mbb{Z}_4^{2n}\ \Big|\ 
r_i \equiv p(\mb{v}_i, \mb{v}_i)\ ( \text{mod } 2)\ \forall i \Big\}.
\end{equation*}
Note that $r_i$ and $p(\mb{v}_i, \mb{v}_i)$ are in $\mbb{Z}_4$ and they are the same in $\GF{2}$ and hence $r_i=p(\mb{v}_i, \mb{v}_i)+2\tilde{r}_i$ for some $\tilde{r}_i\in\GF{2}$.
The sign of the Pauli operator $\cP$ such that $\hat{\ell}(\cP) = [\mb{v}_i\,|\, r_i]$ is $\msf{i}^{r_i-p(\mb{v}_i, \mb{v}_i)}$ or $(-1)^{\tilde{r}_i}$ and $[V\, |\, \tilde{\mb{r}}^{\tr} ]$ is called a tableau. 
\end{theorem}

\begin{proof}
Since $V$ is the unsigned encoding matrix, $V$ is from the symplectic group.
Suppose $\cP_j \defeq \Cal{V} X_j \Cal{V}^{\HP}$ and $\cP_{j+n} \defeq \Cal{V} Z_j \Cal{V}^{\HP}$ for $1\leq j\leq n$.
Then we have $[\mb{v}_i \,|\, r_i] = \hat{\ell}(\cP_i)$.
By Lemma~\ref{lm:addition_binary_rep}, we have 
$\mb{0} = \hat{\ell}(I) = \hat{\ell}(\cP_i^2) = [2\mb{v}_i \,|\, 2 r_i + 2 p(\mb{v}_i,\mb{v}_i)]$, 
which implies $2 r_i + 2 p(\mb{v}_i,\mb{v}_i) = 0 \in\mbb{Z}_4$,
which is equivalent to $ r_i \equiv p(\mb{v}_i,\mb{v}_i)\ ( \text{mod } 2)$.
If two Clifford unitaries $\Cal{U}$ and $\Cal{W}$ have the same signed encoding matrix, then $\Cal{W}^{\mathrm{H}}\Cal{U}$ commutes with every Pauli generator $X_j$ and $Z_j$. Since these generators generate the full matrix algebra, $\Cal{W}^{\mathrm{H}}\Cal{U}$ is a scalar multiple of the identity. Thus, $\Cal{U}$ and $\Cal{W}$ differ only by a global phase, and the signed encoding matrix uniquely determines an element of $\mathrm{Cl}_n$.
The number of Clifford transformations on $n$ qubits and the size of $\hat{G}_n$ are both $2^{n^2+2n}\prod_{i=1}^{n}(2^{2i}-1)$~\cite{aaronson2004improved, maslov2018shorter}.
Consequently, the set of signed encoding matrices is exactly $\widehat G_n$, and the correspondence between $\mathrm{Cl}_n$ and $\widehat G_n$ is bijective. With the composition law specified in Theorem~\ref{thm:closing_box_matrix_op}, this correspondence is a group isomorphism.
\end{proof}

The operator on $\hat{G}_n$ is defined in the following theorem.
\begin{theorem}
\label{thm:closing_box_matrix_op}
Let $\Cal{U}$ and $\Cal{V}$ be two Clifford transformations with signed encoding matrices $\hat{U} = [U \,|\, \mb{r}^{\tr}]$ and $\hat{V} = [V \,|\, \mb{s}^{\tr}]$, respectively, 
where $U, V\in\GF{2}^{2n\times 2n}$ and $\mb{r},\mb{s}\in\mbb{Z}_4^{2n}$. 
Then the signed encoding matrix of $\Cal{W}\defeq\Cal{VU}$ is $\left[W \,|\, \mb{t}^{\tr}\right]$, where 
\begin{equation}
\label{eq:closing_box_matrix_op}
W \defeq UV \text{ and } 
t_i  \defeq r_i + \mb{u}_i \mb{s}^{\tr} + \mb{u}_i M \mb{u}_i^{\tr} - \mb{u}_i \mb{d}^{\tr},
\end{equation}
where 
$\mu_{i,j} \defeq p (\mb{v}_{i^*}, \mb{v}_{j^*} ), i^*\defeq\min\{i,j\}, j^*\defeq\max\{i,j\}$, and 
$d_i \defeq \mu_{i,i}$. 
\end{theorem}

\begin{proof}
It is easy to verify $W = UV$ by considering the equivalence classes of Pauli operators without signs.
Since $\mb{w}_i = \sum_{j=1}^{n}u_{i,j}\mb{v}_j\ \forall i$, let $V'$ be a matrix of size $\left(\sum_{j}u_{i,j}\right) \times (2n+1)$ such that there are $u_{i,j}$ copies of $\mb{v}_j$ $\forall j$, which are listed as rows of $V'$ and then by~\eqref{eq:signed_add} we have 
\begin{align*}
t_i &= r_i + \sum_{j}u_{i,j}s_j   + 2\sum_{j,k:\,j<k} u_{i,j}u_{i,k} p (\mb{v}_{j}, \mb{v}_k) \\
&+ 2\sum_{j}\frac{u_{i,j}(u_{i,j}-1)}{2} p (\mb{v}_j,\mb{v}_j)\\
&= r_i + \mb{u}_i \mb{s}^{\tr} + \sum_{j,k} u_{i,j} u_{i,k} \mu_{j,k} - \sum_{j} u_{i,j} \mu_{j,j}\\
&= r_i + \mb{u}_i \mb{s}^{\tr} + \mb{u}_i M \mb{u}_i^{\tr} - \mb{u}_i \mb{d}^{\tr}\in\mbb{Z}_4.
\end{align*}
\end{proof}

\begin{remark}
Note that~\eqref{eq:closing_box_matrix_op} in Theorem~\ref{thm:closing_box_matrix_op} is well defined for the following reasons. 
\begin{itemize}
\item If we add $2$ to any entry of $V$, i.e., $v_{i_0,j_0} \leftarrow v_{i_0,j_0}+2 $, then we have 
$p(\mb{v}_i,\mb{v}_j) \leftarrow p(\mb{v}_i,\mb{v}_j) + 2 c_{i,j}$ for all $i,j$, where $c_{i,j}$ is some constant integer, and hence 
$t_i\in\mbb{Z}_4$ in~\eqref{eq:closing_box_matrix_op} is not changed for all $i$ according to~\eqref{eq:signed_add}.

\item If we add $2$ to any entry of $U$, i.e., $u_{i_0,j_0} \leftarrow u_{i_0,j_0} + 2 $, then we have 
$t_i \leftarrow t_i$ for $i \neq i_0$ and 
\begin{align*}
t_{i_0} \leftarrow & \ t_{i_0} +2\mb{e}_{j_0} \mb{s}^{\tr} + 2 \mb{e}_{j_0} M \mb{u}_{i_0}^{\tr} + 2 \mb{u}_{i_0} M \mb{e}_{j_0}^{\tr}\\ 
&+ 4 \mb{e}_{j_0} M \mb{e}_{j_0}^{\tr} - 2\mb{e}_{j_0} \mb{d}^{\tr}\\
\stackrel{(a)}{=} & \ t_{i_0} + (2 s_{j_0} + 2 m_{j_0,j_0}) + 4 \mb{e}_{j_0} M \mb{u}_{i_0}^{\tr} 
\stackrel{(b)}{=} t_{i_0} \in \mbb{Z}_4,
\end{align*}
where $\mb{e}_{j_0}$ is the $(j_0)^{\Rm{th}}$ row of $\mb{I}_{2n}$, 
$(a)$ follows from the fact that $A$ is symmetric, and 
$(b)$ follows from Theorem~\ref{thm:clifford_transformation_property}.
\end{itemize}
\end{remark}

\begin{example}
The signed encoding matrices of the Clifford transformations corresponding to the Hadamard, phase, and CNOT gates are, respectively,
\begin{align*}
\left[\begin{array}{cc|c}
0 & 1 & 0\\
1 & 0 & 0
\end{array}\right],\ 
\left[\begin{array}{cc|c}
1 & 1 & 1\\
0 & 1 & 0
\end{array}\right],\text{ and }
\left[\begin{array}{cccc|c}
1 & 1 & 0 & 0 & 0\\
0 & 1 & 0 & 0 & 0\\
0 & 0 & 1 & 0 & 0\\
0 & 0 & 1 & 1 & 0
\end{array}\right].
\end{align*}
The signed encoding matrix of the Clifford transformation consisting a phase gate followed by a Hadamard gate is 
$\left[\begin{array}{cc|c}
1 & 1 & 3 \\
1 & 0 & 0 
\end{array}\right]$ by Theorem~\ref{thm:closing_box_matrix_op}.
The signs of signed encoding matrix can be tracked using Theorem~\ref{thm:closing_box_matrix_op} when applying the three basic quantum gates, i.e., the Hadamard, phase, and CNOT gates.
\exampleend
\end{example}
\section{Normal Factor Graphs}
\label{sec:NFG:3}

\begin{figure}
  \begin{center}
    \begin{tikzpicture}[nodes= {draw=none, fill=none}, 
						vertex/.style={rectangle, fill = white, draw, minimum height=0.5cm, minimum width=1.2cm},
						 labeled edge/.style={ sloped, midway, above=0cm},
						 nearend edge/.style={ sloped, near end, above=0cm},
						node distance=1.9cm]
\pgfmathsetmacro\r{2};
\pgfdeclarelayer{main}
\pgfdeclarelayer{back}
\pgfsetlayers{back,main}
\begin{pgfonlayer}{main}
\node  (m1) {};
\node [vertex] (f1) [right of = m1] {$\Cal{B}_1$};
\node [vertex] (f2) [right of = f1]{$\Cal{B}_2$};
\node  (m2) [right of = f2]{};

\path[draw, thick] (m1) -- (f1) node[labeled edge] {$T_1$};
\path[draw, thick] (f1) -- (f2) node[labeled edge] {$M$};
\path[draw, thick] (f2) -- (m2) node[labeled edge] {$T_2$};
\end{pgfonlayer}

\end{tikzpicture}
  \end{center}
  \caption{Quantum-code NFG $\cG_{\cB}$ for the code $\codeB$.}
  \label{fig:NFG_eg1}
\end{figure}

\begin{figure}
  \begin{center}
    \begin{tikzpicture}[nodes= {draw=none, fill=none}, 
						vertex/.style={rectangle, fill = white, draw, minimum height=0.5cm, minimum width=1.2cm},
						 labeled edge/.style={ sloped, midway, above=0cm},
						 nearend edge/.style={ sloped, near end, above=0cm},
						node distance=1.9cm]
\pgfmathsetmacro\r{2};
\pgfdeclarelayer{main}
\pgfdeclarelayer{back}
\pgfsetlayers{back,main}
\begin{pgfonlayer}{main}
\node  (m1) {};
\node [vertex] (f1) [right of = m1] {$\Cal{B}_1^{\perp}$};
\node [vertex] (f2) [right of = f1]{$\Cal{B}_2^{\perp}$};
\node  (m2) [right of = f2]{};

\path[draw, thick] (m1) -- (f1) node[labeled edge] {$\hat{T}_1$};
\path[draw, thick] (f1) -- (f2) node[labeled edge] {$\hat{M}$};
\path[draw, thick] (f2) -- (m2) node[labeled edge] {$\hat{T}_2$};
\end{pgfonlayer}

\end{tikzpicture}
  \end{center}
  \caption{Quantum-code NFG $\cG_{\hat{\cB}}$ for the code $\hat{\codeB}$.}
  \label{fig:NFG_eg1_dual}
\end{figure}
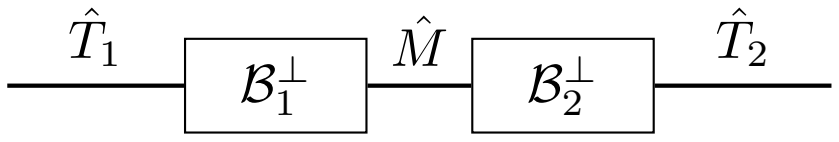

In this section, we use normal factor graphs (NFGs) to express our results, i.e., representing codes and probability for general stabilizer QECCs. 
\subsection{Quantum-code NFG Representation}
\label{sec:QCNFG:3a}
Fig.~\ref{fig:NFG_eg1} shows a type of NFG we use to represent stabilizer label codes of stabilizer QECCs which we call the quantum-code NFG representation of stabilizer QECCs. 
\begin{itemize}
\item All variables are associated with full edges and half edges whose alphabet is $\GF{2}^2$ or $\GF{4}$.
A full edge represents a variable that appears as arguments in two functions, 
while a half edge represents a variable that is argument of only one function.
\item The local functions are the indicator functions of some codes,
e.g., the local function for the function node whose label is $\codeB_1$ is $f_1(\vt_1, \vm) \defeq \big[(\vt_1, \vm) \in \codeB_1\big]$.
\item The global function of the NFG is 
\begin{equation}
f(\vt_1, \vt_2, \vm) \defeq f_1(\vt_1, \vm) \cdot f_2(\vt_2, \vm).
\end{equation}
The code represented by the NFG is obtained by restricting on the half edges of the global function, 
e.g., the code $\cB$ represented by the NFG $\cG_{\cB}$ is
\begin{equation}
\cB \defeq \big\{ \mb{t}\ \big|\ \text{there exists } \mb{m} \text{ s.t. } f(\mb{t}, \mb{m})=1 \big\}.
\end{equation}
\end{itemize}

The NFG $\cG_{\hat{\cB}}$ in Fig.~\ref{fig:NFG_eg1_dual} 
represents the code $\hat{\cB}$,
where the code $\cB_i$ is replaced by its dual code $\cB_i^{\perp}$ for $i=1,2$.
Let the code $\cB^{\perp}$ be the the dual code of $\cB$.
The duality is defined w.r.t. the symplectic inner product if the codes are over $\GF{2}^2$, 
and it is defined w.r.t. the Hermitian inner product if the codes are over $\GF{4}$.
We can show $\hat{\cB} = \cB^{\perp}$ when the alphabets are over $\GF{2}^2$.
The proof of the case when the alphabets are over $\GF{4}$ is similar.
Note that this is the case when there are two blocks in the NFG of the duality result in~\cite{vontobel2008stabilizer} and its proof can be easily extended to the general case using mathematical induction.
The following theorem completes the proof of Theorem 3 of~\cite{vontobel2008stabilizer} by showing $\cB^{\perp} \subseteq \hat{\cB}$.

\begin{theorem}
\label{thm:selforthogonal_thm}
With the above definitions, 
given two binary codes $\cB_1$ and $\cB_2$ which are self-orthogonal under the symplectic inner product, 
let the code $\cB$ be the code represented by the NFG $\cG_{\cB}$ in Fig.~\ref{fig:NFG_eg1} and the code $\hat{\cB}$ be the code represented by the NFG $\cG_{\hat{\cB}}$ in Fig.~\ref{fig:NFG_eg1_dual}, respectively.
Then we have $\hat{\cB} = \cB^{\perp}$,
where $\cB^{\perp}$ is the dual code of $\cB$ under the symplectic inner product.\footnote{The proof in~\cite{vontobel2008stabilizer} is incomplete as only $\hat{\cB} \subseteq \cB^{\perp}$ is shown and the reverse part $\cB^{\perp} \subseteq \ctB$ is not shown.}
\end{theorem}
\begin{proof}
See Appendix~\ref{appendix:thm1}.
\end{proof}

The above theorem can be easily generalized to the case when there are more than two $\cB_i$'s and when $\cB_i$'s are additive code over $\GF{4}$. 
The generalized version of Theorem~\ref{thm:selforthogonal_thm} could imply the following properties of quantum-code NFGs:
\begin{itemize}
\item The stabilizer label code represented by a NFG like Fig.~\ref{fig:NFG_eg1} is guaranteed to be self-orthogonal given that  the local codes, codes represented by the local functions, are all self-orthogonal. 
\item The corresponding normalizer label code, which the dual code of the stabilizer label code, is represented by a NFG having the same structure with each local code replaced by the dual code of the original local code.
\end{itemize}

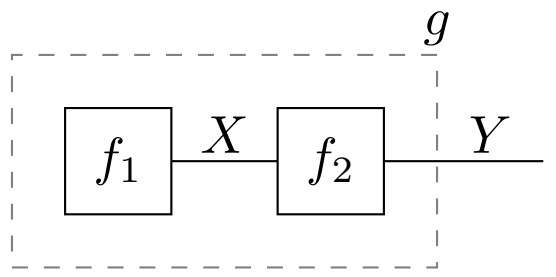
\begin{figure}
  \begin{center}
    \begin{tikzpicture}[x=1cm, y= 1cm, node/.style={draw=none, inner sep = 0pt},
                          block/.style={draw,minimum height = 50pt,minimum width = 30pt},
                          block3/.style={draw,minimum height = 60pt, minimum width = 20pt},
                          block2/.style={draw, minimum height = 40pt, minimum width = 20pt},
                          dashcblock/.style={draw=gray, dashed, fill = none, minimum height = 40pt, minimum width = 80pt, inner sep=0pt},
                          blockw4/.style={draw,fill = white, minimum height = 40pt, minimum width = 20pt, inner sep=0pt},
                          blockw3/.style={draw,fill = white, minimum height = 30pt, minimum width = 20pt, inner sep=0pt},
                          blockw2/.style={draw, fill = white, minimum height = 20pt, minimum width = 20pt, inner sep=0pt},
                          blockw1/.style={draw, fill = white, minimum height = 20pt, minimum width = 28pt, inner sep=0pt},
                          eqblock/.style={draw, fill = white, minimum height = 10pt, minimum width = 10pt, inner sep=0pt},
                          factor/.style={draw,minimum size = 15pt,inner sep = 0pt}
                          ]
	
\draw (0pt, 0pt) -- (80pt, 0pt);

\node[blockw2] at (0pt, 0pt) {};
\node at (0pt, 0pt) {$f_1$};
\node[blockw2] at (40pt, 0pt) {};
\node at (40pt, 0pt) {$f_2$};
\node at (60pt, 25pt) {$g$};
\node at (20pt, 5pt) {$X$};
\node at (70pt, 5pt) {$Y$};

\node[dashcblock] at (20pt, 0pt) {};
\end{tikzpicture}

  \end{center}
  \caption{Closing-the-box and opening-the-box operations.}
  \label{fig:NFG_eg2_OTB_CTB}
\end{figure}

Here we introduce two useful operations, i.e., closing-the-box and opening-the-box operations, which is illustrated by a small example in Fig.~\ref{fig:NFG_eg2_OTB_CTB}.
Closing-the-box is an operation by replacing the functions in the box using a single function whose arguments are the half edges.
For the example in Fig.~\ref{fig:NFG_eg2_OTB_CTB}, closing-the-box is done by replacing the functions $f_1$ and $f_2$ inside the dashed box by the function $g$, where 
\begin{equation*}
g(y) \defeq \sum_{x} f_1(x)f_2(x,y),
\end{equation*} 
where the functions $f_1$ and $f_2$ are local functions, while the global function $g$ is called the partition function. 
Whereas, opening-the-box is the reversed operation by replacing the global function $g$ by the local functions $f_1$ and $f_2$ for this example.

In the following two sections, we discuss the quantum-code NFG representations of a general stabilizer QECC $\cC$ based on two cases. 
If the quaternary stabilizer label code $\cB$ of $\cC$ is not linear over $\GF{4}$, then we consider the binary stabilizer label code $\ctB$ which is a linear code over $\GF{2}$. 
Otherwise we consider the quaternary linear code $\ctB$ directly.

\subsection{Quantum-code NFG representations of stabilizer QECCs with binary linear stabilizer label codes}
\label{sec:QCNFG_binary:3b}
In this section, we discuss the quantum-code NFG representations of a stabilizer QECC $\cC$ based on its binary stabilizer label code $\ctB$. 
W.l.o.g., suppose the generator matrix $\tilde{G}$ of $\ctB$ is in the standard form~\cite[Section 10.5.7]{nielsen2010quantum}, where some CNOT gates may be applied to permute qubits, i.e.,
\begin{equation*}
\tilde{G} = 
\left[\begin{array}{ccc:ccc}
\mb{I}_{r} & A_1 & A_2 & B & \mb{0} & C\\
\mb{0} & \mb{0} & \mb{0} & D & \mb{I}_{n-k-r} & E
\end{array}\right],
\end{equation*}
where 
$A_1 \in \GF{2}^{r \times (n-k-r)}$, 
$A_2 \in \GF{2}^{r \times k}$,
$B \in \GF{2}^{r \times r}$,
$C \in \GF{2}^{r \times k}$,
$D \in \GF{2}^{(n-k-r) \times r}$, and 
$E \in \GF{2}^{(n-k-r) \times k}$.

First, we can find possible encoding and decoding circuits for $\cC$ consisting only Hadamard, phase, and CNOT gates.
\begin{theorem}
\label{thm:enc_dec_circuit_binary_case}
There exists a decoding circuit for $\cC$ consisting 6 rounds in the sequence C-P-C-P-H-C, where H, P, and C stand for a series of Hadamard, phase, and CNOT gates, respectively. 
In the end, some $X$ (=$HP^2H$) gates may be needed to correct phases of Pauli operators.
A possible encoding circuit is obtained by reversing the order of gates in the decoding circuit with suitable sign correction in the beginning.
\end{theorem}

\begin{proof}
Here we ignore the phases of Pauli operators.
The signs of signed encoding matrix can be tracked using Theorem~\ref{thm:closing_box_matrix_op} and corrected using $X$ gates in the end.
The procedure of finding a decoding circuit is as follows. 
\begin{enumerate}
\item Use CNOT gates to eliminate $A_1$ and $A_2$ to obtain
\begin{align*}
\left[\begin{array}{ccc:ccc}
\mb{I}_{r} & \mb{0} & \mb{0} & B_1 & \mb{0} & C\\
\mb{0} & \mb{0} & \mb{0} & D_1 & \mb{I}_{n-k-r} & E
\end{array}\right].
\end{align*}
By the self-orthogonality of $\ctB$, we have $B_1^{\tr} + B_1 = \mb{0}$, i.e., $B_1$ is symmetric.

\item Apply phase gates to obtain
\begin{align*}
\left[\begin{array}{ccc:ccc}
\mb{I}_{r} & \mb{0} & \mb{0} & B_1+\Lambda & \mb{0} & C\\
\mb{0} & \mb{0} & \mb{0} & D_1 & \mb{I}_{n-k-r} & E
\end{array}\right],
\end{align*}
where $\Lambda$ is a diagonal matrix such that $B_1 + \Lambda = M M^{\tr}$ for some invertible binary $M$ by Lemma 7 of~\cite{aaronson2004improved}.

\item Apply CNOT gates to map $\mb{I}_r$ to $M$ and $B_1 + \Lambda$ to $M M^{\tr} \left( M^{\tr} \right)^{-1} = M $, respectively. 
Then we obtain
\begin{align*}
\left[\begin{array}{ccc:ccc}
M & \mb{0} & \mb{0} & M & \mb{0} & C\\
\mb{0} & \mb{0} & \mb{0} & D_2 & \mb{I}_{n-k-r} & E
\end{array}\right].
\end{align*}

\item Apply phase gates to the first $r$ qubits to obtain
\begin{align*}
\left[\begin{array}{ccc:ccc}
M & \mb{0} & \mb{0} & \mb{0} & \mb{0} & C\\
\mb{0} & \mb{0} & \mb{0} & D_2 & \mb{I}_{n-k-r} & E
\end{array}\right].
\end{align*}

\item Apply Hadamard gates to the first $r$ qubits and obtain
\begin{align*}
\left[\begin{array}{ccc:ccc}
\mb{0} & \mb{0} & \mb{0} & M & \mb{0} & C\\
D_2 & \mb{0} & \mb{0} & \mb{0} & \mb{I}_{n-k-r} & E
\end{array}\right].
\end{align*}

\item Apply CNOT gates to diagonalize $M$ and obtain
\begin{align*}
\left[\begin{array}{ccc:ccc}
\mb{0} & \mb{0} & \mb{0} & \mb{I}_r & \mb{0} & C\\
D_3 & \mb{0} & \mb{0} & \mb{0} & \mb{I}_{n-k-r} & E
\end{array}\right].
\end{align*}
By the self-orthogonality of $\ctB$, we have $D_3 = \mb{0}$.

\item Use CNOT gates to eliminate $C$ and $E$ and obtain
\begin{align*}
\left[\begin{array}{ccc:ccc}
\mb{0} & \mb{0} & \mb{0} & \mb{I}_r & \mb{0} & \mb{0}\\
\mb{0} & \mb{0} & \mb{0} & \mb{0} & \mb{I}_{n-k-r} & \mb{0}
\end{array}\right].
\end{align*}
\end{enumerate}
\end{proof}

\begin{corollary}
There exists a decoding circuit for $\cC$ consisting 5 rounds in the sequence C-CZ-P-H-C, where CZ stands for the controlled-Z gate.
\end{corollary}
\begin{proof}
By Theorem 2 of~\cite{maslov2018shorter}, we have P-C-P-C = P-CZ-C which implies C-P-C-P = C-CZ-P.
\end{proof}

The procedure of finding possible encoding and decoding circuits corresponding to a specific Clifford transformation that stabilized a certain quantum state can be found in~\cite{aaronson2004improved,maslov2018shorter}. 
For a stabilizer QECC $\cC$, there are multiple Clifford transformations that can stabilize all the quantum state in $\cC$ and we just need find a possible one.
Therefore, it is possible that we can have fewer rounds in Theorem~\ref{thm:enc_dec_circuit_binary_case} compared with results of procedures in~\cite{aaronson2004improved,maslov2018shorter}. 

\begin{figure}
  \begin{center}
    \begin{tikzpicture}[x=1cm, y= 1cm, node/.style={draw=none, inner sep = 0pt},
                          block/.style={draw,fill=white,minimum height = 10pt,minimum width = 10pt},
                          block2/.style={draw,fill=white,minimum height = 20pt,minimum width = 30pt},
                          eqblock/.style={draw, fill = white, minimum height = 10pt, minimum width = 10pt, inner sep=0pt},
                          factor/.style={draw,minimum size = 15pt,inner sep = 0pt}
                          ]
\def\xx{0pt}
\def\ya{0pt}
\def\yb{\ya-14pt}
\def\yc{\yb-14pt}
\def\yd{\yc-14pt}
\def\ye{\yd-14pt}

\def\xaa{\xx+20pt}
\def\xa{\xaa+20pt}
\def\xb{\xa+10pt}
\def\xc{\xb+10pt}
\def\xd{\xc+10pt}
\def\xe{\xd+20pt}
\def\xf{\xe+20pt}
\def\xg{\xf+20pt}
\def\xh{\xg+10pt}
\def\xi{\xh+20pt}
\def\xj{\xi+20pt}
\def\xk{\xj+10pt}
\def\xl{\xk+10pt}
\def\xm{\xl+10pt}
\def\xxx{\xm+20pt}

\draw (\xx, \ya) 	-- (\xxx, \ya);
\draw (\xx, \yb) 	-- (\xxx, \yb);
\draw (\xx, \yc) 	-- (\xxx, \yc);
\draw (\xx, \yd) 	-- (\xxx, \yd);
\draw (\xx, \ye) 	-- (\xxx, \ye);

\node[block] at (\xaa, \yb) {}; 
\node at (\xaa, \yb) {X};
\node[block] at (\xaa, \yc) {}; 
\node at (\xaa, \yc) {X};
\node[draw,circle,inner sep=1pt,fill] at (\xa, \ye) {}; 
\draw (\xa, \ye) -- (\xa, \yd);
\node at (\xa, \yd) {$\oplus$};
\node[draw,circle,inner sep=1pt,fill] at (\xb, \ye) {}; 
\draw (\xb, \ye) -- (\xb, \ya);
\node at (\xb, \ya) {$\oplus$};
\node[draw,circle,inner sep=1pt,fill] at (\xc, \ya) {}; 
\draw (\xc, \ya) -- (\xc, \yc);
\node at (\xc, \yc) {$\oplus$};
\node[draw,circle,inner sep=1pt,fill] at (\xd, \ya) {}; 
\draw (\xd, \ya) -- (\xd, \yb);
\node at (\xd, \yb) {$\oplus$};
\node[block] at (\xe, \ya) {}; 
\node at (\xe, \ya) {H};
\node[block] at (\xe, \yb) {}; 
\node at (\xe, \yb) {H};
\node[block] at (\xe, \yc) {}; 
\node at (\xe, \yc) {H};
\node[block] at (\xf, \ya) {}; 
\node at (\xf, \ya) {P};
\node[block] at (\xf, \yb) {}; 
\node at (\xf, \yb) {P};
\node[block] at (\xf, \yc) {}; 
\node at (\xf, \yc) {P};
\node[draw,circle,inner sep=1pt,fill] at (\xg, \yc) {}; 
\draw (\xg, \yc) -- (\xg, \ya);
\node at (\xg, \ya) {$\oplus$};
\node[draw,circle,inner sep=1pt,fill] at (\xh, \yb) {}; 
\draw (\xh, \yb) -- (\xh, \ya);
\node at (\xh, \ya) {$\oplus$};
\node[block] at (\xi, \ya) {}; 
\node at (\xi, \ya) {P};
\node[block] at (\xi, \yb) {}; 
\node at (\xi, \yb) {P};
\node[draw,circle,inner sep=1pt,fill] at (\xj, \yc) {}; 
\draw (\xj, \yc) -- (\xj, \yd);
\node at (\xj, \yd) {$\oplus$};
\node[draw,circle,inner sep=1pt,fill] at (\xk, \yb) {}; 
\draw (\xk, \yb) -- (\xk, \ye);
\node at (\xk, \ye) {$\oplus$};
\node[draw,circle,inner sep=1pt,fill] at (\xl, \ya) {}; 
\draw (\xl, \ya) -- (\xl, \ye);
\node at (\xl, \ye) {$\oplus$};
\node[draw,circle,inner sep=1pt,fill] at (\xm, \ya) {}; 
\draw (\xm, \ya) -- (\xm, \yd);
\node at (\xm, \yd) {$\oplus$};

\def\yg{\ya+12pt}
\draw[fill=none, draw=black, dashed] (\xaa-7pt, \ya+7pt) rectangle 	(\xaa+7pt, \ye-7pt);
\node at (\xaa, \yg) {$\Cal{X}_1$};
\draw[fill=none, draw=black, dashed] (\xa-7pt, \ya+7pt) rectangle 		(\xd+7pt, \ye-7pt);
\node at (\xb+5pt, \yg) {$\Cal{C}_1$};
\draw[fill=none, draw=black, dashed] (\xe-7pt, \ya+7pt) rectangle 		(\xe+7pt, \ye-7pt);
\node at (\xe, \yg) {$\Cal{H}_1$};
\draw[fill=none, draw=black, dashed] (\xf-7pt, \ya+7pt) rectangle 		(\xf+7pt, \ye-7pt);
\node at (\xf, \yg) {$\Cal{P}_1$};
\draw[fill=none, draw=black, dashed] (\xg-7pt, \ya+7pt) rectangle 		(\xh+7pt, \ye-7pt);
\node at (\xg+5pt, \yg) {$\Cal{C}_2$};
\draw[fill=none, draw=black, dashed] (\xi-7pt, \ya+7pt) rectangle 		(\xi+7pt, \ye-7pt);
\node at (\xi, \yg) {$\Cal{P}_2$};
\draw[fill=none, draw=black, dashed] (\xj-7pt, \ya+7pt) rectangle 		(\xm+7pt, \ye-7pt);
\node at (\xk+5pt, \yg) {$\Cal{C}_3$};
	
\draw[fill=none, draw=black, dashed] (\xaa-10pt, \yg+7pt) rectangle (\xxx-10pt, \ye-10pt);
\node at (\xxx-5pt, \yg+7pt) {$\Cal{V}$};
\end{tikzpicture}
  \end{center}
  \caption{The encoding circuit for the stabilizer QECC in Example~\ref{eg:Enc_eg1}.}
  \label{fig:QCircuit3}
\end{figure}
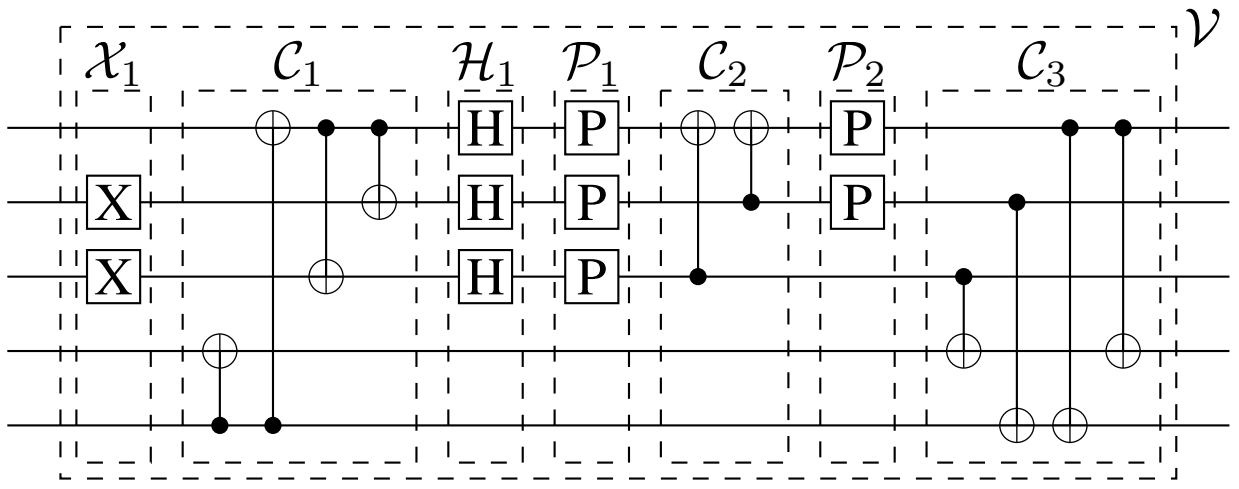

\begin{example}
\label{eg:Enc_eg1}
Consider a $\llbracket 5,1,3 \rrbracket$ QECC $\cC$ and its binary stabilizer label code $\ctB$ with a generator matrix
\begin{equation*}
\left[\begin{array}{ccccc : ccccc}
1 & 0 & 0 & 1 & 1 & 1 & 0 & 1 & 0 & 1\\
0 & 1 & 0 & 0 & 1 & 1 & 1 & 1 & 0 & 0\\
0 & 0 & 1 & 1 & 0 & 1 & 1 & 1 & 0 & 0\\
0 & 0 & 0 & 0 & 0 & 0 & 1 & 1 & 1 & 1
\end{array}\right].
\end{equation*}
Using the procedure described in the proof of Theorem~\ref{thm:enc_dec_circuit_binary_case}, we can obtain a possible encoding circuit corresponding to the Clifford transformation $\Cal{V}$ in Fig.~\ref{fig:QCircuit3}, where the quantum gates and circuit symbols follow from~\cite{nielsen2010quantum} and phase gate is denoted by ``P'' instead of ``S'' here. 
The corresponding encoding matrix is 
\begin{equation*}
V = \left[\begin{array}{ccccc : ccccc}
0 & 0 & 0 & 0 & 0 & 1 & 0 & 0 & 0 & 0\\
0 & 0 & 0 & 0 & 0 & 0 & 1 & 0 & 0 & 0\\
0 & 0 & 0 & 0 & 0 & 0 & 0 & 1 & 0 & 0\\
0 & 0 & 0 & 1 & 0 & 0 & 0 & 0 & 0 & 0\\
0 & 0 & 0 & 1 & 1 & 1 & 0 & 0 & 0 & 0\\
1 & 0 & 0 & 1 & 1 & 1 & 0 & 1 & 0 & 1\\
0 & 1 & 0 & 0 & 1 & 1 & 1 & 1 & 0 & 0\\
0 & 0 & 1 & 1 & 0 & 1 & 1 & 1 & 0 & 0\\
0 & 0 & 0 & 0 & 0 & 0 & 1 & 1 & 1 & 1\\
0 & 0 & 0 & 0 & 0 & 1 & 1 & 0 & 0 & 1
\end{array}\right].
\end{equation*}
\exampleend
\end{example}

\begin{figure}
\begin{center}
\begin{tikzpicture}[x=1cm, y= 1cm, node/.style={draw=none},
                          block/.style={draw,minimum height = 50pt,minimum width = 30pt},
                          block1/.style={draw,fill=white,minimum height = 50pt,minimum width = 16pt},
                          code/.style={draw,minimum height = 20pt,minimum width = 20pt},
                          factor/.style={draw,minimum size = 15pt,inner sep = 0pt},
                          dashblock/.style={draw=gray, dashed, fill = none, minimum height = 60pt, minimum width = 80pt, inner sep=0pt}
                          ]

\def\x{0pt}
\def\y{0pt}
\def\ya{\y+15pt}
\def\yb{\y-15pt}

\def\xx{\x+20pt}
\def\xxa{\xx-10pt}
\def\xxb{\xx+10pt}
\node[code] at (\xx, \ya) {$\ctB_1$};
\node[code] at (\xx, \yb) {$\ctB_2$};

\def\xa{\xxb+8pt+10pt}
\def\xaa{\xa-8pt}
\def\xab{\xa+8pt}
\node[block1] at (\xa, \y) {};
\node at (\xa, \y) {$\tilde{\Cal{X}}_1$};
\draw (\xxb,\ya) -- (\xaa, \ya);
\draw (\xxb,\yb) -- (\xaa, \yb);

\def\xb{\xab+8pt+10pt}
\def\xba{\xb-8pt}
\def\xbb{\xb+8pt}
\node[block1] at (\xb, \y) {};
\node at (\xb, \y) {$\tilde{\Cal{C}}_1$};
\draw (\xab,\y) -- (\xba, \y);

\def\xc{\xbb+8pt+10pt}
\def\xca{\xc-8pt}
\def\xcb{\xc+8pt}
\node[block1] at (\xc, \y) {};
\node at (\xc, \y) {$\tilde{\Cal{H}}_1$};
\draw (\xbb,\y) -- (\xca, \y);

\def\xd{\xcb+8pt+10pt}
\def\xda{\xd-8pt}
\def\xdb{\xd+8pt}
\node[block1] at (\xd, \y) {};
\node at (\xd, \y) {$\tilde{\Cal{P}}_1$};
\draw (\xcb,\y) -- (\xda, \y);

\def\xe{\xdb+8pt+10pt}
\def\xea{\xe-8pt}
\def\xeb{\xe+8pt}
\node[block1] at (\xe, \y) {};
\node at (\xe, \y) {$\tilde{\Cal{C}}_2$};
\draw (\xdb,\y) -- (\xea, \y);

\def\xf{\xeb+8pt+10pt}
\def\xfa{\xf-8pt}
\def\xfb{\xf+8pt}
\node[block1] at (\xf, \y) {};
\node at (\xf, \y) {$\tilde{\Cal{P}}_2$};
\draw (\xeb,\y) -- (\xfa, \y);

\def\xg{\xfb+8pt+10pt}
\def\xga{\xg-8pt}
\def\xgb{\xg+8pt}
\node[block1] at (\xg, \y) {};
\node at (\xg, \y) {$\tilde{\Cal{C}}_3$};
\draw (\xfb,\y) -- (\xga, \y);
\draw (\xgb,\y) -- (\xgb+30pt, \y);

\draw[dashed] (\xaa-5pt, \ya+15pt) rectangle (\xgb+5pt,\yb-15pt);
\node at (\xgb+12pt, \ya) {$\ctB_3$};
\draw[dashed] (\xxa-5pt, \ya+18pt) rectangle (\xgb+20pt,\yb-18pt);
\node at (\xgb+25pt, \ya+10pt) {$\ctB$};
\end{tikzpicture}
\end{center}
\caption{Quantum-code NFG representation of a stabilizer QECC $\cC$ based on its binary stabilizer label code $\ctB$, 
where 
$\ctB_1 \defeq \GF{2}\text{-rowspan}([\mb{0}\ \mb{I}_{n-k}])$,
$\ctB_2 \defeq\{\mb{0}\}$,  
$\ctB_3 \defeq \GF{2}\text{-rowspan}([\mb{I}_{2n}\ |\ V])$, 
$V$ is the encoding matrix and ``$|$'' is used to group variables on the left and right sides. 
}
\label{fig:NFG_eg3_QC_QECC}
\end{figure}
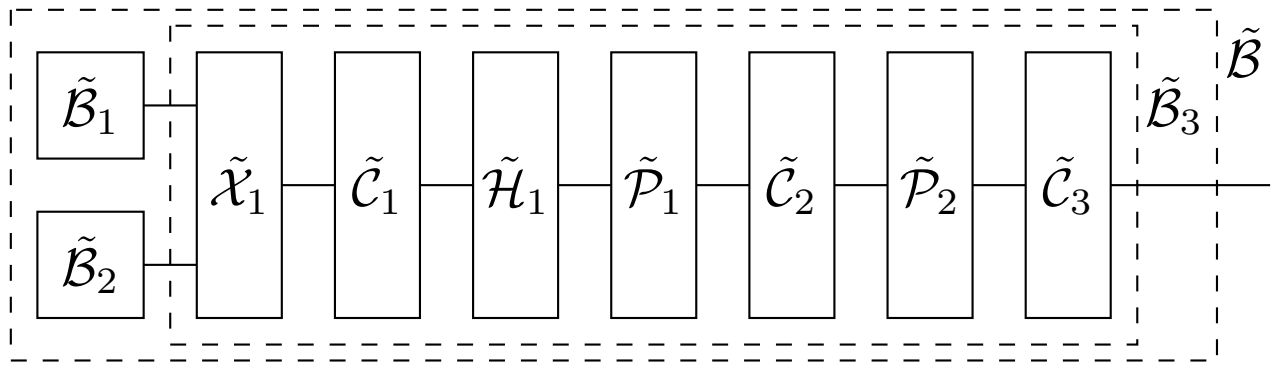

Fig.~\ref{fig:NFG_eg3_QC_QECC} shows the quantum-code NFG representation of the stabilizer QECC $\cC$ with its binary stabilizer label code $\ctB$.
For example, for the stabilizer QECC $\cC$ in Example~\ref{eg:Enc_eg1} we group the same type of gates together and get Clifford transformations $\Cal{X}_1, \cC_1, \ldots, \cC_3$ in Fig.~\ref{fig:QCircuit3}.
Then the codes in Fig.~\ref{fig:NFG_eg3_QC_QECC} are defined in the following way, e.g., $\tilde{\Cal{X}}_1 \defeq \GF{2}\text{-rowspan}([\mb{I}_{2n}\ |\ X_1])$, where $X_1$ is the encoding matrix of $\Cal{X}_1$.

With closing-the-box operation, the global function is the indicator function of $\ctB$ and by Theorem~\ref{thm:selforthogonal_thm} we have $\ctB$ is self-orthogonal under the symplectic inner product since $\ctB_1$, $\ctB_2$, and $\ctB_3$ are all self-orthogonal under the symplectic inner product, i.e., $[\mb{0}\ \mb{I}_{n-k}]\Lambda_{n-k}[\mb{0}\ \mb{I}_{n-k}]^{\tr} = \mb{0}$ and $\mb{I}_{2n}\Lambda_{2n}\mb{I}_{2n}^{\tr} + V \Lambda_{2n}V^{\tr} = \Lambda_{2n}+ \Lambda_{2n} = \mb{0}$ over $\GF{2}$.

With opening-the-box operation, the codes $\tilde{\Cal{X}}_1, \tilde{\Cal{C}}_1, \ldots, \tilde{\Cal{C}}_3$ can be decomposed into the three basic codes with generator matrix $[\mb{I}\ |\ V_0]$, where $V_0$ is the encoding matrix of a Clifford transformation consisting only a single Hadamard, phase, or CNOT gate.

\subsection{Quantum-code NFG representations of stabilizer QECCs with quaternary linear stabilizer label codes}
\label{sec:QCNFG_quaternary:3b}
In this section, we discuss how to decompose the quantum-code NFGs of stabilizer QECCs whose stabilizer label code is linear over $\GF{4}$ into basic blocks representing quaternary linear codes.  

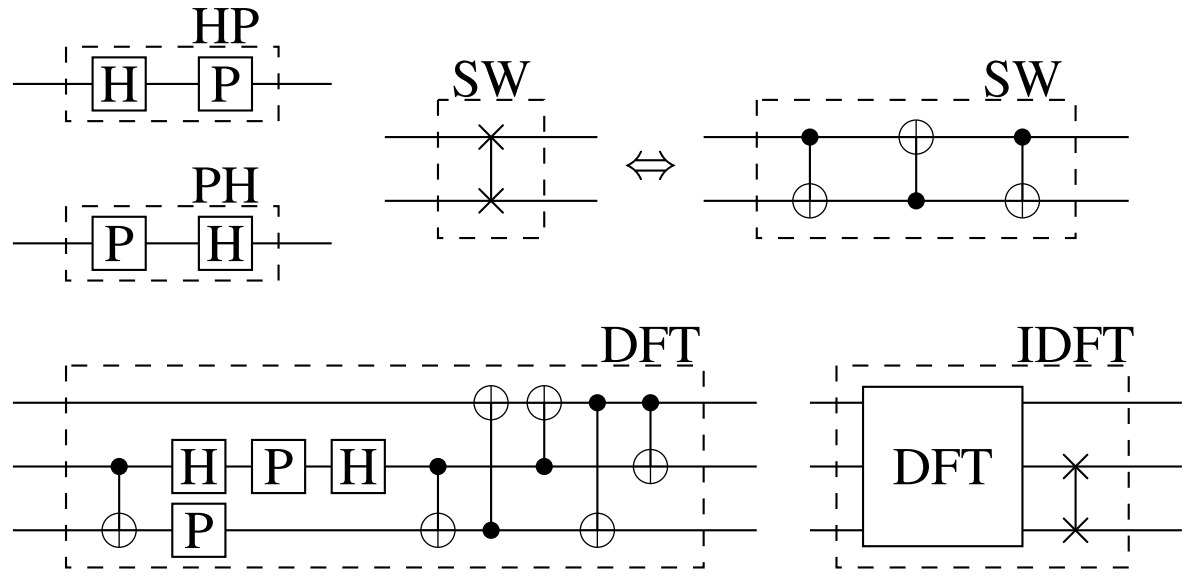
\begin{figure}
  \begin{center}
    \begin{tikzpicture}[x=1cm, y= 1cm, node/.style={draw=none, inner sep = 0pt},
                          block/.style={draw,fill=white,minimum height = 10pt,minimum width = 10pt},
                          block2/.style={draw,fill=white,minimum height = 30pt,minimum width = 30pt},
                          eqblock/.style={draw, fill = white, minimum height = 10pt, minimum width = 10pt, inner sep=0pt},
                          factor/.style={draw,minimum size = 15pt,inner sep = 0pt}
                          ]
\def\x{0pt}
\def\y{0pt}
\def\ya{\y-30pt}

\def\xa{\x+20pt}
\def\xb{\xa+20pt}
\draw (\xa-20pt,\y) -- (\xb+20pt,\y);
\node[block] at (\xa, \y) {};
\node at (\xa, \y) {H};
\node[block] at (\xb, \y) {};
\node at (\xb, \y) {P};
\draw[fill=none, draw=black, dashed] (\xa-10pt, \y+7pt) rectangle (\xb+10pt, \y-7pt);
\node at (\xb,\y+11pt) {HP};

\def\yaa{\y-30pt}
\def\aa{\x+20pt}
\def\ab{\aa+20pt}
\draw (\aa-20pt, \yaa) -- (\ab+20pt, \yaa);
\node[block] at (\aa, \yaa) {};
\node at (\aa, \yaa) {P};
\node[block] at (\ab, \yaa) {};
\node at (\ab, \yaa) {H};
\draw[fill=none, draw=black, dashed] (\aa-10pt, \yaa+7pt) rectangle (\ab+10pt, \yaa-7pt);
\node at (\ab,\yaa+11pt) {PH};

\def\ya{\y-10pt}
\def\yb{\ya-12pt}
\def\ba{\xb+50pt}
\def\bb{\ba+30pt}
\def\bc{\bb+30pt}
\def\bd{\bc+20pt}
\def\be{\bd+20pt}

\draw (\ba-20pt,\ya) -- (\ba+20pt,\ya);
\draw (\ba-20pt,\yb) -- (\ba+20pt,\yb);
\node at (\ba, \ya) {$\times$};
\node at (\ba, \yb) {$\times$};
\draw (\ba,\ya) -- (\ba,\yb);
\draw[fill=none, draw=black, dashed] (\ba-10pt, \ya+7pt) rectangle (\ba+10pt, \yb-7pt);
\node at (\ba,\ya+11pt) {SW};

\node at (\bb, \ya-6pt) {$\Leftrightarrow$};

\draw (\bc-20pt,\ya) -- (\be+20pt,\ya);
\draw (\bc-20pt,\yb) -- (\be+20pt,\yb);
\node[draw,circle,inner sep=1pt,fill] at (\bc, \ya) {}; 
\draw (\bc, \ya) -- (\bc, \yb);
\node at (\bc, \yb) {$\oplus$};
\node[draw,circle,inner sep=1pt,fill] at (\bd, \yb) {}; 
\draw (\bd, \yb) -- (\bd, \ya);
\node at (\bd, \ya) {$\oplus$};
\node[draw,circle,inner sep=1pt,fill] at (\be, \ya) {}; 
\draw (\be, \ya) -- (\be, \yb);
\node at (\be, \yb) {$\oplus$};
\draw[fill=none, draw=black, dashed] (\bc-10pt, \ya+7pt) rectangle (\be+10pt, \yb-7pt);
\node at (\be,\ya+11pt) {SW};

\def\yc{\yaa-30pt}
\def\yd{\yc-12pt}
\def\ye{\yd-12pt}
\def\ca{\x+20pt}
\def\cb{\ca+15pt}
\def\cc{\cb+15pt}
\def\cd{\cc+15pt}
\def\ce{\cd+15pt}
\def\cf{\ce+10pt}
\def\cg{\cf+10pt}
\def\ch{\cg+10pt}
\def\ci{\ch+10pt}

\draw (\ca-20pt,\yc) -- (\ci+20pt,\yc);
\draw (\ca-20pt,\yd) -- (\ci+20pt,\yd);
\draw (\ca-20pt,\ye) -- (\ci+20pt,\ye);
\node[draw,circle,inner sep=1pt,fill] at (\ca, \yd) {}; 
\draw (\ca, \yd) -- (\ca, \ye);
\node at (\ca, \ye) {$\oplus$};

\node[block] at (\cb, \yd) {}; 
\node at (\cb, \yd) {H};
\node[block] at (\cb, \ye) {}; 
\node at (\cb, \ye) {P};
\node[block] at (\cc, \yd) {}; 
\node at (\cc, \yd) {P};
\node[block] at (\cd, \yd) {}; 
\node at (\cd, \yd) {H};

\node[draw,circle,inner sep=1pt,fill] at (\ce, \yd) {}; 
\draw (\ce, \yd) -- (\ce, \ye);
\node at (\ce, \ye) {$\oplus$};
\node[draw,circle,inner sep=1pt,fill] at (\cf, \ye) {}; 
\draw (\cf, \ye) -- (\cf, \yc);
\node at (\cf, \yc) {$\oplus$};
\node[draw,circle,inner sep=1pt,fill] at (\cg, \yd) {}; 
\draw (\cg, \yd) -- (\cg, \yc);
\node at (\cg, \yc) {$\oplus$};
\node[draw,circle,inner sep=1pt,fill] at (\ch, \yc) {}; 
\draw (\ch, \yc) -- (\ch, \ye);
\node at (\ch, \ye) {$\oplus$};
\node[draw,circle,inner sep=1pt,fill] at (\ci, \yc) {}; 
\draw (\ci, \yc) -- (\ci, \yd);
\node at (\ci, \yd) {$\oplus$};

\draw[fill=none, draw=black, dashed] (\ca-10pt, \yc+7pt) rectangle (\ci+10pt, \ye-7pt);
\node at (\ci,\yc+11pt) {DFT};

\def\da{\ci+20pt+15pt+20pt}
\def\db{\da+15pt+10pt}

\draw (\da-25pt,\yc) -- (\db+20pt,\yc);
\draw (\da-25pt,\yd) -- (\db+20pt,\yd);
\draw (\da-25pt,\ye) -- (\db+20pt,\ye);
\node[block2] at (\da,\yd) {};
\node at (\da,\yd) {DFT};
\node at (\db, \yd) {$\times$};
\node at (\db, \ye) {$\times$};
\draw (\db,\yd) -- (\db,\ye);
\draw[fill=none, draw=black, dashed] (\da-20pt, \yc+7pt) rectangle (\db+10pt, \ye-7pt);
\node at (\db,\yc+11pt) {IDFT};

\end{tikzpicture}
  \end{center}
  \caption{Basic quantum gates for linear stabilizer circuit}
  \label{fig:linear_gates}
\end{figure}

\begin{definition}[linear stabilizer circuit]
Any circuit consisting entirely HP, SW, DFT, their inverses, i.e., PH, SW and IDFT, respectively, which are defined in Fig.~\ref{fig:linear_gates}, and measurement gates is called a linear stabilizer circuit. 
Applying HP on qubit $a$, SW on qubits $a$ and $b$, DFT on qubits $a$, $b$, and $c$, are abbreviated as $\Rm{HP}_a$, $\Rm{SW}_{a,b}$, and $\Rm{DFT}_{a,b,c}$, respectively.

The HP, PH, SW, DFT and IDFT gates represent linear quaternary codes with generators, respectively, 
\begin{align*}
G_{\Rm{HP}} &\defeq \left[\begin{array}{c|c} 1 & \om \end{array}\right],\ 
G_{\Rm{PH}} \defeq \left[\begin{array}{c|c} 1 & \ob \end{array}\right],\\
G_{\Rm{SW}} &\defeq \left[\begin{array}{cc|cc} 
1 & 0 & 0 & 1\\
0 & 1 & 1 & 0
\end{array}\right],\\
G_{\Rm{DFT}} &\defeq \left[\begin{array}{ccc|ccc} 
1 & 0 & 0 & 1 & 1 & 1\\
0 & 1 & 0 & 1 & \om & \ob\\
0 & 0 & 1 & 1 & \ob & \om
\end{array}\right],\\
G_{\Rm{IDFT}} &\defeq \left[\begin{array}{ccc|ccc} 
1 & 0 & 0 & 1 & 1 & 1\\
0 & 1 & 0 & 1 & \ob & \om\\
0 & 0 & 1 & 1 & \om & \ob
\end{array}\right].
\end{align*}
\defend
\end{definition}

\begin{theorem}
For any stabilizer QECC $\cC$ with a linear quaternary stabilizer label code $\cB$, there exists a linear stabilizer circuit as encoding or decoding circuit for $\cC$.
\end{theorem}

\begin{proof}
We can use Algorithm~\ref{alg:Gen_dec_circuit_linear_case} to obtain a decoding circuit which is a linear stabilizer circuit.
\end{proof}

\begin{algorithm}
\caption{Algorithm to to obtain a decoding circuit which is a linear stabilizer circuit.
Signs should be tracked and corrected in the end.}
\label{alg:Gen_dec_circuit_linear_case}
\begin{algorithmic}[1]
\REQUIRE the generator matrix of $\cB$ is $G = \left[\begin{array}{cc} \mb{I}_{(n-k)/2} & A \end{array}\right]$, where $A\in\GF{4}^{(n-k)/2 \times (n+k)/2}$.
\ENSURE  a list of quantum gates for a possible decoding circuit 
\FOR {$A$ is not diagonalized}
\STATE Compute the index of the first nonzero entry of the $j^{\Rm{th}}$ column of $A$ is 
$I(j) \defeq \min \left\{i \ |\ a_{i,j}\neq 0\right\} $ and $\Cal{J}_i \defeq \{ j \ |\ I(j) = i\}$, which is the set of index of columns with leading non-zero entry in the $i^{\Rm{th}}$ row.
\STATE Apply SW gates to permute columns of $G$ to have $\{i_j\}_{j=1}^{(n+k)/2}$ in non-decreasing order.
\STATE Apply HP and PH gates to have $a_{i_j,j} = 1$ $\forall j$.
\STATE For each $i$, divide $\Cal{J}_i$ into sets of size $3$ with possible one item left and apply DFT gates on each set of qubits with size $3$.
Repeat this step until $|\Cal{J}_i|<3$ or $|\Cal{J}_i|$ does not decrease by applying DFT gates for all $i$.
\ENDFOR
\end{algorithmic}
\end{algorithm}

We use the following example to show how to generate a decoding circuit which is a linear stabilizer circuit.
\begin{example}
Consider a $\llbracket12, 4\rrbracket$ QECC $\cC$ with a linear quaternary stabilizer label code $\cB$ whose generator matrix is
\begin{equation*}
G = \left[\begin{array}{cccc cc cccccc}
1 & 0 & 0 & 0 & 		1 & 1 & 	1 & 	1 & 	1 & 	0 & 0 & 		0\\
0 & 1 & 0 & 0 & 		0 & 1 & 	1 & 	\om & \om & 1 & 1 & 		1\\
0 & 0 & 1 & 0 & 		1 & 0 & 	\om& 1 & 	\om & 1 & \om & 	\ob\\
0 & 0 & 0 & 1 & 		1 & \om &0 & 	\om & 1 & 	1 & \ob & 	\om
\end{array}\right].
\end{equation*}
\begin{itemize}
\item Step 1 and 2 are skipped. 
For step 3, we have $\Cal{J}_1 = \{5,6,7,8,9\}$ and $\Cal{J}_2 = \{10,11,12\}$.
Apply $\Rm{DFT}_{5,6,7}$ and $\Rm{DFT}_{10,11,12}$ to obtain
\begin{equation*}
G_1 = \left[\begin{array}{cccc cc cccccc}
1 & 0 & 0 & 0 &   1 & 	0 & 	0 &    1 & 		1 & 		0 & 		0 & 		0\\
0 & 1 & 0 & 0 &   0 & 	1 & 	1 &    \om & 		\om &		1 & 		0 & 		0\\
0 & 0 & 1 & 0 &   \ob& 	0 & 	\om& 1 & 		\om & 	0 & 		0 & 		1\\
0 & 0 & 0 & 1 &   \ob& 	\om& 	0 &    \om & 		1 & 		0 & 		1 & 		0
\end{array}\right].
\end{equation*}

\item Update $\Cal{J}_i$: 
$\Cal{J}_1=\{5,8,9\}$, 
$\Cal{J}_2=\{6,7,10\}$, 
$\Cal{J}_3=\{12\}$, and 
$\Cal{J}_4=\{11\}$. 
Apply $\Rm{DFT}_{5,8,9}$ and $\Rm{DFT}_{6,7,10}$ to obtain
\begin{equation*}
G_2 = \left[\begin{array}{cccc cc cccccc}
1 & 0 & 0 & 0 &   1 & 	0 & 	0 &    0 & 		0 & 		0 & 		0 & 		0\\
0 & 1 & 0 & 0 &   0 & 	1 & 	0 &    \om & 		\om &		0 & 		0 & 		0\\
0 & 0 & 1 & 0 &   0 & 	\om & \ob& 	0 & 		\ob & 		1 & 		0 & 		1\\
0 & 0 & 0 & 1 &   0 & 	\om& 	\om&  \ob & 		0 & 		\om & 	1 & 		0
\end{array}\right].
\end{equation*}

\item Repeat step 1 and 2. 
Apply $\Rm{SW}_{7,9}$, $\Rm{SW}_{11,12}$, $\Rm{PH}_7$, $\Rm{PH}_8$, and $\Rm{HP}_9$,
to obtain
\begin{equation*}
G_3 = \left[\begin{array}{cccc cc cccccc}
1 & 0 & 0 & 0 &   1 & 	0 & 	0 &    0 & 		0 & 		0 & 		0 & 		0\\
0 & 1 & 0 & 0 &   0 & 	1 & 	1 &    1 & 		0 &		0 & 		0 & 		0\\
0 & 0 & 1 & 0 &   0 & 	\om & \om& 	0 & 		1 & 		1 & 		1 & 		0\\
0 & 0 & 0 & 1 &   0 & 	\om& 	0&  	\om& 		\ob& 		\om & 	0 & 		1
\end{array}\right].
\end{equation*}

\item Apply $\Rm{DFT}_{6,7,8}$ and $\Rm{DFT}_{9,10,11}$
to obtain
\begin{equation*}
G_4 = \left[\begin{array}{cccc cc cccccc}
1 & 0 & 0 & 0 &   1 & 	0 & 	0 &    0 & 		0 & 		0 & 		0 & 		0\\
0 & 1 & 0 & 0 &   0 & 	1 & 	0 &    0 & 		0 &		0 & 		0 & 		0\\
0 & 0 & 1 & 0 &   0 & 	0 &     \ob& 	1 & 		1 & 		0 & 		0 & 		0\\
0 & 0 & 0 & 1 &   0 & 	0 & 	1 &  	\ob& 		1 & 		0 & 		\om& 		1
\end{array}\right].
\end{equation*}

\item Apply $\Rm{SW}_{7,9}$, $\Rm{SW}_{10,12}$, $\Rm{HP}_9$, and $\Rm{PH}_{11}$
to obtain
\begin{equation*}
G_5 = \left[\begin{array}{cccc cc cccccc}
1 & 0 & 0 & 0 &   1 & 	0 & 	0 &    0 & 		0 & 		0 & 		0 & 		0\\
0 & 1 & 0 & 0 &   0 & 	1 & 	0 &    0 & 		0 &		0 & 		0 & 		0\\
0 & 0 & 1 & 0 &   0 & 	0 &     1 & 	1 & 		1 & 		0 & 		0 & 		0\\
0 & 0 & 0 & 1 &   0 & 	0 & 	1 &  \ob& 		\om& 		1 & 		1& 		0
\end{array}\right].
\end{equation*}

\item Apply $\Rm{DFT}_{7,8,9}$ and $\Rm{DFT}_{8,10,11}$
to obtain
\begin{equation*}
G_6 = \left[\begin{array}{cccc cc cccccc}
1 & 0 & 0 & 0 &   1 & 	0 & 	0 &    	0 & 		0 & 		0 & 		0 & 		0\\
0 & 1 & 0 & 0 &   0 & 	1 & 	0 &    	0 & 		0 &		0 & 		0 & 		0\\
0 & 0 & 1 & 0 &   0 & 	0 &     1 & 	0 & 		0 & 		0 & 		0 & 		0\\
0 & 0 & 0 & 1 &   0 & 	0 & 	0 &    	1 & 		0 & 		0 & 		0& 		0
\end{array}\right].
\end{equation*}
\end{itemize}
The last step of tracking and correcting signs of Pauli operators is omitted.
For this example, instead of having $|\mb{0}\rangle$ whose stabilizers are corresponding to an additive code over $\GF{4}$ we need to have the bell pairs, whose stabilizers are corresponding to a repetition code, at the $i^{\Rm{th}}$ and $(i+4)^{\Rm{th}}$ qubits for $i=1,2,3,4$. 
\exampleend
\end{example}

We have similar results as last section where all basic blocks are representing quaternary linear codes here.
\subsection{Quantum-probability NFG Representation}
\label{sec:QPNFG:3d}
In this section, we use NFGs to represent probability of stabilizer QECCs and follow the notation of~\cite{loeliger2017factor}.
First, we have a brief review of the basics of quantum-probability NFGs.

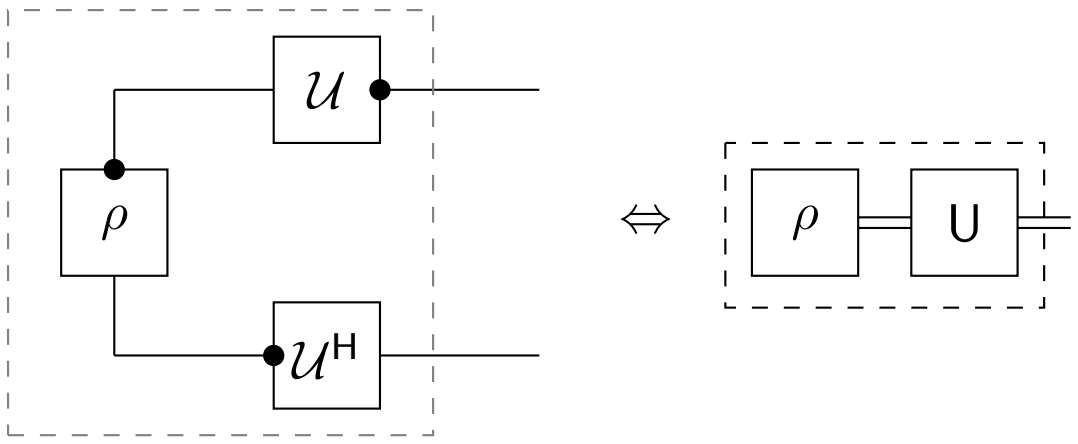
\begin{figure}
  \begin{center}
    \begin{tikzpicture}[x=1cm, y= 1cm, node/.style={draw=none, inner sep = 0pt},
					   block/.style={draw,minimum height = 50pt,minimum width = 30pt},
                          block3/.style={draw,minimum height = 60pt, minimum width = 20pt},
                          block2/.style={draw, minimum height = 40pt, minimum width = 20pt},
                          dashcblock/.style={draw=gray, dashed, fill = none, minimum height = 80pt, minimum width = 80pt, inner sep=0pt},
                          blockw4/.style={draw,fill = white, minimum height = 40pt, minimum width = 20pt, inner sep=0pt},
                          blockw3/.style={draw,fill = white, minimum height = 30pt, minimum width = 20pt, inner sep=0pt},
                          blockw2/.style={draw, fill = white, minimum height = 20pt, minimum width = 20pt, inner sep=0pt},
                          blockw1/.style={draw, fill = white, minimum height = 20pt, minimum width = 28pt, inner sep=0pt},
                          eqblock/.style={draw, fill = white, minimum height = 10pt, minimum width = 10pt, inner sep=0pt},
                          factor/.style={draw,minimum size = 15pt,inner sep = 0pt}
                          ]
	
\draw (0pt, 0pt) -- (0pt, -50pt);
\draw (0pt, 0pt) -- (80pt, 0pt);
\draw (0pt, -50pt) -- (80pt, -50pt);

\node[blockw2] at (0pt, -25pt) {};
\node at (0pt, -25pt) {$\rho$};
\node[blockw2] at (40pt, 0pt) {};
\node at (40pt, 0pt) {$\Cal{U}$};
\node[blockw2] at (40pt, -50pt) {};
\node at (40pt, -50pt) {$\Cal{U}^{\herp}$};
\fill (50pt, 0pt) circle[radius=2pt];
\fill (30pt, -50pt) circle[radius=2pt];
\fill (0pt, -15pt) circle[radius=2pt];

\node[dashcblock] at (20pt, -25pt) {};

\node at (100pt, -25pt) {$\Leftrightarrow$};

\node[blockw2] at (130pt, -25pt) {$\rho$};
\draw (140pt, -24pt) -- (150pt, -24pt);
\draw (140pt, -26pt) -- (150pt, -26pt);
\node[blockw2] at (160pt, -25pt) {$\msf{U}$};
\draw (170pt, -24pt) -- (180pt, -24pt);
\draw (170pt, -26pt) -- (180pt, -26pt);
\draw[fill=none, draw=black, dashed] (130-15pt, -25pt+15pt) rectangle (175pt, -26pt-15pt);
\end{tikzpicture}
  \end{center}
  \caption{Matrix multiplication and double-edge NFG representation.}
  \label{fig:NFG_eg5_QP_UrhoUH}
\end{figure}

\begin{figure}
  \begin{center}
    \begin{tikzpicture}[x=1cm, y= 1cm, node/.style={draw=none, inner sep = 0pt},
                          block/.style={draw,fill = white,minimum height = 10pt,minimum width = 10pt},
                          block3/.style={draw,minimum height = 60pt, minimum width = 20pt},
                          block2/.style={draw, minimum height = 40pt, minimum width = 20pt},
                          dashcblock/.style={draw=gray, dashed, fill = none, minimum height = 80pt, minimum width = 120pt, inner sep=0pt},
                          dashcblocks/.style={draw=gray, dashed, fill = none, minimum height = 80pt, minimum width = 40pt, inner sep=0pt},
                          blockw4/.style={draw,fill = white, minimum height = 40pt, minimum width = 20pt, inner sep=0pt},
                          blockw3/.style={draw,fill = white, minimum height = 30pt, minimum width = 20pt, inner sep=0pt},
                          blockw2/.style={draw, fill = white, minimum height = 20pt, minimum width = 20pt, inner sep=0pt},
                          blockw1/.style={draw, fill = white, minimum height = 20pt, minimum width = 28pt, inner sep=0pt},
                          eqblock/.style={draw, fill = white, minimum height = 10pt, minimum width = 10pt, inner sep=0pt},
                          factor/.style={draw,minimum size = 15pt,inner sep = 0pt}
                          ]
	
\draw (-80pt, 0pt) -- (80pt, 0pt);
\draw (-80pt, -50pt) -- (80pt, -50pt);
\draw (-40pt, 0pt) -- (-40pt, -50pt);
\draw (0pt, 0pt) -- (0pt, -50pt);
\draw (40pt, 0pt) -- (40pt, -50pt);

\draw (-40pt, -25pt) -- (-25pt, -25pt) -- (-25pt, -80pt);
\draw (0pt, -25pt) -- (15pt, -25pt) -- (15pt, -80pt);
\draw (40pt, -25pt) -- (55pt, -25pt) -- (55pt, -80pt);

\node[blockw2] at (-40pt, 0pt) {};
\node at (-40pt, 0pt) {$M_1$};
\node[blockw2] at (-40pt, -50pt) {};
\node at (-40pt, -50pt) {$M_1^{\herp}$};
\node[block] at (-40pt, -25pt) {};
\node at (-40pt, -26pt) {$=$};
\node at (-18pt, -75pt) {$s_1$};

\node[blockw2] at (0pt, 0pt) {};
\node at (0pt, 0pt) {$M_2$};
\node[blockw2] at (0pt, -50pt) {};
\node at (0pt, -50pt) {$M_2^{\herp}$};
\node[block] at (0pt, -25pt) {};
\node at (0pt, -26pt) {$=$};
\node at (22pt, -75pt) {$s_2$};

\node[blockw2] at (40pt, 0pt) {};
\node at (40pt, 0pt) {$M_3$};
\node[blockw2] at (40pt, -50pt) {};
\node at (40pt, -50pt) {$M_3^{\herp}$};
\node[block] at (40pt, -25pt) {};
\node at (40pt, -26pt) {$=$};
\node at (62pt, -75pt) {$s_3$};

\node[dashcblock] at (0pt, -25pt) {};
\node at (85pt, -25pt) {$\Leftrightarrow$};

\draw (90pt, 0pt) -- (150pt, 0pt);
\draw (90pt, -50pt) -- (150pt, -50pt);
\draw (120pt, 0pt) -- (120pt, -50pt);
\draw (120pt, -25pt) -- (135pt, -25pt) -- (135pt, -80pt);

\node[blockw2] at (120pt, 0pt) {};
\node at (120pt, 0pt) {$M$};
\node[blockw2] at (120pt, -50pt) {};
\node at (120pt, -50pt) {$M^{\herp}$};
\node[block] at (120pt, -25pt) {};
\node at (120pt, -26pt) {$=$};
\node at (142pt, -75pt) {$\mb{s}$};

\node[dashcblocks] at (120pt, -25pt) {};
\end{tikzpicture}

  \end{center}
  \caption{Quantum measurements.}
  \label{fig:NFG_eg6_QP_Measurement}
\end{figure}
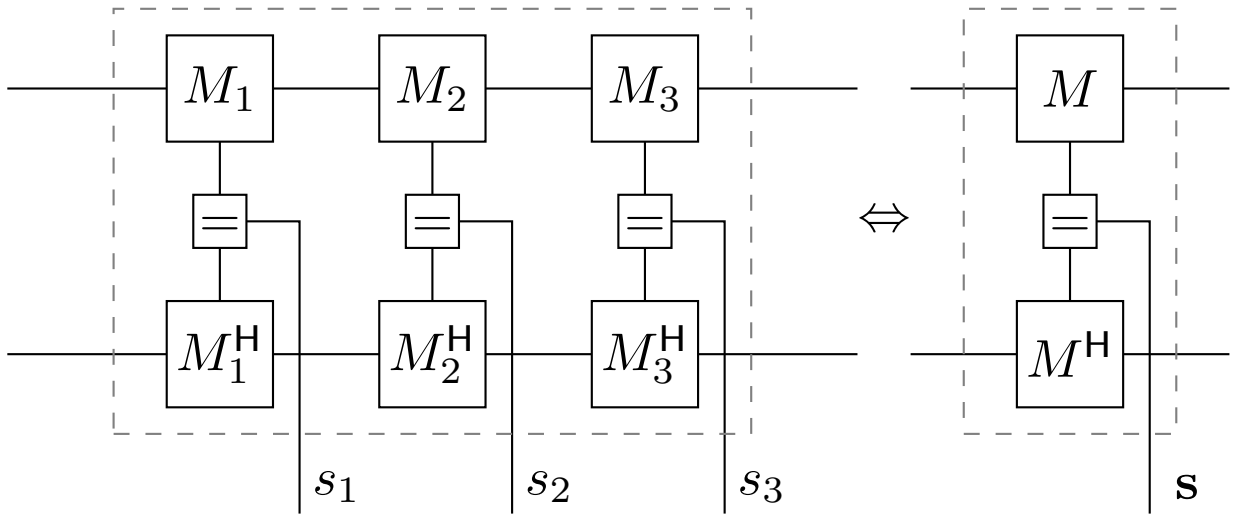

Matrix multiplication can be represented using NFGs when we view a matrix $A\in\mbb{C}^{m \times n}$ as a function
\begin{equation*}
\set{I}_{m+1}^{+} \times \set{I}_{n+1}^{+} \rightarrow \mbb{C}:\ (x,y) \mapsto A(x,y).
\end{equation*}
Fig.~\ref{fig:NFG_eg5_QP_UrhoUH} shows the matrix multiplication $\Cal{U} \rho \Cal{U}^{\herp}$, where dots indicate the first arguments of matrices, $\rho \defeq |\psi\rangle \langle \psi |$ is the density matrix for some quantum state $|\psi\rangle$, and $\Cal{U}$ is a Clifford transformation.
The equivalent double-edge representation of this matrix multiplication $\Cal{U} \rho \Cal{U}^{\herp}$ is on the right hand side, where we use $\msf{U}$ in a different font to represent the Clifford transformation $\Cal{U}$ in the double-edge NFG.

Since the generators in the stabilizer commute with each other, the ordering of the measurements of stabilizer QECCs does not matter. 
For example, the multiple measurements with their syndromes in Fig.~\ref{fig:NFG_eg6_QP_Measurement} can be represented by a general measurement $M$ and a vector syndrome $\mb{s}$,
where $M_i$ is the measurement w.r.t. to the $i^{\Rm{th}}$ generators in the stabilizer and $s_i\in\{\pm 1\}$ is the corresponding syndrome, for $i\in\{1,2,3\}$.
An equal function $f_{=}$ is omitted between each pair of $M_i$ and $M_i^{\herp}$, $i\in\{1,2,3\}$,
where $f_{=}(x_1, \ldots, x_n) \defeq [x_1 = \cdots = x_n]$, for all $n$.

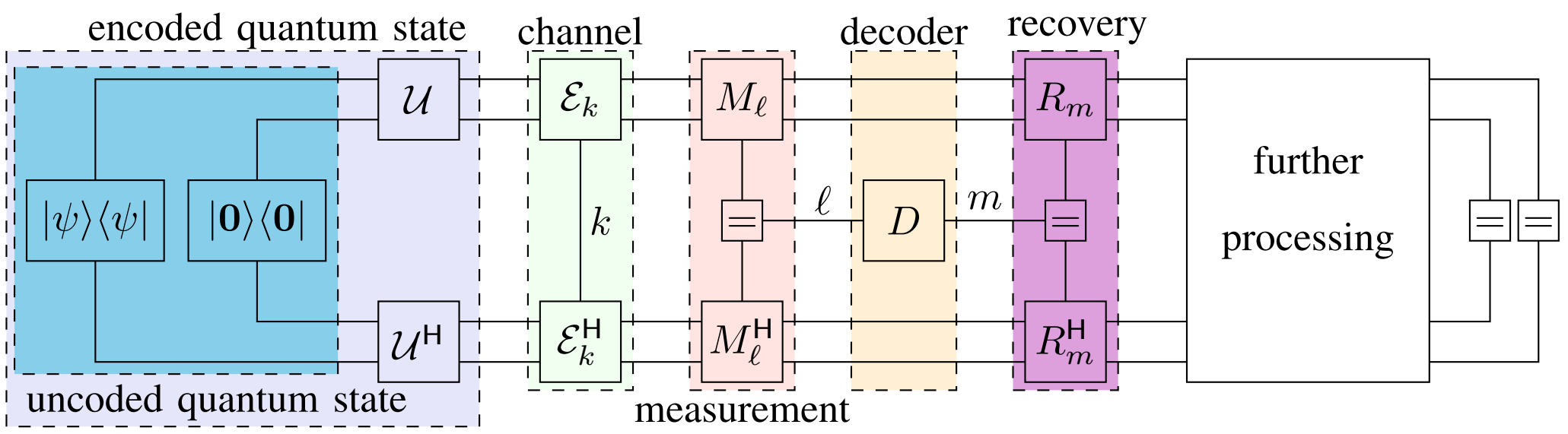
\begin{figure*}
  \begin{center}
    \begin{tikzpicture}[x=1cm, y= 1cm, node/.style={draw=none, inner sep = 0pt},
                          block/.style={draw,minimum height = 50pt,minimum width = 30pt},
                          block3/.style={draw,minimum height = 60pt, minimum width = 20pt},
                          block2/.style={draw, minimum height = 40pt, minimum width = 20pt},
                          procblock/.style={draw=gray,dashed,fill = none, minimum height = 80pt, minimum width = 60pt, inner sep=0pt},
                          sprocblock/.style={draw, fill = none, minimum height = 80pt, minimum width = 60pt, inner sep=0pt},
                          blockw4/.style={draw,fill = none, minimum height = 40pt, minimum width = 20pt, inner sep=0pt},
                          blockw3/.style={draw,fill = none, minimum height = 30pt, minimum width = 20pt, inner sep=0pt},
                          blockw2/.style={draw, fill = none, minimum height = 20pt, minimum width = 20pt, inner sep=0pt},
                          blockw1/.style={draw, fill = none, minimum height = 20pt, minimum width = 34pt, inner sep=0pt},
                          eqblock/.style={draw, fill = none, minimum height = 10pt, minimum width = 10pt, inner sep=0pt},
                          factor/.style={draw,minimum size = 15pt,inner sep = 0pt}
                          ]
\definecolor{color1}{RGB}{0,191,255} 
\definecolor{color2}{RGB}{240,255,255} 
\definecolor{color3}{RGB}{255,182,193} 
\definecolor{color4}{RGB}{255,255,224} 
\definecolor{color5}{RGB}{250,128,114} 
\definecolor{color6}{RGB}{255,165,0} 
\definecolor{color7}{RGB}{255,255,0} 
\definecolor{Lavender}{RGB}{230,230,250} 
\definecolor{SkyBlue}{RGB}{135,206,235} 
\definecolor{PapayaWhip}{RGB}{255,239,213} 
\definecolor{silver}{RGB}{192,192,192} 
\definecolor{pink}{RGB}{255,192,203} 
\definecolor{honeydew}{RGB}{240,255,240} 
\definecolor{MistyRose}{RGB}{255,228,225} 
\definecolor{plum}{RGB}{221,160,221} 

\def\x{0pt}
\def\xa{\x-10pt}
\def\xb{\xa+40pt}
\def\xc{\xb+40pt}
\def\xd{\xc+40pt}
\def\xe{\xd+40pt}
\def\xf{\xe+40pt}
\def\y{-35pt};
\def\ya{\y+30pt};
\def\yb{\y-30pt};

\filldraw[fill=Lavender, draw=black, dashed] (\xa-22pt, 7pt) rectangle (\xc+15pt, -86pt);
\node at (35pt, 12pt) {encoded quantum state};
\filldraw[fill=SkyBlue, draw=black, dashed] (\xa-20pt, 3pt) rectangle (\xb+20pt, -73pt);
\node at (20pt, -80pt) {uncoded quantum state};
\filldraw[fill=honeydew, draw=black, dashed] (97pt, 7pt) rectangle (123pt, -77pt);
\node at (110pt, 12pt) {channel};
\filldraw[fill=MistyRose, draw=black, dashed] (137pt, 7pt) rectangle (163pt, -77pt);
\node at (150pt, -82pt) {measurement};
\filldraw[fill=PapayaWhip, draw=black, dashed] (177pt, 7pt) rectangle (203pt, -77pt);
\node at (190pt, 12pt) {decoder};
\filldraw[fill=plum, draw=black, dashed] (217pt, 7pt) rectangle (243pt, -77pt);
\node at (233pt, 12pt) {recovery};

\node(rho)[blockw1] at (\xa, \y) {};
\node at (\xa, \y) {$| \psi \rangle \langle \psi |$};
\node(rho0)[blockw1] at (\xb, \y) {};
\node at (\xb, \y) {$| \mb{0} \rangle \langle \mb{0} |$};

\node(U)[blockw2] at (\xc, \ya) {};
\node at (\xc, \ya) {$\Cal{U}$};
\node(UH)[blockw2] at (\xc, \yb) {};
\node at (\xc, \yb) {$\Cal{U}^{\herp}$};
\draw (\xa, \y+10pt) -- (\xa, \ya+5pt) -- (\xc-10pt, \ya+5pt);
\draw (\xa, \y-10pt) -- (\xa, \yb-5pt) -- (\xc-10pt, \yb-5pt);
\draw (\xb, \y+10pt) -- (\xb, \ya-5pt) -- (\xc-10pt, \ya-5pt);
\draw (\xb, \y-10pt) -- (\xb, \yb+5pt) -- (\xc-10pt, \yb+5pt);

\node(E)[blockw2] at (\xd, \ya) {};
\node at (\xd, \ya) {$\Cal{E}_{k}$};
\node(EH)[blockw2] at (\xd, \yb) {};
\node at (\xd, \yb) {$\Cal{E}_{k}^{\herp}$};
\draw (\xd, \ya-10pt) -- (\xd, \yb+10pt);
\node at (\xd+5pt, \y) {$k$};
\draw (80pt, 0pt) -- (100pt, 0pt);
\draw (80pt, -10pt) -- (100pt, -10pt);
\draw (80pt, -60pt) -- (100pt, -60pt);
\draw (80pt, -70pt) -- (100pt, -70pt);

\node(M)[blockw2] at (\xe, \ya) {};
\node at (\xe, \ya) {$M_{\ell}$};
\node(MH)[blockw2] at (\xe, \yb) {};
\node at (\xe, \yb) {$M_{\ell}^{\herp}$};
\node[eqblock] at (\xe, \y) {};
\node at (150pt, -36pt) {$=$};
\node at (170pt, -30pt) {$\ell$};
\draw (120pt, 0pt) -- (140pt, 0pt);
\draw (120pt, -10pt) -- (140pt, -10pt);
\draw (120pt, -60pt) -- (140pt, -60pt);
\draw (120pt, -70pt) -- (140pt, -70pt);
\draw (150pt, -15pt) -- (150pt, -30pt);
\draw (150pt, -40pt) -- (150pt, -55pt);
\draw (155pt, -35pt) -- (180pt, -35pt);

\node(D)[blockw2] at (190pt, -35pt) {};
\node at (190pt, -35pt) {$D$};

\node[blockw2] at (230pt, -5pt) {};
\node at (230pt, -5pt) {$R_{m}$};
\node[blockw2] at (230pt, -65pt) {};
\node at (230pt, -65pt) {$R_{m}^{\herp}$};
\node[eqblock] at (230pt, -35pt) {};
\node at (230pt, -36pt) {$=$};
\node at (210pt, -30pt) {$m$};
\draw (160pt, 0pt) -- (220pt, 0pt);
\draw (160pt, -10pt) -- (220pt, -10pt);
\draw (160pt, -60pt) -- (220pt, -60pt);
\draw (160pt, -70pt) -- (220pt, -70pt);
\draw (230pt, -15pt) -- (230pt, -30pt);
\draw (230pt, -40pt) -- (230pt, -55pt);
\draw (200pt, -35pt) -- (225pt, -35pt);

\node[sprocblock] at (290pt, -35pt) {}; 
\node at (290pt, -20pt) {further};
\node at (290pt, -40pt) {processing};
\draw (240pt, 0pt) -- (260pt, 0pt);
\draw (240pt, -10pt) -- (260pt, -10pt);
\draw (240pt, -60pt) -- (260pt, -60pt);
\draw (240pt, -70pt) -- (260pt, -70pt);

\node[eqblock] at (335pt, -35pt) {};
\node at (335pt, -36pt) {$=$};
\node[eqblock] at (347pt, -35pt) {};
\node at (347pt, -36pt) {$=$};
\draw (320pt, 0pt) -- (347pt, 0pt) -- (347pt, -30pt);
\draw (320pt, -70pt) -- (347pt, -70pt) -- (347pt, -40pt);
\draw (320pt, -10pt) -- (335pt, -10pt) -- (335pt, -30pt);
\draw (320pt, -60pt) -- (335pt, -60pt) -- (335pt, -40pt);
\end{tikzpicture}
  \end{center}
  \caption{Quantum-probability NFG representation of a stabilizer QECC.}
  \label{fig:QP1_QECC_color}
\end{figure*}

With the above basics, we introduce the NFG in Fig.~\ref{fig:QP1_QECC_color}, which we call the quantum probability NFG representation of a stabilizer QECC.
The information is stored in the quantum state $|\psi\rangle$.
The uncoded quantum state $|\mb{0}\rangle \otimes |\psi\rangle$ is stabilized by $Z_1,\ldots,Z_{n-k}$, while the encoded quantum state is stabilized by the $\Cal{U}Z_1\Cal{U}^{\HP},\ldots,\Cal{U}Z_{n-k}\Cal{U}^{\HP}$.
For a quantum probability NFG representation, if we close the whole box, then the partition sum is $1$.

\begin{figure*}
  \begin{center}
    \begin{tikzpicture}[x=1cm, y= 1cm, 
node/.style={draw=none, inner sep = 0pt},
blockw1/.style={draw, fill = none, minimum height = 20pt, minimum width = 34pt, inner sep=0pt},
blockh2w1/.style={draw, fill = none, minimum height = 50pt, minimum width = 30pt, inner sep=0pt},
blockh3w1/.style={draw, fill = none, minimum height = 80pt, minimum width = 30pt, inner sep=0pt},
blockh5w1/.style={draw, fill = none, minimum height = 140pt, minimum width = 30pt, inner sep=0pt},
blockh7w1/.style={draw, fill = none, minimum height = 200pt, minimum width = 30pt, inner sep=0pt},
block1/.style={draw, fill = white, minimum height = 20pt, minimum width = 20pt, inner sep=0pt},
block/.style={draw,minimum height = 50pt,minimum width = 30pt},
block3/.style={draw,minimum height = 60pt, minimum width = 20pt},
block2/.style={draw, minimum height = 40pt, minimum width = 20pt},
procblock/.style={draw=gray,dashed,fill = white, minimum height = 80pt, minimum width = 60pt, inner sep=0pt},
sprocblock/.style={draw, fill = white, minimum height = 80pt, minimum width = 60pt, inner sep=0pt},
blockw4/.style={draw,fill = white, minimum height = 40pt, minimum width = 20pt, inner sep=0pt},
blockw3/.style={draw,fill = white, minimum height = 30pt, minimum width = 20pt, inner sep=0pt},
blockw2/.style={draw, fill = white, minimum height = 20pt, minimum width = 20pt, inner sep=0pt},
factor/.style={draw,minimum size = 15pt,inner sep = 0pt},
blockh3w2/.style={draw, fill = white, minimum height = 90pt, minimum width = 40pt, inner sep=0pt},
blockh3w3/.style={draw, fill = white, minimum height = 90pt, minimum width = 60pt, inner sep=0pt},
blockh2w2/.style={draw, fill = white, minimum height = 60pt, minimum width = 50pt, inner sep=0pt},
blockh1w2/.style={draw, fill = white, minimum height = 20pt, minimum width = 50pt, inner sep=0pt},
blockh1w3/.style={draw, fill = white, minimum height = 20pt, minimum width = 110pt, inner sep=0pt},
blockh1w4/.style={draw, fill = white, minimum height = 20pt, minimum width = 150pt, inner sep=0pt}
                          ]
\def\ya{0pt}
\def\yb{-30pt}
\def\yc{-60pt}
\def\yd{-90pt}
\def\ye{-120pt}
\def\yf{-150pt}
\def\yg{-180pt}

\def\xa{0pt}
\def\xab{\xa+17pt}
\node[blockw1] at (\xa, \ya) 	{$| \mb{0} \rangle \langle \mb{0} |$};
\node[blockw1] at (\xa, \yb) 	{$| \mb{0} \rangle \langle \mb{0} |$};
\node[blockw1] at (\xa, \yc) 	{$| \psi \rangle \langle \psi |$};
\node[blockw1] at (\xa, \yd) 	{$| \mb{0} \rangle \langle \mb{0} |$};
\node[blockw1] at (\xa, \ye) 	{$| \phi \rangle \langle \phi |$};
\node[blockw1] at (\xa, \yf) 	{$| \mb{0} \rangle \langle \mb{0} |$};
\node[blockw1] at (\xa, \yg) 	{$| \chi \rangle \langle \chi |$};

\def\xb{\xa+17pt+15pt+20pt}
\def\xba{\xb-15pt}
\def\xbb{\xb+15pt}
\node[blockh3w1] at (\xb, \yb) {$\msf{U}$};
\draw (\xab, \ya+1pt) -- (\xba, \ya+1pt);
\draw (\xab, \ya-1pt)  -- (\xba, \ya-1pt);
\draw (\xab, \yb+1pt) -- (\xba, \yb+1pt);
\draw (\xab, \yb-1pt)  -- (\xba, \yb-1pt);
\draw (\xab, \yc+1pt) -- (\xba, \yc+1pt);
\draw (\xab, \yc-1pt)  -- (\xba, \yc-1pt);
\node at (\xab+10pt,\ya) {$/$};
\node at (\xab+10pt,\ya+8pt) {$m$};
\node at (\xab+10pt,\yb) {$/$};
\node at (\xab+8pt,\yb+14pt) {$n\!-\!k$};
\node at (\xab+10pt,\yc) {$/$};
\node at (\xab+10pt,\yc+10pt) {$k$};

\def\xc{\xb+15pt+15pt+20pt}
\def\xca{\xc-15pt}
\def\xcb{\xc+15pt}
\node[blockh3w1] at (\xc, \yd) {$\msf{U}$};
\draw (\xbb, \yc+1pt) -- (\xca, \yc+1pt);
\draw (\xbb, \yc-1pt)  -- (\xca, \yc-1pt);
\draw (\xab, \yd+1pt) -- (\xca, \yd+1pt);
\draw (\xab, \yd-1pt)  -- (\xca, \yd-1pt);
\draw (\xab, \ye+1pt) -- (\xca, \ye+1pt);
\draw (\xab, \ye-1pt)  -- (\xca, \ye-1pt);
\node at (\xbb+10pt,\yc) {$/$};
\node at (\xbb+10pt,\yc+10pt) {$m$};
\node at (\xb,\yd) {$/$};
\node at (\xb,\yd+10pt) {$n\!-\!k$};
\node at (\xb,\ye) {$/$};
\node at (\xb,\ye+10pt) {$k$};

\def\xd{\xc+15pt+15pt+20pt}
\def\xda{\xd-15pt}
\def\xdb{\xd+15pt}
\node[blockh3w1] at (\xd, \yf) {$\msf{U}$};
\draw (\xcb, \ye+1pt) -- (\xda, \ye+1pt);
\draw (\xcb, \ye-1pt)  -- (\xda, \ye-1pt);
\draw (\xab, \yf+1pt) -- (\xda, \yf+1pt);
\draw (\xab, \yf-1pt)  -- (\xda, \yf-1pt);
\draw (\xab, \yg+1pt) -- (\xda, \yg+1pt);
\draw (\xab, \yg-1pt)  -- (\xda, \yg-1pt);
\node at (\xcb+10pt,\ye) {$/$};
\node at (\xcb+10pt,\ye+10pt) {$m$};
\node at (\xbb+10pt,\yf) {$/$};
\node at (\xbb+10pt,\yf+10pt) {$n\!-\!k$};
\node at (\xbb+10pt,\yg) {$/$};
\node at (\xbb+10pt,\yg+10pt) {$k$};

\def\xe{\xd+15pt+15pt+20pt}
\def\xea{\xe-15pt}
\def\xeb{\xe+15pt}
\node[blockh7w1] at (\xe, \yd) {};
\node[rotate=-90] at (\xe, \yd) {channel};
\draw (\xbb, \ya+1pt) -- (\xea, \ya+1pt);
\draw (\xbb, \ya-1pt)  -- (\xea, \ya-1pt);
\draw (\xcb, \yc+1pt) -- (\xea, \yc+1pt);
\draw (\xcb, \yc-1pt)  -- (\xea, \yc-1pt);
\draw (\xdb, \ye+1pt) -- (\xea, \ye+1pt);
\draw (\xdb, \ye-1pt)  -- (\xea, \ye-1pt);
\draw (\xdb, \yg+1pt) -- (\xea, \yg+1pt);
\draw (\xdb, \yg-1pt)  -- (\xea, \yg-1pt);
\node at (\xcb+10pt,\ya) {$/$};
\node at (\xcb+10pt,\ya+10pt) {$n$};
\node at (\xd+10pt,\yc) {$/$};
\node at (\xd+10pt,\yc+10pt) {$n$};
\node at (\xdb+10pt,\ye) {$/$};
\node at (\xdb+10pt,\ye+10pt) {$n$};
\node at (\xdb+10pt,\yg) {$/$};
\node at (\xdb+10pt,\yg+10pt) {$n$};

\def\xf{\xe+15pt+15pt+20pt}
\def\xfa{\xf-15pt}
\def\xfb{\xf+15pt}
\node[blockh3w1] at (\xf, \yf) {$\msf{U}^{-1}$};
\draw (\xeb, \ye+1pt) -- (\xfa, \ye+1pt);
\draw (\xeb, \ye-1pt)  -- (\xfa, \ye-1pt);
\draw (\xeb, \yg+1pt) -- (\xfa, \yg+1pt);
\draw (\xeb, \yg-1pt)  -- (\xfa, \yg-1pt);
\node at (\xeb+10pt,\ye) {$/$};
\node at (\xeb+10pt,\ye+10pt) {$n$};
\node at (\xeb+10pt,\yg) {$/$};
\node at (\xeb+10pt,\yg+10pt) {$n$};

\def\xg{\xf+15pt+15pt+20pt}
\def\xga{\xg-15pt}
\def\xgb{\xg+15pt}
\node[blockh3w1] at (\xg, \yd) {$\msf{U}^{-1}$};
\draw (\xeb, \yc+1pt) -- (\xga, \yc+1pt);
\draw (\xeb, \yc-1pt)  -- (\xga, \yc-1pt);
\draw (\xfb, \ye+1pt) -- (\xga, \ye+1pt);
\draw (\xfb, \ye-1pt)  -- (\xga, \ye-1pt);
\node at (\xf,\yc) {$/$};
\node at (\xf,\yc+10pt) {$n$};
\node at (\xfb+10pt,\ye) {$/$};
\node at (\xfb+10pt,\ye+10pt) {$m$};

\def\xh{\xg+15pt+15pt+20pt}
\def\xha{\xh-15pt}
\def\xhb{\xh+15pt}
\node[blockh3w1] at (\xh, \yb) {$\msf{U}^{-1}$};
\draw (\xeb, \ya+1pt) -- (\xha, \ya+1pt);
\draw (\xeb, \ya-1pt)  -- (\xha, \ya-1pt);
\draw (\xgb, \yc+1pt) -- (\xha, \yc+1pt);
\draw (\xgb, \yc-1pt)  -- (\xha, \yc-1pt);
\node at (\xf+10pt,\ya) {$/$};
\node at (\xf+10pt,\ya+10pt) {$n$};
\node at (\xgb+10pt,\yc) {$/$};
\node at (\xgb+10pt,\yc+10pt) {$m$};

\def\xi{\xh+15pt+10pt+20pt}
\def\xia{\xi-10pt}
\def\xib{\xi+10pt}
\node[block1] at (\xi, \ya) {$\msf{M}$};
\draw (\xhb, \ya+1pt) -- (\xia, \ya+1pt);
\draw (\xhb, \ya-1pt)  -- (\xia, \ya-1pt);
\node at (\xhb+10pt,\ya) {$/$};
\node at (\xhb+10pt,\ya+10pt) {$m$};
\node[block1] at (\xi, \yb) {$\msf{M}$};
\draw (\xhb, \yb+1pt) -- (\xia, \yb+1pt);
\draw (\xhb, \yb-1pt)  -- (\xia, \yb-1pt);
\node at (\xhb+10pt,\yb) {$/$};
\node at (\xhb+12pt,\yb+14pt) {$n\!-\!k$};
\node[block1] at (\xh, \yd) {$\msf{M}$};
\draw (\xgb, \yd+1pt) -- (\xh-10pt, \yd+1pt);
\draw (\xgb, \yd-1pt)  -- (\xh-10pt, \yd-1pt);
\node at (\xgb+12pt,\yd) {$/$};
\node at (\xgb+12pt,\yd+10pt) {$n\!-\!k$};
\node[block1] at (\xg, \yf) {$\msf{M}$};
\draw (\xfb, \yf+1pt) -- (\xg-10pt, \yf+1pt);
\draw (\xfb, \yf-1pt)  -- (\xg-10pt, \yf-1pt);
\node at (\xfb+12pt,\yf) {$/$};
\node at (\xfb+12pt,\yf+10pt) {$n\!-\!k$};

\def\xj{\xi+15pt+15pt+20pt}
\def\xja{\xj-15pt}
\def\xjb{\xj+15pt}
\node[blockh2w1] at (\xj, \ya-15pt) {$D$};
\draw (\xib, \ya)  -- (\xja, \ya);
\draw (\xib, \yb)  -- (\xja, \yb);
\draw (\xh+10pt, \yd)  -- (\xja, \yb-2pt);
\draw (\xg+10pt, \yf)  -- (\xja, \yb-5pt);
\node[blockh5w1] at (\xj, \ye) {$R$};
\draw (\xhb, \yc+1pt) -- (\xja, \yc+1pt);
\draw (\xhb, \yc-1pt)  -- (\xja, \yc-1pt);
\draw (\xgb, \ye+1pt) -- (\xja, \ye+1pt);
\draw (\xgb, \ye-1pt)  -- (\xja, \ye-1pt);
\draw (\xfb, \yg+1pt) -- (\xja, \yg+1pt);
\draw (\xfb, \yg-1pt)  -- (\xja, \yg-1pt);
\node at (\xhb+10pt,\yc) {$/$};
\node at (\xhb+10pt,\yc+10pt) {$k$};
\node at (\xhb+10pt,\ye) {$/$};
\node at (\xhb+10pt,\ye+10pt) {$k$};
\node at (\xhb+10pt,\yg) {$/$};
\node at (\xhb+10pt,\yg+10pt) {$k$};
\draw (\xj, \yb-10pt) -- (\xj, \yc+10pt);

\def\xk{\xj+15pt+10pt+20pt}
\def\xka{\xk-10pt}
\node[block1] at (\xk, \yc) {$I$};
\node[block1] at (\xk, \ye) {$I$};
\node[block1] at (\xk, \yg) {$I$};
\draw (\xjb, \yc+1pt) -- (\xka, \yc+1pt);
\draw (\xjb, \yc-1pt)  -- (\xka, \yc-1pt);
\node at (\xjb+10pt,\yc) {$/$};
\node at (\xjb+10pt,\yc+10pt) {$k$};
\draw (\xjb, \ye+1pt) -- (\xka, \ye+1pt);
\draw (\xjb, \ye-1pt)  -- (\xka, \ye-1pt);
\node at (\xjb+10pt,\ye) {$/$};
\node at (\xjb+10pt,\ye+10pt) {$k$};
\draw (\xjb, \yg+1pt) -- (\xka, \yg+1pt);
\draw (\xjb, \yg-1pt)  -- (\xka, \yg-1pt);
\node at (\xjb+10pt,\yg) {$/$};
\node at (\xjb+10pt,\yg+10pt) {$k$};

\end{tikzpicture}
  \end{center}
  \caption{Double-edge quantum-probability NFG representation of a convolutional stabilizer QECC.}
  \label{fig:QP6_QCC_DE}
\end{figure*}
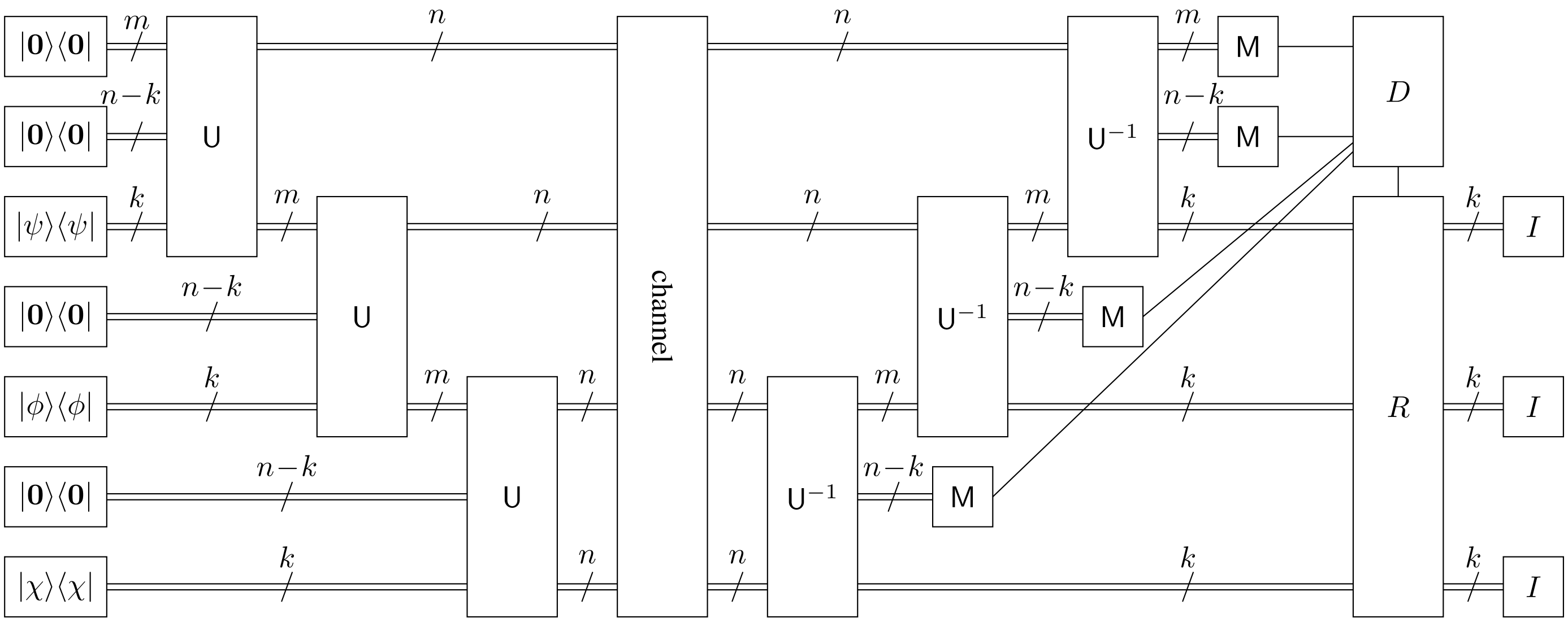

The double-edge quantum-probability NFG representation in Fig.~\ref{fig:QP2_QECC_DE} is equivalent to the NFG in Fig.~\ref{fig:QP1_QECC_color}.
The NFG in Fig.~\ref{fig:QP3_QECC_DE} is another double-edge quantum-probability NFG representation for a stabilizer QECC.
The double-edge quantum-probability NFGs have similar structures as the quantum-code NFGs except that doubled edges are for quantum qubits while single edges are for classical variables.
The double-edge quantum-probability NFGs have much more freedom and are easier to plot when the situation is complicated, e.g. the one in Fig.~\ref{fig:QP6_QCC_DE}, compared with the NFG in Fig.~\ref{fig:QP2_QECC_DE}.
As for notations, the label ``$/$'' is indicating that multiple qubits are represented by a doubled edge, 
``$R$'' stands for recovery, 
``$D$'' stands for decoder,
``$M_i$'' stands for the measurement w.r.t. the $i^{\Rm{th}}$ generator in the stabilizer, and 
``$M$'' stands for the measurement projecting on the basis consisting $|0\rangle$ and $|1\rangle$ for a single qubit.
The NFGs in Fig.~\ref{fig:QP2_QECC_DE} and~\ref{fig:QP3_QECC_DE} both are double-edge quantum-probability NFG representations for a general stabilizer QECC and the difference is that the NFG in Fig.~\ref{fig:QP2_QECC_DE} shows how to maintain the encoded quantum state while the NFG in Fig.~\ref{fig:QP3_QECC_DE} shows how to protect the information in the information qubits.

\begin{figure*}
  \begin{center}
    \begin{tikzpicture}[x=1cm, y= 1cm, node/.style={draw=none, inner sep = 0pt},
                          block/.style={draw,minimum height = 50pt,minimum width = 30pt},
                          block3/.style={draw,minimum height = 60pt, minimum width = 20pt},
                          block2/.style={draw, minimum height = 40pt, minimum width = 20pt},
                          procblock/.style={draw=gray,dashed,fill = white, minimum height = 80pt, minimum width = 60pt, inner sep=0pt},
                          sprocblock/.style={draw, fill = white, minimum height = 80pt, minimum width = 60pt, inner sep=0pt},
                          blockw4/.style={draw,fill = white, minimum height = 40pt, minimum width = 20pt, inner sep=0pt},
                          blockw3/.style={draw,fill = white, minimum height = 30pt, minimum width = 20pt, inner sep=0pt},
                          blockw2/.style={draw, fill = white, minimum height = 20pt, minimum width = 20pt, inner sep=0pt},
                          factor/.style={draw,minimum size = 15pt,inner sep = 0pt},
                          blockw1/.style={draw, fill = white, minimum height = 20pt, minimum width = 34pt, inner sep=0pt},
                          block1/.style={draw, fill = white, minimum height = 20pt, minimum width = 20pt, inner sep=0pt},
                          blockh3w1/.style={draw, fill = white, minimum height = 90pt, minimum width = 30pt, inner sep=0pt},
                          blockh3w2/.style={draw, fill = white, minimum height = 90pt, minimum width = 40pt, inner sep=0pt},
                          blockh3w3/.style={draw, fill = white, minimum height = 90pt, minimum width = 60pt, inner sep=0pt},
                          blockh2w2/.style={draw, fill = white, minimum height = 60pt, minimum width = 44pt, inner sep=0pt},
                          blockh1w2/.style={draw, fill = white, minimum height = 20pt, minimum width = 44pt, inner sep=0pt},
                          blockh1w3/.style={draw, fill = white, minimum height = 20pt, minimum width = 110pt, inner sep=0pt},
                          blockh1w4/.style={draw, fill = white, minimum height = 20pt, minimum width = 150pt, inner sep=0pt},
                          dashh1w2/.style={draw,dashed, fill = white, minimum height = 20pt, minimum width = 60pt, inner sep=0pt},
                          dashh3w2/.style={draw,dashed, fill = white, minimum height = 90pt, minimum width = 30pt, inner sep=0pt}
                          ]
\def\ya{0pt}
\def\yaa{-1pt}
\def\yab{1pt}
\def\yb{-30pt}
\def\yba{-29pt}
\def\ybb{-31pt}
\def\yc{-35pt}
\def\yd{-46pt}
\def\ye{-70pt}
\def\yea{-69pt}
\def\yeb{-71pt}

\def\xa{0pt}
\node[blockw1] at (\xa, \ya) 	{$| \psi \rangle \langle \psi |$};
\node[blockw1] at (\xa, \yb) 	{$| {0} \rangle \langle {0} |$};
\node 		 	   at (\xa, \yd) 	{$\vdots$};
\node[blockw1] at (\xa, \ye) 	{$| {0} \rangle \langle {0} |$};

\def\xab{17pt}

\def\xb{52pt}
\def\xba{37pt}
\def\xbb{67pt}
\node[blockh3w1] at (\xb, \yc) {$\msf{U}$};
\draw     (\xab, \yaa) 	-- (\xba, \yaa);
\draw     (\xab, \yab)  	-- (\xba, \yab);
\draw     (\xab, \yba) 	-- (\xba, \yba);
\draw     (\xab, \ybb)  	-- (\xba, \ybb);
\node at (27pt, \yd) 	{$\vdots$};
\draw     (\xab, \yea) 	-- (\xba, \yea);
\draw     (\xab, \yeb)  	-- (\xba, \yeb);
\node at (27pt,\ya) {$/$};

\def\xc{107pt}
\def\xca{87pt}
\def\xcb{127pt}
\node[blockh3w2] at (\xc, \yc) {channel};
\draw     (\xbb, \yaa) 	-- (\xca, \yaa);
\draw     (\xbb, \yab)  	-- (\xca, \yab);
\draw     (\xbb, \yba) 	-- (\xca, \yba);
\draw     (\xbb, \ybb)  	-- (\xca, \ybb);
\node at (77pt, \yd) 	{$\vdots$};
\draw     (\xbb, \yea) 	-- (\xca, \yea);
\draw     (\xbb, \yeb)  	-- (\xca, \yeb);
\node at (77pt,\ya) {$/$};

\def\xd{162pt}
\def\xda{147pt}
\def\xdb{177pt}
\def\xdc{187pt}
\def\xdd{197pt}
\def\xde{207pt}
\node[blockh3w1] at (\xd, \yc) {$\msf{M}_1$};
\draw     (\xcb, \yaa) 	-- (\xda, \yaa);
\draw     (\xcb, \yab)  	-- (\xda, \yab);
\draw     (\xcb, \yba) 	-- (\xda, \yba);
\draw     (\xcb, \ybb)  	-- (\xda, \ybb);
\node at (137pt, \yd) 	{$\vdots$};
\draw     (\xcb, \yea) 	-- (\xda, \yea);
\draw     (\xcb, \yeb)  	-- (\xda, \yeb);
\node at (137pt,\ya) {$/$};
\node at (182pt,\ya) {$/$};

\def\xe{232pt}
\def\xea{217pt}
\def\xeb{247pt}
\node[blockh3w1] at (\xe, \yc) {$\msf{M}_{n\!-\!k}$};
\node at (212pt,\ya) {$/$};

\draw     (\xdb, \yaa) 	-- (\xdc, \yaa);
\draw     (\xdb, \yab)  	-- (\xdc, \yab);
\draw     (\xdb, \yba) 	-- (\xdc, \yba);
\draw     (\xdb, \ybb)  	-- (\xdc, \ybb);
\draw     (\xdb, \yea) 	-- (\xdc, \yea);
\draw     (\xdb, \yeb)  	-- (\xdc, \yeb);

\node at (\xdd, \ya) 	{$\cdots$};
\node at (\xdd, \yb) 	{$\cdots$};
\node at (\xdd, \yd) 	{$\vdots$};
\node at (\xdd, \ye) 	{$\cdots$};

\draw     (\xde, \yaa) 	-- (\xea, \yaa);
\draw     (\xde, \yab)  	-- (\xea, \yab);
\draw     (\xde, \yba) 	-- (\xea, \yba);
\draw     (\xde, \ybb)  	-- (\xea, \ybb);
\draw     (\xde, \yea) 	-- (\xea, \yea);
\draw     (\xde, \yeb)  	-- (\xea, \yeb);

\def\xf{282pt}
\def\xfa{267pt}
\def\xfb{297pt}
\node[blockh3w1] at (\xf, \yc) {${R}$};
\draw     (\xeb, \yaa) 	-- (\xfa, \yaa);
\draw     (\xeb, \yab)  	-- (\xfa, \yab);
\draw     (\xeb, \yba) 	-- (\xfa, \yba);
\draw     (\xeb, \ybb)  	-- (\xfa, \ybb);
\node at (257pt, \yd) 	{$\vdots$};
\draw     (\xeb, \yea) 	-- (\xfa, \yea);
\draw     (\xeb, \yeb)  	-- (\xfa, \yeb);
\node at (257pt,\ya) {$/$};

\def\xg{222pt}
\def\yec{-80pt}
\def\yed{-90pt}
\def\yfa{-100pt}
\def\yf{-110pt}
\node[blockh1w4] at (\xg, \yf) {decoder};
\draw     (\xd, \yec) 	-- (\xd, \yfa);
\node at (\xdd, \yed) 	{$\cdots$};
\draw     (\xe, \yec)  	-- (\xe, \yfa);
\draw     (\xf, \yec)  	-- (\xf, \yfa);
\node at (171pt, \yed) 	{$s_1$};
\node at (245pt, \yed) 	{$s_{n\!-\!k}$};

\def\xh{347pt}
\def\xha{317pt}
\def\xhb{377pt}
\node[blockh3w3] at (\xh, \yc) {};
\node at (\xh, -20pt) {further};
\node at (\xh, -40pt) {processing};
\draw     (\xfb, \yaa) 	-- (\xha, \yaa);
\draw     (\xfb, \yab)  	-- (\xha, \yab);
\draw     (\xfb, \yba) 	-- (\xha, \yba);
\draw     (\xfb, \ybb)  	-- (\xha, \ybb);
\node at (307pt, \yd) 	{$\vdots$};
\draw     (\xfb, \yea) 	-- (\xha, \yea);
\draw     (\xfb, \yeb)  	-- (\xha, \yeb);
\node at (307pt,\ya) {$/$};
\node at (387pt,\ya) {$/$};

\def\xi{407pt}
\def\xia{397pt}
\node[block1] at (\xi, \ya) 	{$I$};
\node[block1] at (\xi, \yb) 	{$I$};
\node at (\xi, \yd) 	{$\vdots$};
\node[block1] at (\xi, \ye) 	{$I$};
\draw     (\xhb, \yaa) 	-- (\xia, \yaa);
\draw     (\xhb, \yab)  	-- (\xia, \yab);
\draw     (\xhb, \yba) 	-- (\xia, \yba);
\draw     (\xhb, \ybb)  	-- (\xia, \ybb);
\node at (387pt, \yd) 	{$\vdots$};
\draw     (\xhb, \yea) 	-- (\xia, \yea);
\draw     (\xhb, \yeb)  	-- (\xia, \yeb);
\end{tikzpicture}
  \end{center}
  \caption{Double-edge quantum-probability NFG representation of a stabilizer QECC.}
  \label{fig:QP2_QECC_DE}
\end{figure*}

\begin{figure*}
  \begin{center}
    \begin{tikzpicture}[x=1cm, y= 1cm, node/.style={draw=none, inner sep = 0pt},
                          block/.style={draw,minimum height = 50pt,minimum width = 30pt},
                          block3/.style={draw,minimum height = 60pt, minimum width = 20pt},
                          block2/.style={draw, minimum height = 40pt, minimum width = 20pt},
                          procblock/.style={draw=gray,dashed,fill = white, minimum height = 80pt, minimum width = 60pt, inner sep=0pt},
                          sprocblock/.style={draw, fill = white, minimum height = 80pt, minimum width = 60pt, inner sep=0pt},
                          blockw4/.style={draw,fill = white, minimum height = 40pt, minimum width = 20pt, inner sep=0pt},
                          blockw3/.style={draw,fill = white, minimum height = 30pt, minimum width = 20pt, inner sep=0pt},
                          blockw2/.style={draw, fill = white, minimum height = 20pt, minimum width = 20pt, inner sep=0pt},
                          factor/.style={draw,minimum size = 15pt,inner sep = 0pt},
                          blockw1/.style={draw, fill = white, minimum height = 20pt, minimum width = 34pt, inner sep=0pt},
                          block1/.style={draw, fill = white, minimum height = 20pt, minimum width = 20pt, inner sep=0pt},
                          blockh3w1/.style={draw, fill = white, minimum height = 90pt, minimum width = 30pt, inner sep=0pt},
                          blockh3w2/.style={draw, fill = white, minimum height = 90pt, minimum width = 40pt, inner sep=0pt},
                          blockh3w3/.style={draw, fill = white, minimum height = 90pt, minimum width = 60pt, inner sep=0pt},
                          blockh2w2/.style={draw, fill = white, minimum height = 60pt, minimum width = 50pt, inner sep=0pt},
                          blockh1w2/.style={draw, fill = white, minimum height = 20pt, minimum width = 50pt, inner sep=0pt},
                          blockh1w3/.style={draw, fill = white, minimum height = 20pt, minimum width = 110pt, inner sep=0pt},
                          blockh1w4/.style={draw, fill = white, minimum height = 20pt, minimum width = 150pt, inner sep=0pt},
                          dashh1w2/.style={draw,dashed, fill = white, minimum height = 20pt, minimum width = 60pt, inner sep=0pt},
                          dashh3w2/.style={draw,dashed, fill = white, minimum height = 90pt, minimum width = 30pt, inner sep=0pt}
                          ]
\def\ya{0pt}
\def\yaa{1pt}
\def\yab{-1pt}
\def\yat{-10pt}
\def\ybh{-20pt}
\def\yb{-30pt}
\def\yba{-29pt}
\def\ybb{-31pt}
\def\yc{-35pt}
\def\yd{-50pt}
\def\yda{-47pt}
\def\ye{-70pt}
\def\yea{-69pt}
\def\yeb{-71pt}

\def\xa{0pt}
\def\xab{17pt}
\node[blockw1] at (\xa, \ya) 	{$| \psi \rangle \langle \psi |$};
\node[blockw1] at (\xa, \yb) 	{$| {0} \rangle \langle {0} |$};
\node 		 	   at (\xa, \yda) 	{$\vdots$};
\node[blockw1] at (\xa, \ye) 	{$| {0} \rangle \langle {0} |$};

\def\xb{52pt}
\def\xba{37pt}
\def\xbb{67pt}
\node[blockh3w1] at (\xb, \yc) {$\msf{U}$};

\draw     (\xab, \yaa) 	-- (\xba, \yaa);
\draw     (\xab, \yab)  	-- (\xba, \yab);
\draw     (\xab, \yba) 	-- (\xba, \yba);
\draw     (\xab, \ybb)  	-- (\xba, \ybb);
\node at (27pt, \yda) 	{$\vdots$};
\draw     (\xab, \yea) 	-- (\xba, \yea);
\draw     (\xab, \yeb)  	-- (\xba, \yeb);
\node at (27pt,\ya) {$/$};

\def\xc{107pt}
\def\xca{87pt}
\def\xcb{127pt}
\node[blockh3w2] at (\xc, \yc) {channel};
\node at (77pt,\ya) {$/$};

\draw     (\xbb, \yaa) 	-- (\xca, \yaa);
\draw     (\xbb, \yab)  	-- (\xca, \yab);
\draw     (\xbb, \yba) 	-- (\xca, \yba);
\draw     (\xbb, \ybb)  	-- (\xca, \ybb);
\node at (77pt, \yda) 	{$\vdots$};
\draw     (\xbb, \yea) 	-- (\xca, \yea);
\draw     (\xbb, \yeb)  	-- (\xca, \yeb);

\def\xd{162pt}
\def\xda{147pt}
\def\xdb{177pt}
\node[blockh3w1] at (\xd, \yc) {$\msf{U}^{-1}$};

\draw     (\xcb, \yaa) 	-- (\xda, \yaa);
\draw     (\xcb, \yab)  	-- (\xda, \yab);
\draw     (\xcb, \yba) 	-- (\xda, \yba);
\draw     (\xcb, \ybb)  	-- (\xda, \ybb);
\node at (137pt, \yda) 	{$\vdots$};
\draw     (\xcb, \yea) 	-- (\xda, \yea);
\draw     (\xcb, \yeb)  	-- (\xda, \yeb);
\node at (137pt,\ya) {$/$};

\def\xe{207pt}
\def\xea{197pt}
\def\xeb{217pt}
\def\xec{232pt}
\node[block1] at (\xe, \yb) {$\msf{M}$};
\node at (\xe, \yda) {$\vdots$};
\node[block1] at (\xe, \ye) {$\msf{M}$};
\draw     (\xdb, \yba)  	-- (\xea, \yba);
\draw     (\xdb, \ybb)  	-- (\xea, \ybb);
\draw     (\xdb, \yea)  	-- (\xea, \yea);
\draw     (\xdb, \yeb)  	-- (\xea, \yeb);
\node at (\xec, -25pt) {$s_1$};
\node at (\xec, -65pt) {$s_{n\!-\!k}$};
\node at (\xe,\ya) {$/$};

\def\xf{272pt}
\def\xfa{247pt}
\def\xfb{297pt}
\node[blockh1w2] at (\xf, \ya) {recovery}; 
\node[blockh2w2] at (\xf, \yd) {decoder};

\draw     (\xdb, \yaa)  	-- (\xfa, \yaa);
\draw     (\xdb, \yab)  	-- (\xfa, \yab);
\draw     (\xeb, \yb)  	-- (\xfa, \yb);
\draw     (\xeb, \ye)  	-- (\xfa, \ye);
\draw     (\xf, \yat)  	-- (\xf, \ybh);
\node at (307pt,\ya) {$/$};

\def\xh{347pt}
\def\xha{317pt}
\def\xhb{377pt}
\node[blockh3w3] at (\xh, \yc) {}; 
\node at (\xh, -20pt) {further};
\node at (\xh, -40pt) {processing};
\draw     (\xfb, \yaa) 	-- (\xha, \yaa);
\draw     (\xfb, \yab)  	-- (\xha, \yab);
\node at (387pt,\ya) {$/$};

\def\xi{407pt}
\def\xia{397pt}
\node[block1] at (\xi, \ya) 	{$I$};
\draw     (\xhb, \yaa) 	-- (\xia, \yaa);
\draw     (\xhb, \yab)  	-- (\xia, \yab);

\end{tikzpicture}
  \end{center}
  \caption{Double-edge quantum-probability NFG representation of a stabilizer QECC.}
  \label{fig:QP3_QECC_DE}
\end{figure*}

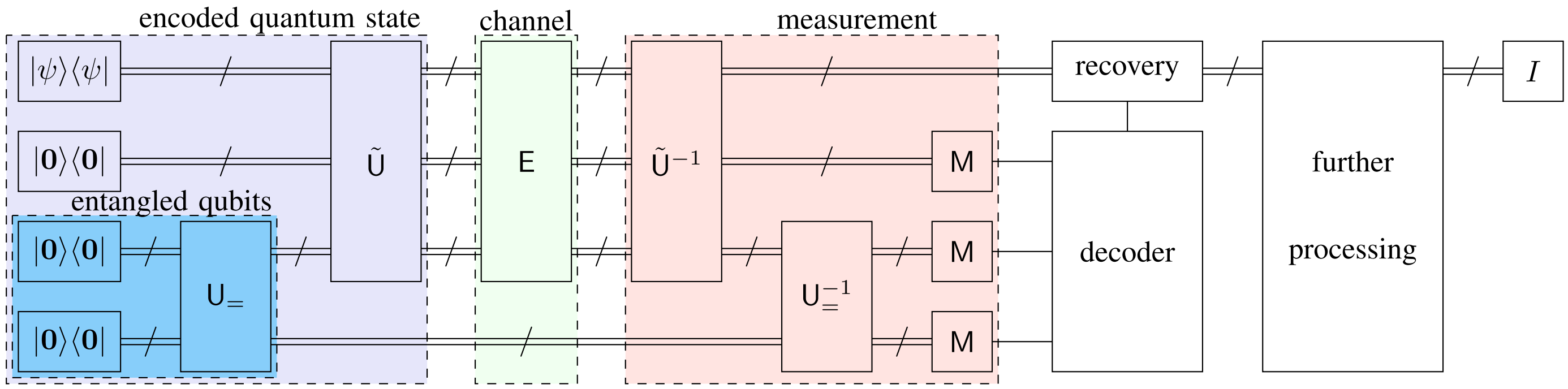
\begin{figure*}
  \begin{center}
    \begin{tikzpicture}[x=1cm, y= 1cm, node/.style={draw=none, inner sep = 0pt},
                          block/.style={draw,minimum height = 50pt,minimum width = 30pt},
                          block3/.style={draw,minimum height = 60pt, minimum width = 20pt},
                          block2/.style={draw, minimum height = 40pt, minimum width = 20pt},
                          procblock/.style={draw=gray,dashed,fill = white, minimum height = 80pt, minimum width = 60pt, inner sep=0pt},
                          sprocblock/.style={draw, fill = none, minimum height = 80pt, minimum width = 60pt, inner sep=0pt},
                          blockw4/.style={draw,fill = none, minimum height = 40pt, minimum width = 20pt, inner sep=0pt},
                          blockw3/.style={draw,fill = none, minimum height = 30pt, minimum width = 20pt, inner sep=0pt},
                          blockw2/.style={draw, fill = none, minimum height = 20pt, minimum width = 20pt, inner sep=0pt},
                          factor/.style={draw,minimum size = 15pt,inner sep = 0pt},
                          blockw1/.style={draw, fill = none, minimum height = 20pt, minimum width = 34pt, inner sep=0pt},
                          block1/.style={draw, fill = none, minimum height = 20pt, minimum width = 20pt, inner sep=0pt},
                          blockh4w1/.style={draw, fill = none, minimum height = 110pt, minimum width = 30pt, inner sep=0pt},
                          blockh4w2/.style={draw, fill = none, minimum height = 110pt, minimum width = 40pt, inner sep=0pt},
                          blockh4w3/.style={draw, fill = none, minimum height = 110pt, minimum width = 60pt, inner sep=0pt},
                          blockh3w1/.style={draw, fill = none, minimum height = 80pt, minimum width = 30pt, inner sep=0pt},
                          blockh3w4/.style={draw, fill = none, minimum height = 80pt, minimum width = 40pt, inner sep=0pt},
                          blockh3w5/.style={draw, fill = none, minimum height = 80pt, minimum width = 50pt, inner sep=0pt},
                          blockh3w3/.style={draw, fill = none, minimum height = 80pt, minimum width = 60pt, inner sep=0pt},
                          blockh2w1/.style={draw, fill = none, minimum height = 50pt, minimum width = 30pt, inner sep=0pt},
                          blockh2w2/.style={draw, fill = none, minimum height = 60pt, minimum width = 50pt, inner sep=0pt},
                          blockh1w2/.style={draw, fill = none, minimum height = 20pt, minimum width = 50pt, inner sep=0pt},
                          blockh1w3/.style={draw, fill = none, minimum height = 20pt, minimum width = 110pt, inner sep=0pt},
                          blockh1w4/.style={draw, fill = none, minimum height = 20pt, minimum width = 150pt, inner sep=0pt},
                          dashh1w2/.style={draw,dashed, fill = none, minimum height = 20pt, minimum width = 60pt, inner sep=0pt},
                          dashh3w2/.style={draw,dashed, fill = white, minimum height = 90pt, minimum width = 30pt, inner sep=0pt}
                          ]

\definecolor{silver}{RGB}{192,192,192} 
\definecolor{Lavender}{RGB}{230,230,250} 
\definecolor{SkyBlue}{RGB}{135,206,235} 
\definecolor{PapayaWhip}{RGB}{255,239,213} 
\definecolor{silver}{RGB}{192,192,192} 
\definecolor{pink}{RGB}{255,192,203} 
\definecolor{honeydew}{RGB}{240,255,240} 
\definecolor{MistyRose}{RGB}{255,228,225} 
\definecolor{plum}{RGB}{221,160,221} 
\definecolor{orchid}{RGB}{218,112,214} 
\definecolor{LightSkyBlue}{RGB}{135,206,250} 
\filldraw[fill=Lavender, draw=black, dashed] (-71pt, 12pt) rectangle (69pt, -104pt);
\node at (20pt, 17pt) {encoded quantum state};
\filldraw[fill=LightSkyBlue, draw=black, dashed] (-69pt, -48pt) rectangle (19pt, -102pt);
\node at (-16pt, -44pt) {entangled qubits};
\filldraw[fill=honeydew, draw=black, dashed] (85pt, 12pt) rectangle (119pt, -104pt);
\node at (102pt, 17pt) {channel};
\filldraw[fill=MistyRose, draw=black, dashed] (135pt, 12pt) rectangle (259pt, -104pt);
\node at (212pt, 17pt) {measurement};

\def\ya{0pt}
\def\yaa{-1pt}
\def\yab{1pt}
\def\yb{-30pt}
\def\yba{-29pt}
\def\ybb{-31pt}
\def\y{-45pt}
\def\yc{-60pt}
\def\yca{-59pt}
\def\ycb{-61pt}
\def\yd{-90pt}
\def\yda{-89pt}
\def\ydb{-91pt}
\def\ye{-70pt}
\def\yea{-69pt}
\def\yeb{-71pt}

\def\xxa{-50pt}
\def\xxab{-33pt}
\node[blockw1] at (\xxa, \ya) 	{$| \psi \rangle \langle \psi |$};
\node[blockw1] at (\xxa, \yb) 	{$| \mb{0} \rangle \langle \mb{0} |$};
\node[blockw1] at (\xxa, \yc) 	{$| \mb{0} \rangle \langle \mb{0} |$};
\node[blockw1] at (\xxa, \yd) 	{$| \mb{0} \rangle \langle \mb{0} |$};

\def\xa{2pt}
\def\xaa{-13pt}
\def\xab{17pt}
\node[blockh2w1] at (\xa, -75pt) {$\msf{U}_{=}$};
\draw     (\xxab, \yca) -- (\xaa, \yca);
\draw     (\xxab, \ycb) -- (\xaa, \ycb);
\draw     (\xxab, \yda) -- (\xaa, \yda);
\draw     (\xxab, \ydb) -- (\xaa, \ydb);
\node at (-23pt,\yc) {$/$};
\node at (-23pt,\yd) {$/$};

\def\xb{52pt}
\def\xba{37pt}
\def\xbb{67pt}
\def\xbc{77pt}
\node[blockh3w1] at (\xb, \yb) {$\tilde{\msf{U}}$};
\draw     (\xxab, \yaa) -- (\xba, \yaa);
\draw     (\xxab, \yab) -- (\xba, \yab);
\draw     (\xxab, \yba) -- (\xba, \yba);
\draw     (\xxab, \ybb) -- (\xba, \ybb);
\draw     (\xab, \yca) -- (\xba, \yca);
\draw     (\xab, \ycb) -- (\xba, \ycb);
\node at (\xa,\ya) {$/$};
\node at (\xa,\yb) {$/$};
\node at (27pt,\yc) {$/$};

\def\xc{102pt}
\def\xca{87pt}
\def\xcb{117pt}
\def\xcc{127pt}
\node[blockh3w1] at (\xc, \yb) {$\msf{E}$};
\draw     (\xbb, \yaa) 	-- (\xca, \yaa);
\draw     (\xbb, \yab)  -- (\xca, \yab);
\draw     (\xbb, \yba) -- (\xca, \yba);
\draw     (\xbb, \ybb)  -- (\xca, \ybb);
\draw     (\xbb, \yca) 	-- (\xca, \yca);
\draw     (\xbb, \ycb)  -- (\xca, \ycb);
\node at (\xbc,\ya) {$/$};
\node at (\xbc,\yb) {$/$};
\node at (\xbc,\yc) {$/$};
\node at (\xc,\yd) {$/$};

\def\xd{152pt}
\def\xda{137pt}
\def\xdb{167pt}
\node[blockh3w1] at (\xd, \yb) {$\tilde{\msf{U}}^{-1}$};
\draw     (\xcb, \yaa) 	-- (\xda, \yaa);
\draw     (\xcb, \yab)  	-- (\xda, \yab);
\draw     (\xcb, \yba) 	-- (\xda, \yba);
\draw     (\xcb, \ybb)  	-- (\xda, \ybb);
\draw     (\xcb, \yca) 	-- (\xda, \yca);
\draw     (\xcb, \ycb)  	-- (\xda, \ycb);
\node at (\xcc,\ya) {$/$};
\node at (\xcc,\yb) {$/$};
\node at (\xcc,\yc) {$/$};

\def\xe{202pt}
\def\xea{187pt}
\def\xeb{217pt}
\node[blockh2w1] at (\xe, -75pt) {$\msf{U}_{=}^{-1}$};
\draw     (\xdb, \yca) 	-- (\xea, \yca);
\draw     (\xdb, \ycb)  	-- (\xea, \ycb);
\draw     (\xab, \yda) 	-- (\xea, \yda);
\draw     (\xab, \ydb)  	-- (\xea, \ydb);
\node at (\xe,\ya) {$/$};
\node at (\xe,\yb) {$/$};
\node at (177pt,\yc) {$/$};
\node at (227pt,\yc) {$/$};
\node at (227pt,\yd) {$/$};

\def\xf{247pt}
\def\xfa{237pt}
\def\xfb{257pt}
\node[block1] at (\xf, \yb) {$\msf{M}$};
\node[block1] at (\xf, \yc) {$\msf{M}$};
\node[block1] at (\xf, \yd) {$\msf{M}$};
\draw     (\xdb, \yba) 	-- (\xfa, \yba);
\draw     (\xdb, \ybb)  	-- (\xfa, \ybb);
\draw     (\xeb, \yca) 	-- (\xfa, \yca);
\draw     (\xeb, \ycb)  	-- (\xfa, \ycb);
\draw     (\xeb, \yda) 	-- (\xfa, \yda);
\draw     (\xeb, \ydb)  	-- (\xfa, \ydb);

\def\xh{302pt}
\def\xha{277pt}
\def\xhb{327pt}
\node[blockh1w2] at (\xh, \ya) {recovery}; 
\node[blockh3w5] at (\xh, \yc) {decoder};
\draw     (\xdb, \yaa) 	-- (\xha, \yaa);
\draw     (\xdb, \yab)  	-- (\xha, \yab);
\draw     (\xfb, \yb) 	-- (\xha, \yb);
\draw     (\xfb, \yc) 		-- (\xha, \yc);
\draw     (\xfb, \yd) 	-- (\xha, \yd);
\draw     (\xh, -10pt)  	-- (\xh, -20pt);
\node at (337pt,\ya) {$/$};

\def\xi{377pt}
\def\xia{347pt}
\def\xib{407pt}
\node[blockh4w3] at (\xi, \y) {}; 
\node at (\xi, \yb) {further};
\node at (\xi, \yc) {processing};
\draw     (\xhb, \yaa) 	-- (\xia, \yaa);
\draw     (\xhb, \yab)  	-- (\xia, \yab);
\node at (417pt,\ya) {$/$};

\def\xj{437pt}
\def\xja{427pt}
\node[block1] at (\xj, \ya) 	{$I$};
\draw     (\xib, \yaa) 	-- (\xja, \yaa);
\draw     (\xib, \yab)  	-- (\xja, \yab);
\end{tikzpicture}
  \end{center}
  \caption{Double-edge quantum-probability NFG representation of an entanglement-assisted QECC.}
  \label{fig:QP4_EAQECC_DE}
\end{figure*}


\section{(Tail-bitting) Convolutional \\ Stabilizer QECCs}
\label{sec:TBC_QECC:4}
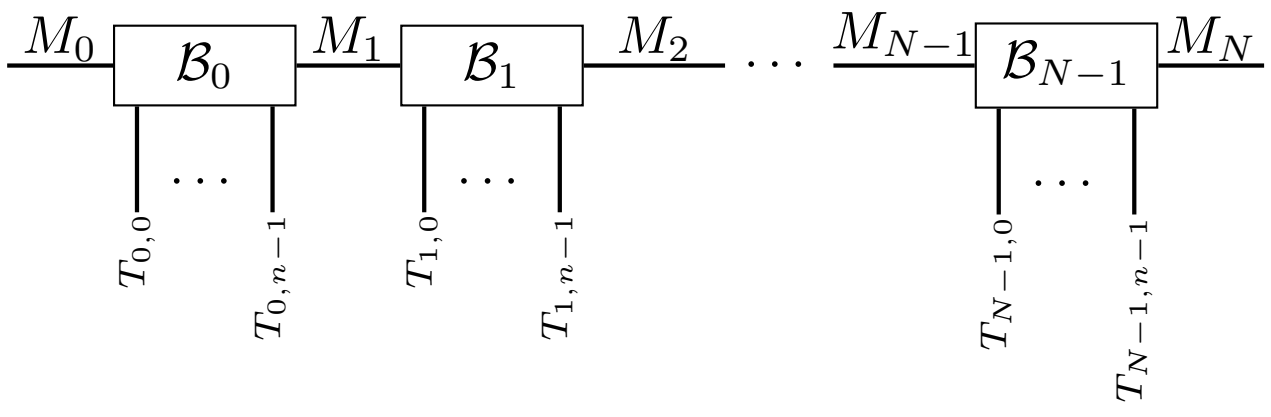
\begin{figure}
\begin{center}
\begin{tikzpicture}[nodes= {draw=none, fill=none}, 
						vertex/.style={rectangle, fill = white, draw, minimum height=0.5cm, minimum width=1.2cm},
						 labeled edge/.style={ sloped, midway, above=-0.1cm},
						 nearend edge/.style={ sloped, near end, above=-0.1cm},
						node distance=1.9cm]
\pgfmathsetmacro\r{2};
\pgfdeclarelayer{main}
\pgfdeclarelayer{back}
\pgfsetlayers{back,main}
\begin{pgfonlayer}{main}
\node  (m1) {};
\node [vertex] (f1) [right = 0.7cm of m1] {$\Cal{B}_0$};
\node (t100) [below =0.3cm of f1]{$\cdots$};
\node (t10) [below =0.7cm  of f1]{};
\node (t11) [left = 0.2cm of t10]{};\vtag{t11}{T_{0,0}};
\node (t12) [right = 0.2cm of t10]{};\vtag{t12}{T_{0,n-1}};
\node [vertex] (f2) [right of = f1]{$\Cal{B}_1$};
\node (t200) [below =0.3cm of f2]{$\cdots$};
\node (t20) [below =0.7cm  of f2]{};
\node (t21) [left = 0.2cm of t20]{};\vtag{t21}{T_{1,0}};
\node (t22) [right = 0.2cm of t20]{};\vtag{t22}{T_{1,n-1}};
\node (f3) [right of = f2, fill = white]{$\cdots$};
\node [vertex] (fm) [right of = f3]{$\Cal{B}_{N-1}$};
\node (tm00) [below =0.3cm of fm]{$\cdots$};
\node (tm0) [below =0.7cm  of fm]{};
\node (tm1) [left = 0.2cm of tm0]{};\vtag{tm1}{T_{N-1,0}};
\node (tm2) [right = 0.2cm of tm0]{};\vtag{tm2}{T_{N-1,n-1}};
\node (fm1) [right = 0.7cm of fm]{};

\path[draw, thick] (t11|-f1.south)--(t11) {};
\path[draw, thick] (t12|-f1.south)--(t12) {};
\path[draw, thick] (t21|-f2.south)--(t21) {};
\path[draw, thick] (t22|-f2.south)--(t22) {};
\path[draw, thick] (tm1|-fm.south)--(tm1) {};
\path[draw, thick] (tm2|-fm.south)--(tm2) {};

\path[draw, thick] (m1) -- (f1) node[labeled edge] {$M_0$};
\path[draw, thick] (f1) -- (f2) node[labeled edge] {$M_1$};
\path[draw, thick] (f2) -- (f3) node[labeled edge] {$M_2$};
\path[draw, thick] (f3) -- (fm) node[labeled edge] {$M_{N-1}$};
\path[draw, thick] (fm) -- (fm1) node[labeled edge] {$M_{N}$};
\end{pgfonlayer}


\end{tikzpicture}
\end{center}
\caption{Quantum-code NFG representation of a convolutional stabilizer QECC or a tail-biting convolutional stabilizer QECC when $M_0 = M_N$.}
\label{fig:Code-design_NFG_conv_QECC}
\end{figure}

In this section, we first show how to obtain the quantum-code representation of a (tail-biting) convolutional stabilizer QECC $\code{C}$ as in Fig.~\ref{fig:Code-design_NFG_conv_QECC}, 
and then demonstrate how to make use of the NFG representation to get the encoding and decoding circuits of $\code{C}$.
We provide the details for the case when the stabilizer label code is additive over $\GF{4}$ and has a single basic generator, which can be easily extended to the case when it has multiple generators and when it is linear over $\GF{4}$.

For (tail-bitting) convolutional stabilizer QECCs, we follow the notations of~\cite{forney2007convolutional},~\cite{poulin2009quantum}, and~\cite{pelchat2013degenerate}.
A convolutional stabilizer QECC is defined on a quantum system with a countably infinite ordered sequence of qubits.
A rate-$k/n$ convolutional stabilizer QECC $\tcode{C}$ is defined by its stabilizer $\set{S}$, 
where $\set{S}$ is the set of all shifts of the basic generators in the basic stabilizer $\set{S}_{\Rm{b}}$ by integer multiples of $n$. 
The basic stabilizer $\set{S}_{\Rm{b}}$ is a group generated by $n\!-\! k$ independent, commuting, and finite support Pauli operators such that $-I \notin \set{S}_{\Rm{b}}$.
Let the quaternary stabilizer label code $\codeB$ and basic quaternary stabilizer label code $\codeB_{\Rm{b}}$ of $\code{C}$ be defined as $\codeB \defeq L(\set{S})$ and $\codeB_{\Rm{b}} \defeq L(\set{S}_{\Rm{b}})$.
If $\codeB_{\Rm{b}}$ is the additive code over $\GF{4}$ generated by $\tmb{g}^{(i)}$, $i\in\set{I}_{n-k}$, then we refer $\tmb{g}^{(i)}$, $i\in\set{I}_{n-k}$, as the {\em basic generators} of $\code{C}$.
For any $\tmb{g} \in \codeB$, we group $n$ symbols to form small blocks and denote the {\em $n$-symbol form} of $\tmb{g}$ as
\begin{equation}
\label{eq:g_conv}
\tmb{g} \defeq ( \ldots, \vect{g}{0},\ \vect{g}{1},\ \cdots,\ \vect{g}{\nu}, \ldots ),
\end{equation}
where $\vect{g}{i} \in \GF{4}^{n}$, $i \in \mbb{Z}$, and $\nu\in\mbb{Z}_{\geq 0}$, and 
where we use ``$( \ )$'' to indicate that there are countably infinitely many symbols and only non-zero blocks are shown.

A rate-$k/n$ tail-biting convolutional (TBC) stabilizer QECC $\code{C}$ of length $N \cdot n$ with its stabilizer $\set{S}$ and basic stabilizer $\set{S}_{\Rm{b}}$ can be obtained from the convolutional stabilizer QECC $\tcode{C}$ by fixing the block length $N \cdot n$ and wrapping all the symbols at the end to the beginning.
The quaternary stabilizer label code and quaternary basic stabilizer label code are $\codeB \defeq L(\set{S})$ and $\codeB_{\Rm{b}} \defeq L(\set{S}_{\Rm{b}})$, respectively.
For any $\vg \in \codeB$, denote the {\em $n$-symbol form} of $\vg$ as
\begin{equation}
\label{eq:g_TBC}
\vg \defeq [\vect{g}{0},\ \vect{g}{1},\ \ldots,\ \vect{g}{N-1} ],
\end{equation}
where $\vect{g}{i} \in \GF{4}^{n}\ \forall i \in \mbb{Z}_{N}$,
and let the {\em $n$-symbol cyclic left shift} of $\mb{g}$ by $p\in\mbb{Z}_{N}$ times be the left shift of $\mb{g}$ by $p \cdot n$ symbols, i.e., 
\begin{equation}
\label{eq:g_p_TBC}
\vg{(p)} \defeq [\vect{g}{p},\ \vect{g}{p+1},\ \ldots,\ \vect{g}{p+N-1} ],
\end{equation}
where the index $i$ of $\vect{g}{i}$ is taken modulo $N$.
Similarly, if $\codeB_{\Rm{b}}$ is the additive code over $\GF{4}$ generated by $\vg^{(i)}$, $i\in\set{I}_{n-k}$, then we refer to $\vg^{(i)}$, $i\in\set{I}_{n-k}$, as the {\em basic generators} of $\code{C}$ and $\codeB$ is the $\GF{2}$-span of $\mb{G}_{\codeB} \defeq \{ \vg^{(i)}(p)\ |\ i\in\set{I}_{n-k}, p\in\mbb{Z}_{N} \}$.
We now introduce the constraint length of a generator and the overall constraint length of a TBC stabilizer QECC which are related the number of memory qubits we need for the quantum-code NFG representation.
\begin{definition}[Constraint length and overall constraint length]
The {\em constrain length} $\nu(\vg)$ of a generator $\vg$ with its $n$-symbol form $[\vect{g}{0},\ \vect{g}{1},\ \ldots,\ \vect{g}{N-1} ]$ is defined to be the minimum non-negative integer $\nu$ such that 
\begin{equation*}
\vect{g}{0} \neq \mb{0},\ 
\vect{g}{\nu} \neq \mb{0}, \text{ and } 
\vect{g}{i}=\mb{0}\ \forall\nu < i < N.
\end{equation*}
As for a basic generator, we can always choose the generator $\vg$ which has the minimum constraint length among its $n$-qubit cyclic shifts $\vg{(p)}$, $p \in \mbb{Z}_{N}$, since they span the same set of generators in the stabilizer. 
For a rate-$k/n$ TBC stabilizer QECC $\code{C}$ with basic generators $\vg^{(i)}$ and constrain lengths $\nu_{i} \defeq \nu(\vg^{(i)})$, $i\in\set{I}_{n-k}$, the {\em overall constraint length} $\nu$ and the {\em maximum constraint length} $\nu_{\Rm{max}}$ of the code $\code{C}$ are defined as, respectively,
\begin{equation*}
\nu \defeq \sum_{i=0}^{n-k-1} \nu_{i}\ \text{ and }\ \nu_{\Rm{max}} \defeq \max_{i\in\set{I}_{n-k}} \nu_i.
\end{equation*}
\defend
\end{definition}

In the following three sections, we discuss how we find the quantum-code NFG representation of a TBC stabilizer QECC  $\cC$ for three cases w.r.t. its stabilizer label code $\cB$, i.e., 
$\cB$ is additive over $\GF{4}$ with a single basic generator;
$\cB$ is additive over $\GF{4}$ with multiple basic generators; or  
$\cB$ is linear over $\GF{4}$.

\subsection{Quantum-code NFG Representation: additive stabilizer label code over $\GF{4}$ with a single basic generator}
\label{sec:TBC_QECC_single_gen:4a}
Consider a rate-$k/n$ TBC stabilizer QECC $\overline{\code{C}}$ of block length $N \cdot n$ which has a single basic generator
\begin{equation*}
\vg \defeq [\vect{g}{0},\ \vect{g}{1},\ \ldots,\ \vect{g}{N-1} ],
\end{equation*}
where $\vect{g}{i} \in \GF{4}^{n}\ \forall i \in \mbb{Z}_{N}$.

\begin{example}(This example is inspired by an example in~\cite[Table VI]{forney2007convolutional})
\label{eg: running_eg_TBC_QECC}
Consider a rate-$2/3$ convolutional stabilizer QECC $\cC$ whose generator set $\cS$ is the set of all shifts of integer multiples of three symbols of the Pauli operators in $\cS_{\Rm{b}}$ which is generated by 
\begin{align*}
\mb{P}=(\ldots|YYY|IYX|IYZ|YZX|\ldots),
\end{align*}
where only non-identity blocks are shown.
The quaternary stabilizer label code $\cB$ of $\cC$ is the additive code over $\GF{4}$ which is the $\GF{2}$-span of the set of all shifts by integer multiples of three symbols of the single generator 
\begin{equation*}
\tmb{g}=(\ldots|111|01\om|01\ob|1\ob\om|\ldots).
\end{equation*}
Note that $\mb{g}=L(\mb{P})$.

Consider the rate-$1/3$ tail-biting code $\code{C}$ of block length $21$ which is obtained from $\tcode{C}$. 
The quaternary stabilizer label code $\codeB$ of $\code{C}$ is the additive code over $\GF{4}$ which is the $\GF{2}$-span of the set of all $3$-symbol cyclic left shifts of $\vg$, i.e., 
\begin{align*}
    \vg(0)
      &= [111|01\om|01\ob|1\ob\om|000|000|000] \ , \\
    \vg(1)
      &= [01\om|01\ob|1\ob\om|000|000|000|111] \ , \\
    \vg(2)
      &= [01\ob|1\ob\om|000|000|000|111|01\om] \ , \\
    \vdots\quad
      &\hskip0.20cm \vdots
       \hskip2.5cm \vdots \\
    \vg(6)
      &= [000|111|01\om|01\ob|1\ob\om|000|000] \ .
  \end{align*}
Note that here $N \defeq 7$, $n \defeq 3$, and $\nu = 3$ following our notations.
One can check that the code $\codeB$ is indeed self-orthogonal under the trace inner product by checking $\langle \vg(0), \vg(p) \rangle_{\tr} = 0\ \forall p\in\mbb{Z}_{7}$.
\exampleend
\end{example}

Here follows how we find the quantum-code representations of the TBC stabilizer QECC $\code{C}$, 
i.e., how to find the codes $\codeB_i$, $i\in\mbb{Z}_{N}$, in Fig.~\ref{fig:Code-design_NFG_conv_QECC} such that the NFG represents the stabilizer label code $\codeB$ of $\code{C}$.
In fact, we use an algorithm (see Algorithm~\ref{alg:TBC_QECC_single_generator}) to generate a code $\codeBe$ which is assigned to $\codeB_i$, $i\in\mbb{Z}_{N}$, where $\codeBe$ is the $\GF{2}$-span of the generator matrix
\begin{align}
  \label{eq:single_generator_matrix_form}
  \vect{G}{\codeBe} = 
  \vect{G}{\vg}
    &\defeq 
       \left[ 
         \begin{array}{c|c|c}
           \vect{s}{0} & \vect{g}{0} & \vect{s}{1}\\
           \vect{s}{1} & \vect{g}{1} & \vect{s}{2}\\
           \vdots & \vdots & \vdots\\
           \vect{s}{\nu} & \vect{g}{\nu} & \vect{s}{\nu+1}
         \end{array}\right],
\end{align}
where $\vect{s}{0} \defeq \mb{0}$ and $\vect{s}{\nu+1} \defeq \mb{0}$, and where $\vect{s}{\ell}$, $\ell\in\set{I}_{\nu+1}^{+}$, are determined by our algorithm. 
Note that such a generator matrix implies that the codeword $\mb{g}$ corresponds to the following state $\rightarrow$ $n$-symbol codeword $\rightarrow$ state $\rightarrow$ $n$-symbol codeword $\rightarrow$ $\cdots$ $\rightarrow$ state sequence:
\begin{align*}
  \mb{0} \! = \! \vect{s}{0}
    &\rightarrow 
  \vect{g}{0} 
    \rightarrow 
  \vect{s}{1} 
    \rightarrow 
  \vect{g}{1} 
    \rightarrow 
  \cdots 
    \rightarrow 
  \vect{g}{\nu} 
    \rightarrow 
  \vect{s}{\nu+1} \! = \! \mb{0} \\
    &\rightarrow 
  \vect{g}{\nu+1} \! = \! \mb{0} 
    \rightarrow 
  \vect{s}{\nu+2} \! = \! \mb{0} 
    \rightarrow 
  \cdots \\
    &\rightarrow 
  \vect{g}{N-1} \! = \! \mb{0} 
    \rightarrow 
  \vect{s}{0} \! = \! \mb{0}.
\end{align*}
We call the code $\codeBe$ the {\em elementary code} of $\code{C}$ and it is designed to be self-orthogonal under the trace inner product.

\begin{algorithm}
\caption{Generate states for a TBC stabilizier QECC with a single basic generator 
(let $[\mb{v}]_{i}$ denote the $i^{\Rm{th}}$ symbol of the vector $\mb{v}$ for any vector $\mb{v}$).}
\label{alg:TBC_QECC_single_generator}
\begin{algorithmic}[1]
\REQUIRE $\mb{g}=[\vect{g}{0},\ \vect{g}{1},\ \ldots,\ \vect{g}{\nu},\ \mb{0},\ \ldots,\ \mb{0}]$ 
\ENSURE $\tilde{\set{S}} \defeq \left\{\vect{s}{\ell}\in\GF{4}^{\nu} \ \middle|\ \ell\in\set{I}_{\nu+1}^{+} \right\}$
\FOR {$\ell, \ell'\in\set{I}_{\nu+2}$}
	\STATE $\gamma_{\ell, \ell'} :=\sum_{\tell=1}^{\min\{\ell,\ell'\}}{\langle \vect{g}{\ell - \tell}, \vect{g}{\ell' - \tell} \rangle_{\tr} }$

\ENDFOR
\FOR {$i:=1$ to $\nu$}
	\STATE $\mb{s}_i:=\mb{0}$
\ENDFOR
\FOR {$i:=1$ to $\nu$}
	\STATE $[{\mb{s}}_{i}]_i := 1$
	\FOR {$j:=i+1$ to $\nu$}
			\STATE $[{\mb{s}}_{j}]_i := \om \cdot \big({\gamma}_{j,i} - \sum_{\ell=1}^{i-1}{\langle {[\mb{s}}_{j+1}]_{\ell}, {[\mb{s}}_{i}]_{\ell} \rangle_{\tr} } \big)\in\GF{4}$
	\ENDFOR
\ENDFOR

\end{algorithmic}
\end{algorithm}

\begin{remark}
\label{rm:Alg1}
Algorithm~\ref{alg:TBC_QECC_single_generator} is designed such that
\begin{equation}
\label{eq:gamma_state_relation}
\gamma_{j,i} = \langle \vect{s}{j}, \vect{s}{i} \rangle_{\msf{T}}\ \forall j,i\in\set{I}_{\nu+2}.
\end{equation}
Since $\gamma_{j,i} = \gamma_{i,j}$, w.l.o.g., we assume $0 \leq i \leq j \leq \nu+1$.
\begin{itemize}
\item For $i=0$, we have $\gamma_{j,i} = 0$ according to its definition.
Thus
\begin{equation*}
\langle \vect{s}{j}, \vect{s}{i} \rangle_{\tr} 
= \langle \vect{s}{j}, \vect{s}{0} \rangle_{\tr} 
= \langle \vect{s}{j}, \mb{0} \rangle_{\msf{T}} 
= 0 
= \gamma_{j,i}.
\end{equation*}
\item For $1 \leq i \leq j \leq \nu$, according to line 10 of Algorithm~\ref{alg:TBC_QECC_single_generator}, we have that
$\langle [\vect{s}{j}]_{i}, [\vect{s}{i}]_{i} \rangle_{\tr} 
= \gamma_{j,i} - \sum_{\ell=1}^{i-1} \langle [\vect{s}{j}]_{\ell}, [\vect{s}{i}]_{\ell} \rangle_{\tr}$
which implies
\begin{equation*}
\gamma_{j,i} 
= \sum_{\ell=1}^{i}{\langle [\vect{s}{j}]_{\ell}, [\vect{s}{i}]_{\ell} \rangle_{\tr} }
\stackrel{(a)}{=} \langle \vect{s}{j}, \vect{s}{i} \rangle_{\tr},
\end{equation*}
where step $(a)$ follows from the fact that $[\vect{s}{i}]_{\ell} = 0$ for all $i < \ell \leq \nu$.
\item For $j=\nu+1$, on the one hand, we have
\begin{equation*}
\langle \vect{s}{j}, \vect{s}{i} \rangle_{\tr}
= \langle \vect{s}{\nu+1}, \vect{s}{i} \rangle_{\tr}
= \langle \mb{0}, \vect{s}{i} \rangle_{\msf{T}}
= 0.
\end{equation*}
On the other hand, we have
\begin{equation*}
\gamma_{j,i} 
= \sum_{\ell = 1}^{i} \langle \vect{g}{\nu+1-\ell}, \vect{g}{i-\ell} \rangle_{\tr} 
= \langle \vg(\nu+1-i), \vg(0) \rangle_{\tr} 
= 0.
\end{equation*}
\end{itemize}

The set $\tilde{\set{S}}$ is linearly independent over $\GF{2}$ since the horizontal/vertical stacking of the state vectors $\vect{s}{\ell}$, $\ell\in\set{I}_{\nu+1}^{+}$, forms a lower/upper triangle matrix with non-zero diagonal entries.
\end{remark}

\begin{theorem}
\label{thm:TBC_QECC_code_design_NFG_single_g}
Consider a TBC stabilizer QECC $\code{C}$ with a single basic generator $\vg$, quaternary stabilizer label code $\codeB$, and overall constraint length $\nu$.
Let $\codeB_{\graph{G}}$ be the code represented by the quantum-code NFG $\graph{G}$ as in Fig.~\ref{fig:Code-design_NFG_conv_QECC} with $\codeB_i = \codeBe$, $i \in\mbb{Z}_{N}$, and $\codeBe$'s generator matrix is in the form of~\eqref{eq:single_generator_matrix_form} where the state vectors are generated by Algorithm~\ref{alg:TBC_QECC_single_generator}.
Then we have 
\begin{itemize}
\item the code $\codeBe$ is self-orthogonal under the trace inner product,
\item the code $\codeB_{\graph{G}}$ represented by the NFG $\graph{G}$ is exactly $\codeB$,
\item the number of memory qubits required for the state vectors in $\codeBe$ is $\nu$. 
\end{itemize}
\end{theorem}
\begin{proof}
See Appendix~\ref{appendix:thm2}.
\end{proof}

\begin{example}
\label{eg: running_eg_TBC_QECC2}
We continue to consider the code in Example~\ref{eg: running_eg_TBC_QECC}.
The state vectors generated by Algorithm~\ref{alg:TBC_QECC_single_generator} are
\begin{align*}
    \vect{s}{1} = [100], 
    \quad 
    \vect{s}{2} = [\om 10],
    \quad 
    \vect{s}{3} = [\om 01].
\end{align*}
The elementary code $\codeBe$ to be assigned to $\codeB_i$, $i \in \mbb{Z}_{N}$, in Fig.~\ref{fig:Code-design_NFG_conv_QECC} is the $\GF{2}$-span of the generator matrix
\begin{align*}
    \setlength{\dashlinegap}{2pt}
    \vect{G}{\codeBe}
      &= \left[
           \begin{array}{ccc | ccc | ccc}
             0&0&0& 1&1&1& 1&0&0\\
             1&0&0& 0&1&\om&  {\om}&1&0\\
             {\om}&1&0&  0&1&\ob&  \om&0&1\\
             \om&0&1&  1&\ob&\om& 0&0&0
           \end{array}
         \right].
\end{align*}
\exampleend
\end{example}
\subsection{Quantum-code NFG Representation: additive stabilizer label code over $\GF{4}$ with multiples basic generators}
\label{sec:TBC_QECC_multiple_gen:4b}
We first introduce some index sets used in this section.
For any vector $\mb{v} = [v_0,\ v_2,\ \ldots,\ v_{\ell-1}]$ of length $\ell$, we define the following index sets
\begin{align*}
\set{I}_{\mb{v}}^{+} &\defeq \left\{ (i,j)\ \middle|\ i\in\set{I}_{\ell},\ j\in\set{I}_{v_i+1}^{+} \right\},\\
\set{I}_{\mb{v}} &\defeq \left\{ (i,j)\ \middle|\ i\in\set{I}_{\ell},\ j\in\set{I}_{v_i+1} \right\}.
\end{align*}
Consider a rate-$k/n$ TBC stabilizer QECC $\code{C}$ of block length $N \cdot n$ which has multiple basic generators $\vg^{(i)}$ and constraint lengths $\nu_{i} \defeq \nu(\vg^{(i)})$ for $i\in\set{I}_{n-k}$.
Then the overall constraint length of $\code{C}$ is $\nu = \sum_{i=0}^{n-k-1}\nu_i$. 
Let the constraint length vector be $\bm{\nu} \defeq [\nu_0,\ \nu_1,\ \ldots,\ \nu_{n-k-1}]$.
For all $i\in\set{I}_{n-k}$, suppose $\vg^{(i)}$ has the $n$-qubit form 
\begin{equation*}
\vg^{(i)} \defeq [\vect{g}{0}^{(i)},\ \vect{g}{1}^{(i)},\ \ldots,\ \vect{g}{N-1}^{(i)} ],
\end{equation*}
where $\vect{g}{j}^{(i)} \in \GF{4}^{n}\ \forall j \in \mbb{Z}_{N}$
and where $\vect{g}{j}^{(i)} = \mb{0}\ \forall \nu_{i}<j<N$.

We can make use of Algorithm~\ref{alg:TBC_QECC_single_generator} with suitable re-labeling.
The elementary code $\codeBe$ to be assigned to $\codeB_i$, $i \in \mbb{Z}_{N}$, in Fig.~\ref{fig:Code-design_NFG_conv_QECC} is the $\GF{2}$-span of the generator matrix
\begin{equation}
\label{eq:multiple_generator_matrix_form}
\vect{G}{\codeB_\Rm{e}} \defeq \left[\begin{array}{c}
\vect{G}{\vg^{(0)}}\\
\vect{G}{\vg^{(1)}}\\
\vdots\\
\vect{G}{\vg^{(n-k-1)}}
\end{array}\right],
\end{equation}
where
\begin{align*}
  \vect{G}{\vg^{(i)}}
    &\defeq 
       \left[ 
         \begin{array}{c|c|c}
           \vect{s}{0}^{(i)} & \vect{g}{0}^{(i)} & \vect{s}{1}^{(i)}\\
           \vect{s}{1}^{(i)} & \vect{g}{1}^{(i)} & \vect{s}{2}^{(i)}\\
           \vdots & \vdots & \vdots\\
           \vect{s}{\nu_i}^{(i)} & \vect{g}{\nu_i}^{(i)} & \vect{s}{\nu_i+1}^{(i)}
         \end{array}\right],
\end{align*}
and where $\vect{s}{0}^{(i)} \defeq \mb{0}$ and $\vect{s}{\nu_i+1}^{(i)} \defeq \mb{0}$, and where $\vect{s}{j}^{(i)}$, $(i,j)\in\set{I}_{\bm{\nu}}$, are state vectors generated in the following procedure.

We now introduce the generalized procedure to generate the elementary code $\codeBe$ for the code design NFG representation:
\begin{itemize}
\item Let the index mapping $\Cal{I}:\ \set{I}_{\bm{\nu}}^{+} \rightarrow \set{I}_{\nu+1}^{+}$ be defined as
\begin{equation}
\label{eq:relabel_mapping_I}
\Cal{I}(i,j) \defeq \sum_{\ell=0}^{i-1} \nu_{\ell} + j.
\end{equation}
Since the mapping $\Cal{I}$ is bijective, we denote its inverse mapping as $\Cal{I}^{-1}$, 
i.e., $\Cal{I}^{-1}(p) = (i,j)$ if and only if $\Cal{I}(i,j) = p$.
\item First, we define the $\tilde{\gamma}$-values as
\begin{equation*}
\tilde{\gamma}_{(i,j), (i',j')} \defeq 
\sum_{\ell = 1}^{\min\{j,j'\}} \langle \vect{g}{j - \ell}^{(i)}, \vect{g}{j'-\ell}^{(i')} \rangle_{\tr}\ \forall (i,j), (i',j')\in\set{I}_{\bm{\nu}}
\end{equation*}
and $\tilde{\gamma}_{(i,j), (i',j')} \defeq 0$ otherwise.
\item Then, we re-define the $\gamma$-values in line 2 of Algorithm~\ref{alg:TBC_QECC_single_generator} as
\begin{equation}
\label{eq:gamma_multiple_case}
\gamma_{\ell,\ell'} \defeq 
\tilde{\gamma}_{(i,j), (i',j')}\ \forall \ell,\ell'\in\set{I}_{\nu+1}^{+},
\end{equation}
where $(i,j) \defeq \Cal{I}^{-1}(\ell)$, $(i',j') \defeq \Cal{I}^{-1}(\ell')$;
and $\gamma_{\ell,\ell'} \defeq 0$ otherwise.
\item Finally, with Algorithm~\ref{alg:TBC_QECC_single_generator} and the $\gamma$-values defined in~\eqref{eq:gamma_multiple_case}, we can obtain $\tilde{\set{S}} = \left\{ \vect{s}{\ell} \in\GF{4}^{\nu}\ \middle|\ \ell\in\set{I}_{\nu+1}^{+}\right\}$.
By the label mapping $\Cal{I}$, we get the state vectors $\vect{s}{j}^{(i)} \defeq \vect{s}{p}$, where $p \defeq \Cal{I}(i,j)$.
Hence we can obtain the generator matrix as in~\eqref{eq:multiple_generator_matrix_form} and the elementary code $\codeBe$.
\end{itemize}

\begin{theorem}
\label{thm:TBC_QECC_code_design_NFG_multiple_g}
Consider a TBC stabilizer QECC $\code{C}$ with multiple basic generators, quaternary stabilizer label code $\codeB$, and overall constraint length $\nu$.
Let $\codeB_{\graph{G}}$ be the code represented by the quantum-code NFG $\graph{G}$ as in Fig.~\ref{fig:Code-design_NFG_conv_QECC} with $\codeB_i = \codeBe$, $i \in\mbb{Z}_{N}$, and $\codeBe$'s generator matrix is in the form of~\eqref{eq:multiple_generator_matrix_form} where the state vectors are generated by Algorithm~\ref{alg:TBC_QECC_single_generator} together with the relabel mapping $\set{I}$ in~\eqref{eq:relabel_mapping_I}.
Then we have 
\begin{itemize}
\item the code $\codeBe$ is self-orthogonal under the trace inner product,
\item the code $\codeB_{\graph{G}}$ represented by the NFG $\graph{G}$ is exactly $\codeB$,
\item the number of memory qubits required for the state vectors in $\codeBe$ is $\nu$. 
\end{itemize}
\end{theorem}
\begin{proof}
All the equations in the previous section still hold after relabeling using $\Cal{I}$.
If we undo the relabeling process,~\eqref{eq:gamma_state_relation} becomes
\begin{equation}
\tilde{\gamma}_{(i,j),(i',j')} = \big\langle \vect{s}{j}^{(i)}, \vect{s}{j'}^{(i')} \big\rangle_{\tr}\ \ \ \forall (i,j),(i',j')\in\set{I}_{\bm{\nu}+\mb{1}}.
\end{equation}
For any $(i,j),(i',j')\in\set{I}_{\bm{\nu}}$, we now have
\begin{align*}
&\Big\langle \left[\vect{s}{j}^{(i)},\ \vect{g}{j}^{(i)},\ \vect{s}{j+1}^{(i)}\right], \left[\vect{s}{j'}^{(i')},\ \vect{g}{j'}^{(i')},\ \vect{s}{j'+1}^{(i')}\right] \Big\rangle_{\tr}\\
&=\ \big\langle \vect{s}{j}^{(i)}, \vect{s}{j'}^{(i')}\big\rangle_{\tr} + \big\langle \vect{g}{j}^{(i)}, \vect{g}{j'}^{(i')}\big\rangle_{\tr} +  \big\langle \vect{s}{j+1}^{(i)}, \vect{s}{j'+1}^{(i')}\big\rangle_{\tr}\\
&=\ \tilde{\gamma}_{(i,j),(i',j')} + \big\langle \vect{g}{j}^{(i)}, \vect{g}{j'}^{(i')}\big\rangle_{\tr} + \tilde{\gamma}_{(i,j+1),(i',j'+1)}\\
&=\ 2 \cdot \tilde{\gamma}_{(i,j+1),(i',j'+1)}\\
&=\ 0\ (\text{mod }2).
\end{align*}
The total number of state vectors which are outputs of Algorithm~\ref{alg:TBC_QECC_single_generator} is $\nu$ and the number of memory qubits for the state vectors is the same as the number of state vectors.
Hence, we require $\nu$ memory qubits.
The other parts of the proof remain the same compared with the proof of Theorem~\ref {thm:TBC_QECC_code_design_NFG_single_g}.
\end{proof}
\begin{example}[A TBC stabilizer QECC in~\cite{lidar2013quantum},\cite{wilde2009quantum}]
Consider a rate-$1/3$ TBC stabilizer QECC $\code{C}$ with two basic generators  
\begin{align*}
g_0 &= [\ldots | \ob 0 \ob | 0 \ob \ob | \ldots],\\ 
g_1 &= [\ldots | 0 \om \om | \om 0 \om | \ldots].
\end{align*}
Note that the overall constraint length is $\nu = \nu_0 + \nu_1 = 1+1 = 2$.

Using the above mentioned method, we obtain the generator matrix of the elementary code $\codeBe$ to be assigned to $\cB_i$, $i\in\mbb{Z}_{N}$, in Fig.~\ref{fig:Code-design_NFG_conv_QECC}, where $\codeBe$ is the $\GF{2}$-span of the following matrix
\begin{equation*}
\vect{G}{\codeB_\Rm{e}} = \left[\begin{array}{cc | ccc | cc}
0&0 & \ob &0&\ob & 1&0\\
1&0 & 0&\ob&\ob & 0&0\\
0&0 & 0&\om&\om & \om&1\\
\om&1 & \om&0&\om & 0&0
\end{array}\right].
\end{equation*}
\end{example}
\subsection{Quantum-code NFG Representation: linear stabilizer label code over $\GF{4}$}
\label{sec:TBC_QECC_linear:4c}
Consider a rate-$k/n$ TBC stabilizer QECC $\code{C}$ of block length $N \cdot n$ whose stabilizer label code $\codeB$ and basic stabilizer label code $\codeB_{\Rm{b}}$ are linear over $\GF{4}$.
Suppose the code $\codeB_{\Rm{b}}$ is the $\GF{4}$-span of $\frac{n-k}{2}\in\mbb{Z}_{\geq 0}$ basic generators $\vg^{(i)}$, $i\in\set{I}_{(n-k)/2}$.
For a TBC stabilizer QECC whose quaternary stabilizer label code is linear, we use the following modified procedure to generate the elementary code $\codeBe$ for the code design NFG representation:
\begin{itemize}
\item replace ``$n\!-\!k$'' by ``$(n\!-\!k)/2$'' everywhere;
\item replace the trace inner product by the Hermitian inner product everywhere;
\item at the end of Algorithm~\ref{alg:TBC_QECC_single_generator}, for all $i$, if $\gamma_{i,i} - \langle \vect{s}{i}, \vect{s}{i} \rangle_{\herp} \neq 0$, add one symbol to all $\vect{s}{\ell}\in\tset{S}$ and assign value ``$\gamma_{i,i} - \langle \vect{s}{i}, \vect{s}{i} \rangle_{\herp}$'' to $\vect{s}{i}$ and ``$0$'' to other vectors for the new symbol.
This step makes sure that the elementary code $\codeBe$ is even since any quaternary linear code which is self-orthogonal under the Hermitian inner product must be even, i.e., all codewords are of even weights.
\end{itemize}
Note that the overall constraint length of $\code{C}$ is $\nu = 2\cdot\left(\sum_{i=0}^{(n-k)/2-1}\nu_i\right)$, where $\nu_i \defeq \nu(\vg^{(i)})$.
With the above modified steps, the size of $\tset{S}$ is $\nu/2$ and the number of qubits for memory states is at lease $\nu/2$ when no symbol is added at the last step and at most $\nu$ when one symbol is added for each state vector $\vect{s}{j}$, $j\in\set{I}_{\nu/2+1}^{+}$, at the last step.
\begin{example}[Rate-$1/3$ TBC stabilizer QECC]
\label{eg:linear_3_1_3_code}
Consider a rate-$1/3$ TBC stabilizer QECC $\code{C}$ constructed from~\cite[Table VI]{forney2007convolutional} whose basic quaternary stabilizer label code $\codeB_{\Rm{b}}$ is linear which is generated by the single basic generator
\begin{equation*}
\vg =  [\ldots | 111 | 1 \om \ob | \ldots].
\end{equation*}
The code $\code{C}$ has two basic generators $\om \vg$ and $\ob \vg$ since the code $\codeB_{\Rm{b}}$ can be viewed as the $\GF{2}$-span of $\om \vg$ and $\ob \vg$.
The stabilizer label code $\codeB$ of $\code{C}$ is the $\GF{2}$-span of the $3$-symbol shifts of $\om \vg$ and $\ob \vg$.
Using the above modified version of Algorithm~\ref{alg:TBC_QECC_single_generator}, we obtain the generator matrix of the elementary code $\codeBe$ to be assigned to $\cB_i$, $i\in\mbb{Z}_{N}$, in Fig.~\ref{fig:Code-design_NFG_conv_QECC}, where $\codeBe$ is the $\GF{4}$-span of the following matrix
\begin{equation*}
\vect{G}{\codeBe} = \left[\begin{array}{c | ccc | c}
0 & 1 & 1 & 1 & 1\\
1 & 1 & \om & \ob & 0
\end{array}\right].
\end{equation*}
Since all the rows $\vect{G}{\codeBe}$ are already even, there is no need to add extra symbols in the last step.
\end{example}
\subsection{Encoding and Decoding Circuits of Convolutional QECCs}
\label{sec:TBC_QECC_enc_dec:4d}
We follow the notations and encoding circuit diagram in~\cite{poulin2009quantum} and~\cite{pelchat2013degenerate}.
Consider an $\llbracket n, k, d; m \rrbracket$-convolutional stabilizer QECC $\cC$ with minimum distance $d$ where each encoding block takes $k$-qubit information input, $m$-qubit memory input and produces $n$-qubit output and $m$-qubit memory output.
The parameters of $\code{C}$ can also be specified as $\llbracket n, k; m \rrbracket$ when the minimum distance is not known. 
The $j^{\Rm{th}}$ encoding block $\msf{U}$ in Fig.~\ref{fig:QP6_QCC_DE} shows the $j^{\Rm{th}}$ block of the encoding circuit.
There are $(N+t)$ such identical encoding blocks $V$ in total for the circuit diagram of a quantum convolutional encoder where the $j^{\Rm{th}}$ and $(j+1)^{\Rm{th}}$ blocks are connected by the $m$-qubit memory $M_j$.
The first $m$-qubit memory $M_0$ is set to be $|0\rangle^{\otimes m}$ and for the last $t$ blocks the $k$-qubit information $L_j$ is set to be $|0\rangle^{\otimes k}$.

We construct of encoding and decoding circuits of the convolutional stabilizer QECC $\cC$ by the following algorithm.

\begin{algorithm}
\caption{Algorithm to find a decoding circuit given some rows of an encoding matrix (signs tracking and correction are omitted). 
Applying the Hadamard, phase gate on the $i^{\Rm{th}}$ qubit, and CNOT gate on the $j^{\Rm{th}}$ qubit controlled by the $i^{\Rm{th}}$ qubit are abbreviated as $\Rm{H}_i$, $\Rm{P}_i$, and $\Rm{C}_{i,j}$, respectively.}
\label{alg:Gen_dec_circuit_Gaussian_elimination}
\begin{algorithmic}[1]
\REQUIRE an encoding matrix $V$ and a set of indices of the known rows $\Cal{I}$

\ENSURE  a list of quantum gates for a possible decoding circuit 
\FOR {$\Cal{I} \neq \emptyset$}
	\STATE Find the smallest $i$ such that $i\in\Cal{I}$ or ${i+n}\in\Cal{I}$. 
	\STATE Let $j:=i$ or $j:=i+n$ such that $j\in\Cal{I}$.
    \STATE Use Hadamard and phase gates to eliminate the nonzeros in the $Z$ part of $\mb{v}_{j}$.
    \STATE Use CNOT gates eliminate nonzeros in the $X$ part of $\mb{v}_{j}$ with only one nonzero entry left and swap it to the $i^{\Rm{th}}$ qubit if necessary.
    \STATE If $j>n$, then use Hadamard gate to move the nonzero entry of $\mb{v}_{j}$ from $X$ part to $Z$ part.
    \STATE Remove $j$ from $\Cal{I}$.
\ENDFOR
\end{algorithmic}
\end{algorithm}

\begin{example}[$\llbracket 3,1,3;1 \rrbracket$-convolutional stabilizer QECC] 
Consider the $\llbracket 3,1,3;1 \rrbracket$-convolutional stabilizer QECC $\code{C}$ in Example~\ref{eg:linear_3_1_3_code},
whose minimum distance is $3$ and number of memory qubits for each encoding block is $1$.

Suppose the matrix $V$ is the encoding matrix associated with a Clifford transformation $\Cal{V}$ for each encoding block.
Our target is to find the encoding matrix $V$ for each of $\codeB_{i}$, $i\in\mbb{Z}_{N}$.

From the generator matrix $\vect{G}{\codeBe}$, we know there are some input-output pairs as listed below
\begin{align*}
&\hspace{10pt} M_{j-1} \hspace{10pt} P_j \hspace{23pt} M_j\\
&\left[\begin{array}{ c | ccc | c }
\om & \om & \ob & 1 & 0\\
\ob & \ob & 1 & \om & 0\\
0 & \om & \om & \om & \om\\
0 & \ob & \ob & \ob & \ob
\end{array}\right]
\end{align*}
We can assume that the following row-vectors after converting into binary vectors are in the $\GF{2}$-span of $[\vect{I}{4}\ |\ V]$:
\begin{equation*}
\left[\begin{array}{cccc | cccc}
\om & 0 & 0 & 0 & \om & \ob & 1 & 0\\
\ob & 0 & 0 & 0 & \ob & 1 & \om & 0\\
0 & \ob & 0 & 0 & \om & \om & \om & \om\\
0 & 0 & \ob & 0 & \ob & \ob & \ob & \ob
\end{array}\right].
\end{equation*}
Some rows of the encoding matrix $V$ can be filled in according to the above matrix
\begin{equation*}
V = \left[\begin{array}{cccc : cccc}
1&0&1&0& 0&1&1&0\\
*&*&*&* & *&*&*&*\\
*&*&*&* & *&*&*&*\\
*&*&*&* & *&*&*&*\\
0&1&1&0& 1&1&0&0\\
1&1&1&1& 0&0&0&0\\
0&0&0&0& 1&1&1&1\\
*&*&*&* & *&*&*&*
\end{array}\right],
\end{equation*}
where ``$*$'' marks an entry to be determined.
We can apply Algorithm~\ref{alg:Gen_dec_circuit_Gaussian_elimination} with the known rows of $V$ to obtain a possible decoding circuit implemented using Hadamard, phase, and CNOT gates.
The encoding circuit is obtained by reversing the order of gates in the decoding circuit and the remaining rows of $V$ can be filled in according to the encoding circuit.
The encoding matrix $V$ of this example is
\begin{equation*}
V = \left[\begin{array}{cccc : cccc}
1&0&1&0& 0&1&1&0\\
0&0&0&0& 0&0&0&1\\
1&1&1&0& 1&1&1&1\\
1&1&0&0& 0&1&1&0\\
0&1&1&0& 1&1&0&0\\
1&1&1&1& 0&0&0&0\\
0&0&0&0& 1&1&1&1\\
1&0&1&0& 1&1&0&0
\end{array}\right].
\end{equation*}

For the same code $\code{C}$, another method called the quantum-shift-register circuit construction attempted to find possible encoding circuits using $5$ frames of memory qubits, i.e., $15$ memory qubits, for each encoding block in~\cite{lidar2013quantum} and using $1$ frame of memory qubits, i.e., $3$ memory qubits, which is the optimal number of memory qubits, for each encoding block in~\cite{wilde2009quantum}.
Our method can reduce the minimum number of memory qubits per encoding block from $3$ to $1$ for this specific example.
\exampleend
\end{example}
\subsection{Memory Requirements}
\label{sec:TBC_QECC_memory:4e}
In general, we require $\nu$ memory qubits in the quantum-code representation and encoding and decoding circuits of a (tail-biting) convolutional stabilizer QECC $\code{C}$, where $\nu$ is the overall constraint length of $\code{C}$.
If the quaternary stabilizer label code of $\code{C}$ is linear, the number of memory qubits can be reduced up to ${\nu}/{2}$ using the modified procedure to generate the elementary code $\codeBe$.

Earlier approaches to construct the encoding and decoding circuits for convolutional stabilizer QECCs include the Grassl-R{\"o}tteler encoding algorithm in~\cite{grassl2007constructions} and the quantum-shift-register circuit based on the Grassl-R{\"o}tteler encoding algorithm in~\cite{wilde2009quantum}.
The quantum-shift-register circuit construction requires $m$ frames of memory qubits where each frame consists of $n$ memory qubits.
The number $m$ is shown to be upper bounded by some finite number in~\cite{wilde2009quantum}.
The minimum number of memory qubits required by the quantum-shift-register circuit is $n \cdot \nu_{\Rm{max}}$ to have sufficient delay, where $\nu_{\Rm{max}}$ is the degree of the quantum check matrix $H(D) = [Z(D)|X(D)]$ in~\cite{wilde2009quantum}.
Our construction method improves the minimum number of memory qubits to be $\nu$ which is also the maximum number of memory qubits we need.
Note that $\nu \leq (n\!-\!k) \cdot \nu_{\Rm{max}} \leq n \cdot \nu_{\Rm{max}}$.
Furthermore, when the quaternary stabilizer label code is linear, the minimum number of memory qubits is $\nu/2$.
 
Efficient decoding algorithms for the encoding and decoding circuits we constructed include the max-product algorithm in~\cite{poulin2009quantum} and a generalized Viterbi algorithm in~\cite{pelchat2013degenerate}.
The complexity of these decoding algorithms is related to the number of memory qubits, which is at most the overall constraint length of the (tail-biting) convolutional QECC.
Instead of randomly generating the encoding and decoding circuits which may not have good parameters or performance, now we can choose (tail-biting) convolutional QECCs with good parameters for example the codes found in~\cite{forney2007convolutional} to construct the encoding and decoding circuits and then apply efficient decoding algorithms.
\subsection{Examples}
\label{sec:TBC_QECC_eg:4f}
Here are more examples using the above construction method.
\begin{example}[A CSS quantum convolutional code in~\cite{wilde2009quantum},\cite{lidar2013quantum}]
Consider a $\llbracket 3,1;2 \rrbracket$ CSS convolutional QECC with two basic generators  
\begin{align*}
g_1 &= [\ldots | \ob 0 \ob | 0 \ob \ob | \ldots],\\ 
g_2 &= [\ldots | 0 \om \om | \om 0 \om | \ldots].
\end{align*}

Using the generalized procedure to generate the elementary code $\codeBe$, a possible $\cB_i$ for the $i^{\Rm{th}}$ block of the NFG in Fig.~\ref{fig:Code-design_NFG_conv_QECC} is the $\GF{2}$-span of the following matrix
\begin{equation*}
\vect{G}{\codeB_{\Rm{e}}}
 = \left[\begin{array}{cc | ccc | cc}
0&0 & \ob&0&\ob & 1&0\\
1&0 & 0&\ob&\ob & 0&0\\
0&0 & 0&\om&\om & \om&1\\
\om&1 & \om&0&\om & 0&0
\end{array}\right].
\end{equation*}
We extend the self-orthogonal code $\cB_{\Rm{e}}$ by adding a vector from its dual code at a time and obtain the extended code $\codeB_{\Rm{e}}'$ with a systematic generator matrix
\begin{equation*}
\label{eq:eg1_extend_A}
\vect{G}{\codeB_{\Rm{e}}'} = 
\left[\begin{array}{cc | ccc | cc}
\om&0 & 0&1&0 & 0&\ob\\
0&\om & 1&1&\om & 0&\ob\\
\ob&0 & 0&\om&\ob & 0&\ob\\
0&\ob & \ob&0&0 & 0&0\\
0&0 & \ob&0&\ob & 1&0\\
0&0 & 0&\om&\om & \om&1
\end{array}\right].
\end{equation*}
We can determine some rows of $V$ according to~\eqref{eq:eg1_extend_A}. 
Then apply Algorithm~\ref{alg:Gen_dec_circuit_Gaussian_elimination} to obtain the decoding circuit and then fill in the remaining rows of $V$, i.e.,
\begin{equation*}
V = \left[\begin{array}{ccccc : ccccc}
\uO&\uI&\uO&\uO&\uO&      \uO&\uI&\uO&\uO&\uI\\
\uI&\uI&\uI&\uO&\uO&     \uI&\uI&\uO&\uO&\uI\\
0&1&1&0&0& 0&1&0&0&1\\
0&0&0&1&0& 0&0&0&1&0\\
0&0&0&1&0& 0&0&0&1&1\\
\uO&\uI&\uO&\uO&\uO&     \uO&\uO&\uI&\uO&\uI\\
\uO&\uO&\uO&\uO&\uO&     \uI&\uO&\uO&\uO&\uO\\
\uO&\uO&\uO&\uI&\uO&     \uI&\uO&\uI&\uI&\uO\\
\uO&\uI&\uI&\uI&\uI&     \uO&\uO&\uO&\uO&\uI\\
0&0&0&0&1& 0&1&0&0&0
\end{array}\right],
\end{equation*}
where the underlined rows are determined before applying Algorithm~\ref{alg:Gen_dec_circuit_Gaussian_elimination}.
The rows of $V$ without underlining are filled according to the encoding circuit found using Algorithm~\ref{alg:Gen_dec_circuit_Gaussian_elimination}.

Compared with the minimum number of memory qubits for each encoding block using the quantum-shift-register circuit construction in~\cite{wilde2009quantum} and \cite{lidar2013quantum} which is $3$, we save $1$ qubit for each encoding block.
\exampleend
\end{example}
\begin{example}[$\llbracket 3,1,5;3 \rrbracket$-convolutional stabilizer QECC] 
This $\llbracket 3,1,5;3 \rrbracket$-convolutional stabilizer QECC $\code{C}$ is constructed from \cite[Table VI]{forney2007convolutional}.  
The quaternary stabilizer label code $\codeB$ of $\code{C}$ is linear with the single basic generator
\begin{equation*}
\vg = [\ldots | 111 | 01\om | 01\ob | 1\ob\om | \ldots].
\end{equation*}
According to the modified procedure to generate the elementary code $\codeBe$,
each $\codeB_i$ for the $i^{\Rm{th}}$ block of the NFG in Fig.~\ref{fig:Code-design_NFG_conv_QECC} is the $\GF{4}$-span of the generator matrix 
\begin{equation*}
\vect{G}{\codeB_{\Rm{e}}}
 = \left[\begin{array}{ccc | ccc | ccc}
0&0&0 & 1&1&1 & 1&1&1\\
1&1&1 & 0&1&\om & 0&\ob&0\\
0&\ob&0 & 0&1&\ob & 0&0&\om\\
0&0&\om & 1&\ob&\om & 0&0&0
\end{array}\right],
\end{equation*}
which is a matrix obtained by slightly modifying the outputs of our algorithm.
This example demonstrates that it is possible to use fewer memory qubits by some small tricks.
Eventually, the number of memory qubits is always between $\nu/2$ and $\nu$.
We can determine some rows of the encoding matrix $V$ which are underlined.
Then we apply Algorithm~\ref{alg:Gen_dec_circuit_Gaussian_elimination} to obtain the decoding circuit and then fill in the remaining rows of $V$, i.e.,
\begin{equation*}
V = \left[\begin{array}{cccccc : cccccc}
\uI&\uI&\uO&\uO&\uI&\uI     &     \uI&\uO&\uI&\uO&\uI&\uI\\
\uO&\uO&\uI&\uO&\uO&\uI     &     \uO&\uI&\uO&\uO&\uO&\uI\\
\uI&\uO&\uI&\uO&\uO&\uO      &     \uI&\uI&\uO&\uO&\uO&\uO\\
0&0&0&0&0&0 & 0&0&0&1&0&0\\
1&1&0&0&0&1 & 1&0&1&1&0&1\\
0&0&1&0&1&0 & 0&1&0&0&1&0\\
\uI&\uO&\uI&\uO&\uI&\uI     &     \uO&\uI&\uI&\uO&\uO&\uO\\
\uO&\uI&\uO&\uO&\uO&\uI     &     \uO&\uI&\uI&\uO&\uO&\uO\\
\uI&\uI&\uO&\uO&\uO&\uO     &     \uO&\uI&\uI&\uO&\uO&\uO\\
\uI&\uI&\uI&\uI&\uI&\uI     &     \uO&\uO&\uO&\uO&\uO&\uO\\
\uO&\uO&\uO&\uO&\uO&\uO     &     \uI&\uI&\uI&\uI&\uI&\uI\\
0&1&1&0&0&0 & 1&1&0&1&0&1\\
\end{array}\right].
\end{equation*}
\exampleend
\end{example}
\section{LDPC Stabilizer QECCs}
\label{sec:LDPC_QECC:5}

In this section we show that the stabilizer label code $\cB$ is self-orthogonal with the help of NFG reformulations based on NFG duality results and switching between image-representation-type NFGs and kernel-representation-type NFGs. 
The latter one borrows some insights from the Kramers--Wannier duality type results in terms of NFGs in~\cite{al2018factor}.

The codes in this section and next section (e.g., LDPC stabilizer QECCs and toric codes) are both Calderbank-Shor-Steane (CSS) stabilizer QECCs (see for example~\cite{calderbank1996good,calderbank1998quantum,steane1996multiple}).
In this section, we discuss the class of LDPC stabilizer QECCs introduced by Tillich and Z\'emor~\cite{tillich2014quantum}.
\subsection{LDPC Stabilizer QECCs by Tillich and Z\'emor}
\label{sec:LDPC_QECC_TZ:5a}
\begin{definition}[CSS stabilizer QECCs]
Let $\cC_{X}$ and $\cC_{Z}$ be two binary linear codes of length $n$ defined via parity-check matrices $\vect{H}{X}$ and $\vect{H}{Z}$, respectively.
If $\cC_{Z}^{\perp} \subseteq \cC_{X}$, the CSS stabilizer QECC based on $\cC_{X}$ and $\cC_{Z}$ has the stabilizer label code $\cB$ whose codewords $(\vt_X, \vt_Z) \in \GF{2}^n \times \GF{2}^n$ are given by the $\GF{2}$-rowspace of the stabilizer matrix
\begin{align*}
\mb{A} &\defeq \left[ \begin{array}{cc}
\vect{H}{X} & \vect{0}{n} \\
\vect{0}{n} & \vect{H}{Z}
\end{array}\right].
\end{align*}
Note that the condition $\cC_{Z}^{\perp} \subseteq \cC_{X}$ is equivalent to the condition $\cC_{X}^{\perp} \subseteq \cC_{Z}$ or the condition $\cB \subseteq \cB^{\perp}$, 
where the dual code is defined w.r.t. the standard inner product for $\cC_{X}$ and $\cC_{Z}$ while the dual code is defined w.r.t. the symplectic inner product for $\cB$.
\end{definition}

\begin{definition}[LDPC stabilizer QECCs by Tillich and Z\'emor]
\label{def:TZ:codes:1}
For $i \in \{1,2\}$, let $\cC_i$ be a binary linear $[n_i, k_i, d_i]$ code defined by the parity-check matrix $\matr{H}_i$ and let $r_i$ be the number of rows of $\matr{H}_i$ and $k_i^{\perp}$ be the dimension of the code with parity-check matrix $\matr{H}_i^{\msf{T}}$.
Tillich and Z\'emor defined a CSS stabilizer QECC $\cC$ with parameters $\llbracket n_1 n_2
  + r_1 r_2, \ k_1 k_2 + k_1^\perp k_2^\perp \rrbracket$ by choosing $\matr{H}_X$ and $\matr{H}_Z$ to be, respectively,
  \begin{align*}
    \matr{H}_{X} 
      &= \left[ 
           \matr{H}_1 \otimes \matr{I}_{n_2}\ |\ 
           \matr{I}_{r_1} \otimes \matr{H}_2^{\msf{T}}
         \right] \ , \\
    \matr{H}_{Z} 
      &= \left[ 
           \matr{I}_{n_1} \otimes \matr{H}_2\ |\  
           \matr{H}_1^{\msf{T}} \otimes \matr{I}_{r_2} 
         \right] \ .
  \end{align*}
This class of codes is referred to as LDPC CSS stabilizer QECCs since the parity-check matrices $\matr{H}_1$ and $\matr{H}_2$ are chosen as sparse matrices.
This code construction is of especial interest as it yields an asymptotic family of CSS stabilizer QECCs with fixed rate and minimum distance proportional to the square root of the block length.
\end{definition}

We now introduce an NFG-based framework for showing that $\cC_{Z}^{\perp} \subseteq \cC_{X}$ for the codes defined in Definition~\ref{def:TZ:codes:1}. 
Although this condition can be verified rather straightforwardly based on the definitions of $\matr{H}_{X}$ and $\matr{H}_{Z}$ stated in Definition~\ref{def:TZ:codes:1}, our approach yields additional insights into this code constructions. In particular, it suggests approaches for constructing other, more general classes of stabilizer QECCs with favorable NFG representations.
In order to proceed, it turns out to be useful to recall the definition of a product code. 

\begin{definition}[Product codes]
Let $\cC_{a}$ and $\cC_{b}$ be two binary linear codes of length $n_c$ and $n_r$, respectively. 
The product code based on $\cC_{a}$ and $\cC_{b}$ is defined to be the binary code of length $n_a \cdot n_b$ where each codeword $\mb{c} \in \GF{2}^{n_a \times n_b}$ can be viewed as a matrix whose columns are codewords of $\cC_{a}$ and rows are codewords of $\cC_{b}$.
\end{definition}

\begin{figure}
\begin{center}
\begin{tikzpicture}[x=1.25cm, y= 1cm, node/.style={draw=none},
                          varNode/.style={draw,minimum size = 5pt, fill = none, circle, inner sep = 0pt },
                          factor/.style={draw,minimum size = 9pt},
                          factorH/.style={draw,minimum height = 30pt,minimum width = 18pt},
                          factorW/.style={draw,minimum height = 18pt,minimum width = 30pt}]
	\foreach \i in {0,1,2}{
		\foreach \j in {0,1,2,3,4}{
			\node[factor] (var\i\j) at (\j,-\i) {};
			\node at ([yshift=-.7pt]var\i\j) {\small{$=$}}; 
			\path ([yshift=-2pt]var\i\j.north east) edge[draw] ([xshift=7pt, yshift=5pt]var\i\j.north east);
	}}
	\node[factorW, above = .9cm of var02] (F1) {$\Cal{C}_{b}$};
	\node[factorH, left = .9cm of var10] (F2) {$\Cal{C}_{a}$};
	\foreach \j in {0,1,2,3,4}{
		\draw (F1)--(var0\j);
	}
	\foreach \i in {0,1,2}{
		\draw (F2)--(var\i0);
	}
\end{tikzpicture}
\end{center}
\caption{The (standard) factor graph $\cG$ represents the product code of $\cC_{a}$ and $\cC_{b}$. (For simplicity, the $\cC_{a}$-based function node is only shown for the first column, and the $\cC_{b}$-based function node is only shown for the first row.)}
\label{fig:product_code_factor_graph}
\end{figure}
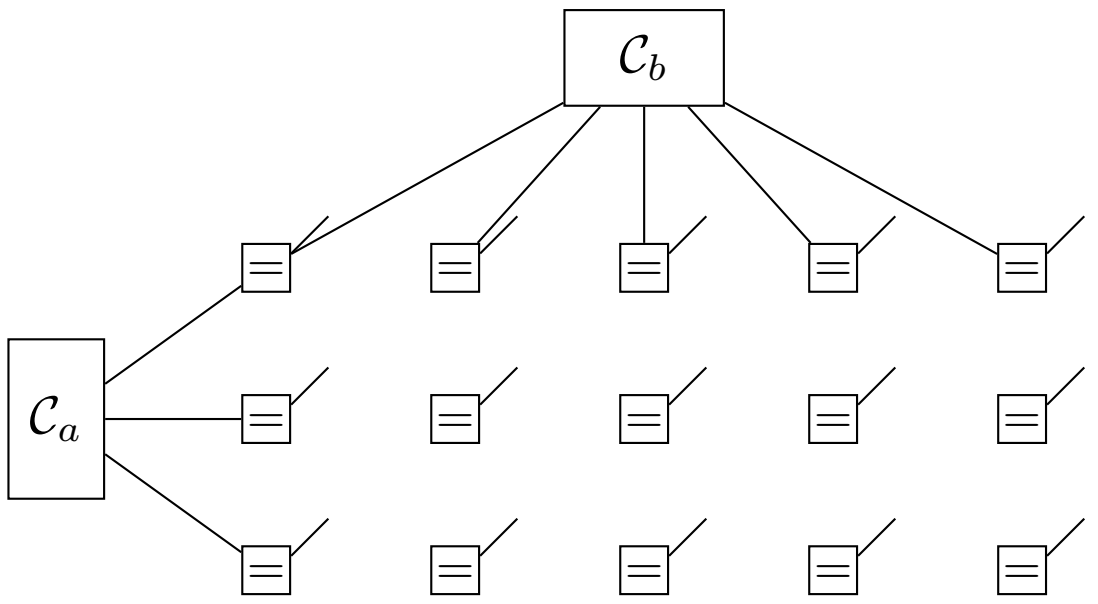

It turns out that $\cC_X^{\perp} \defeq \Rm{rowspace}(\matr{H}_X)$ and $\cC_Z^{\perp}\defeq\Rm{rowspace}(\matr{H}_Z)$ of the stabilizer QECC from Definition~\ref{def:TZ:codes:1} are both punctured product codes explained as follows.
\begin{itemize}
\item Let the code $\cC_{a}$ of length $(n_1+r_1)$ and the code $\cC_{b}$ of length $(n_2+r_2)$ and their dual codes be defined as
\begin{align*}
\cC_{a} &\defeq \Rm{rowspace} [\ \matr{H}_1\ |\ \matr{I}_{r_1}\ ],\\
\cC_{a}^{\perp} &\defeq \Rm{rowspace} [\ \matr{I}_{n_1}\ |\ \matr{H}_1^{\msf{T}}\ ],\\
\cC_{b} &\defeq \Rm{rowspace} [\ \matr{I}_{n_2}\ |\ \matr{H}_2^{\msf{T}}\ ],\\
\cC_{b}^{\perp} &\defeq \Rm{rowspace} [\ \matr{H}_2\ |\ \matr{I}_{r_2}\ ].
\end{align*}
\item One can show that $\cC_X^{\perp}$ is the product code of $\Ca$ and $\Cb$, where all the bits corresponding to the lower left array of size $r_1 \times n_2$ and the upper left array of size $n_1 \times r_2$ are punctured when viewing each codeword in the product code as a matrix.
An NFG presenting $\cC_X^{\perp}$ is shown in Fig.~\ref{fig:NFG_for_C_X_full}, where $\vect{t}{1}$ collects the bits corresponding to the upper left array of size $n_1 \times n_2$ and $\vect{t}{2}$ collects the bits corresponding to the lower right array of size $r_1 \times r_2$.
The upper index of $\Ca$ and $\Cb$ indicate the number of independent copies of them, e.g., $\Ca^{n_2}$ denotes $n_2$  independent copies of $\Ca$.
\item An NFG representing $\cC_{Z}^{\perp}$ is obtained in a similar way by replacing $\Ca$ and $\Cb$ in Fig.~\ref{fig:NFG_for_C_X_full} with $\Ca^{\perp}$ and $\Cb^{\perp}$, respectively.
\end{itemize} 

Because of the special structure of the generator matrices of $\Ca$ and $\Cb$ (i.e., they are of systematic form), the NFG in Fig.~\ref{fig:NFG_for_C_X_simplified} also represents $\cC_{X}^{\perp}$.
In Fig.~\ref{fig:NFG_for_C_X_full}, the punctured bits corresponding to the lower left array can be viewed as information bits and the functions on the upper right part are imposing redundant constraints which can be omitted and simplified as in Fig.~\ref{fig:NFG_for_C_X_simplified}.
Similarly, the NFG in Fig.~\ref{fig:NFG_for_C_Z_dual_simplified} represents $\cC_{Z}^{\perp}$.

Now we have a simple proof of the self-orthogonality of the stabilizer label code of the LDPC stabilizer QECC using the NFG representations, i.e., $\cC_{X}^{\perp} \subseteq \cC_{Z}$. 
The proof is sketched as follows. 
\begin{itemize}
\item Because of the special structure of the generator matrices of $\Ca$ and $\Cb$, the NFG in Fig.~\ref{fig:NFG_for_C_X_simplified} is an image representation of $\cC_{X}^{\perp} \defeq \Rm{rowspan}(\vect{H}{X})$.
\item Let the code $\cS$ be defined via the checks we omitted in the process of simplifying the NFG in Fig.~\ref{fig:NFG_for_C_X_full} as the NFG in Fig.~\ref{fig:NFG_for_C_Z_dual_simplified}.
Because these omitted checks are redundant, the code $\cS$ is a superset of $\cC_{X}^{\perp}$, i.e., $\cC_{X}^{\perp} \subseteq \cS$.
The NFG in Fig.~\ref{fig:NFG_for_C_S} is the kernel representation of $\cS$ which is obtained by making use of the special structure of the generator matrices of $\Ca$ and $\Cb$.
\item Switching between kernel and image representations, the NFGs in Fig.~\ref{fig:NFG_for_C_Z_dual_simplified} and Fig.~\ref{fig:NFG_for_C_S} are dual of each other. 
Sign converters on the edges are omitted since the alphabets are over $\GF{2}$.
The NFG in Fig.~\ref{fig:NFG_for_C_Z_dual_simplified} represents $\cS^{\perp}$, which happens to the NFG representing the code $\cC_{Z}^{\perp} \defeq \Rm{rowspan}(\vect{H}{Z})$ implying $\cS^{\perp} = \cC_{Z}^{\perp}$.
By combining the fact that $\cC_{X}^{\perp} \subseteq \cS$, we can show $\cC_{X}^{\perp} \subseteq \cS = \cC_{Z}$.
\end{itemize}
As a summary, the above proof mainly has two steps.
First, we go from an image-representation-type NFG to a kernel-representation-type NFG which may result in a larger code. 
Then we apply NFG duality results to obtain the result.
Similar ideas are applied to derive Kramers--Wannier duality type results~\cite{al2018factor}.
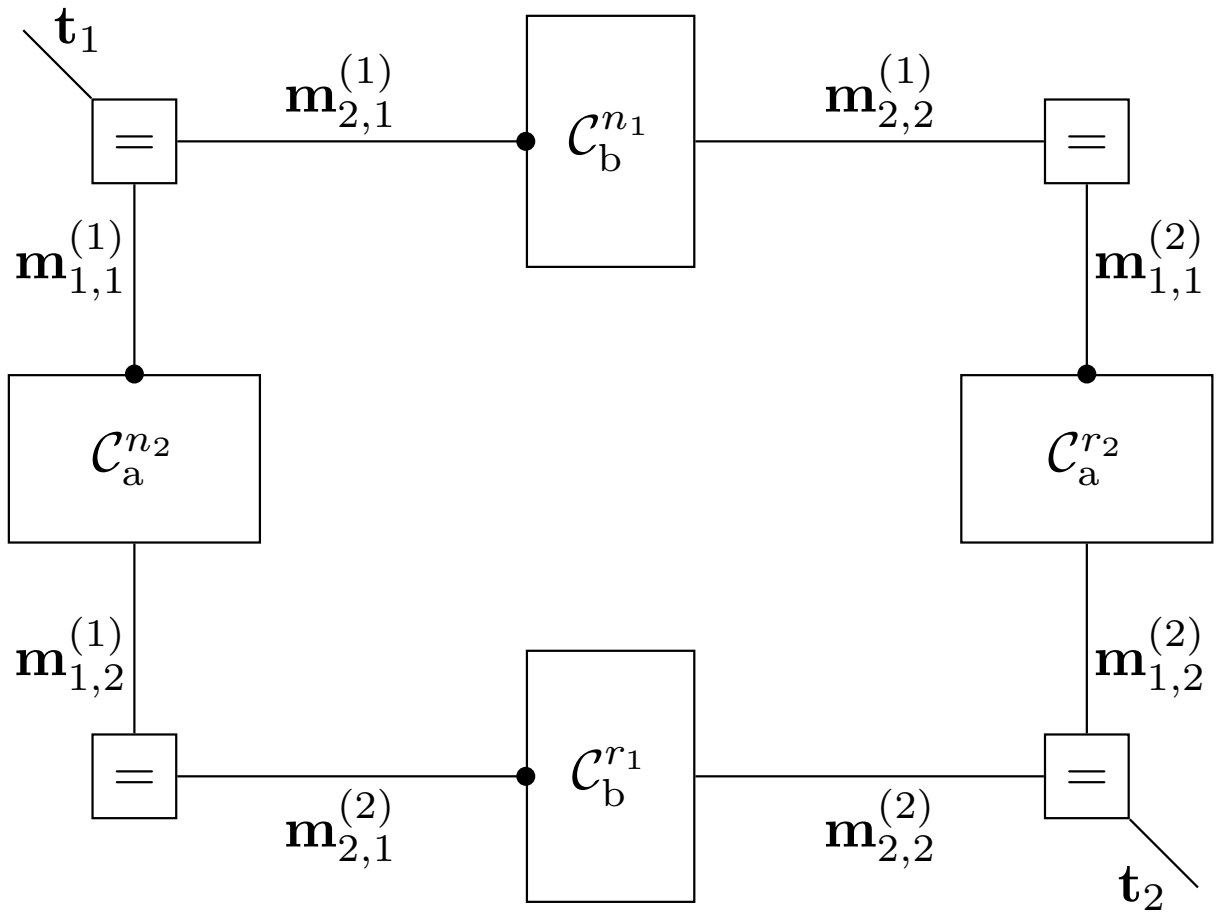
\begin{figure}
  \begin{center}
    \scalebox{1.05}{\begin{tikzpicture}[x=1cm, y= 1cm, node/.style={draw=none},
                          codeH/.style={draw,minimum height = 45pt,minimum width = 30pt},
                          codeW/.style={draw,minimum height = 30pt,minimum width = 45pt},
                          factor/.style={draw,minimum size = 15pt,inner sep = 0pt}
                          ]
	\node[factor] (E11) at (0,0) {};
	\node at ([yshift=-.75pt]E11) {$=$};
	\node[factor] (E13) at (6,0) {};
	\node at ([yshift=-.75pt]E13) {$=$};
	\node[factor] (E31) at (0,-4) {};
	\node at ([yshift=-.75pt]E31) {$=$};
	\node[factor] (E33) at (6,-4) {};
	\node at ([yshift=-.75pt]E33) {$=$};
	
	\node[codeW] (C21) at (0,-2) {$\Ca^{n_2}$};
	\node[codeH] (C12) at (3,0) {$\Cb^{n_1}$};
	\node[codeW] (C23) at (6,-2) {$\Ca^{r_2}$};
	\node[codeH] (C32) at (3,-4) {$\Cb^{r_1}$};
	
	\draw[-*] (E11) -- (C12);
	\draw (C12) -- (E13);
		
	\draw[-*] (E11) -- (C21);
	\draw (C21) -- (E31);
	
	\draw[-*] (E31) -- (C32);
	\draw (C32) -- (E33);
	
	\draw[-*] (E13) -- (C23);
	\draw (C23) -- (E33);
	
	\path (E11) edge[draw] ([xshift=-.7cm, yshift=.7cm]E11);
	\path (E33) edge[draw] ([xshift=.7cm, yshift=-.7cm]E33);
	
	\node at ([xshift=-10pt, yshift=20pt]E11) {$\mathbf{t}_{1}$};
	\node at ([xshift=10pt, yshift=-20pt]E33) {$\mathbf{t}_{2}$};
	\node at (1.3, 0.3) {$\mathbf{m}_{2,1}^{(1)}$};
	\node at (4.7, 0.3) {$\mathbf{m}_{2,2}^{(1)}$};
	\node at (-0.4, -0.75) {$\mathbf{m}_{1,1}^{(1)}$};
	\node at (-0.4, -3.25) {$\mathbf{m}_{1,2}^{(1)}$};
	\node at (1.3, -4.33) {$\mathbf{m}_{2,1}^{(2)}$};
	\node at (4.7, -4.33) {$\mathbf{m}_{2,2}^{(2)}$};
	\node at (6.4, -0.75) {$\mathbf{m}_{1,1}^{(2)}$};
	\node at (6.4, -3.25) {$\mathbf{m}_{1,2}^{(2)}$};
\end{tikzpicture}}
  \end{center}
  \caption{NFG representing $\cC_{X}^{\perp} \defeq \Rm{rowspace}(\matr{H}_X)$. The edge with a dot denotes a variable which is the first argument of the corresponding local function.}
  \label{fig:NFG_for_C_X_full}
\end{figure}
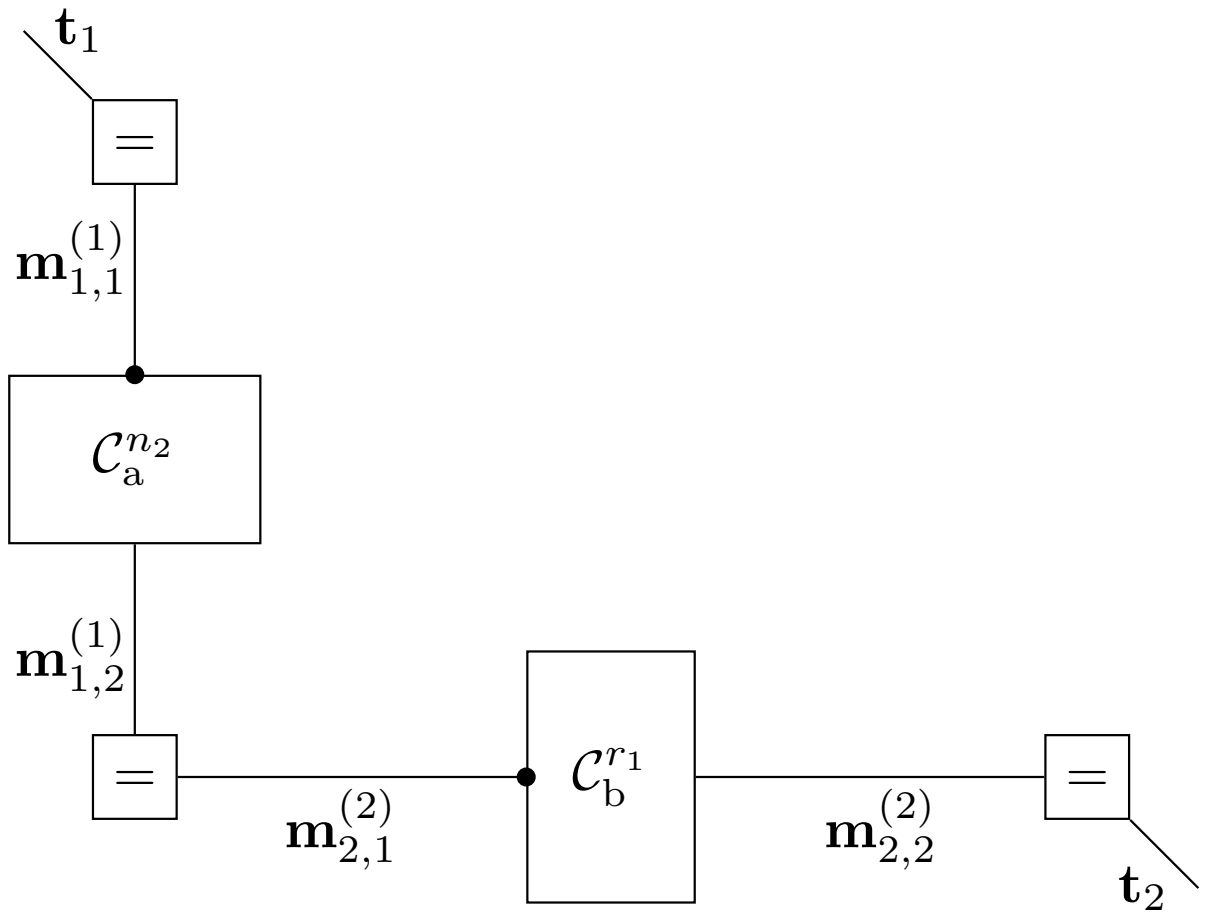
\begin{figure}
  \begin{center}
    \scalebox{1.05}{\begin{tikzpicture}[x=1cm, y= 1cm, node/.style={draw=none},
                          codeH/.style={draw,minimum height = 45pt,minimum width = 30pt},
                          codeW/.style={draw,minimum height = 30pt,minimum width = 45pt},
                          factor/.style={draw,minimum size = 15pt}]
	\node[factor] (E11) at (0,0) {};
	\node at ([yshift=-.75pt]E11) {$=$};
	\node[factor] (E31) at (0,-4) {};
	\node at ([yshift=-.75pt]E31) {$=$};
	\node[factor] (E33) at (6,-4) {};
	\node at ([yshift=-.75pt]E33) {$=$};
	
	\node[codeW] (C21) at (0,-2) {$\Ca^{n_2}$};
	\node[codeH] (C32) at (3,-4) {$\Cb^{r_1}$};
	
	\draw[-*] (E11) -- (C21);
	\draw (C21) -- (E31);
	
	\draw[-*] (E31) -- (C32);
	\draw (C32) -- (E33);
	
	\path (E11) edge[draw] ([xshift=-.7cm, yshift=.7cm]E11);
	\path (E33) edge[draw] ([xshift=.7cm, yshift=-.7cm]E33);
	
	\node at ([xshift=-10pt, yshift=20pt]E11) {$\mathbf{t}_{1}$};
	\node at ([xshift=10pt, yshift=-20pt]E33) {$\mathbf{t}_{2}$};
	\node at (-0.4, -0.75) {$\mathbf{m}_{1,1}^{(1)}$};
	\node at (-0.4, -3.25) {$\mathbf{m}_{1,2}^{(1)}$};
	\node at (1.3, -4.33) {$\mathbf{m}_{2,1}^{(2)}$};
	\node at (4.7, -4.33) {$\mathbf{m}_{2,2}^{(2)}$};
\end{tikzpicture}}
  \end{center}
  \caption{Another NFG representing $\cC_{X}^{\perp} \defeq \Rm{rowspace}(\matr{H}_X)$.}
  \label{fig:NFG_for_C_X_simplified}
\end{figure}
\begin{figure}
  \begin{center}
    \scalebox{1.05}{\begin{tikzpicture}[x=1cm, y= 1cm, node/.style={draw=none},
                          codeH/.style={draw,minimum height = 45pt,minimum width = 30pt},
                          codeW/.style={draw,minimum height = 30pt,minimum width = 45pt},
                          factor/.style={draw,minimum size = 15pt}]
	\node[factor] (E11) at (0,0) {};
	\node at ([yshift=-.75pt]E11) {$=$};
	\node[factor] (E13) at (6,0) {};
	\node at ([yshift=-.75pt]E13) {$=$};
	\node[factor] (E33) at (6,-4) {};
	\node at ([yshift=-.75pt]E33) {$=$};
	
	\node[codeH] (C12) at (3,0) {$(\Cb^{\perp})^{n_1}$};
	\node[codeW] (C23) at (6,-2) {$(\Ca^{\perp})^{r_2}$};
	
	\draw[-*] (E11) -- (C12);
	\draw (C12) -- (E13);
	
	\draw[-*] (E13) -- (C23);
	\draw (C23) -- (E33);
	
	\path (E11) edge[draw] ([xshift=-.7cm, yshift=.7cm]E11);
	\path (E33) edge[draw] ([xshift=.7cm, yshift=-.7cm]E33);
	
	\node at ([xshift=-10pt, yshift=20pt]E11) {$\mathbf{t}_{1}$};
	\node at ([xshift=10pt, yshift=-20pt]E33) {$\mathbf{t}_{2}$};
	\node at (1.3, 0.3) {$\mathbf{m}_{2,1}^{(1)}$};
	\node at (4.7, 0.3) {$\mathbf{m}_{2,2}^{(1)}$};
	\node at (6.4, -0.75) {$\mathbf{m}_{1,1}^{(2)}$};
	\node at (6.4, -3.25) {$\mathbf{m}_{1,2}^{(2)}$};
\end{tikzpicture}}
  \end{center}
  \caption{NFG representing $\cC_{Z}^{\perp} \defeq \Rm{rowspace}(\matr{H}_Z)$.}
  \label{fig:NFG_for_C_Z_dual_simplified}
\end{figure}
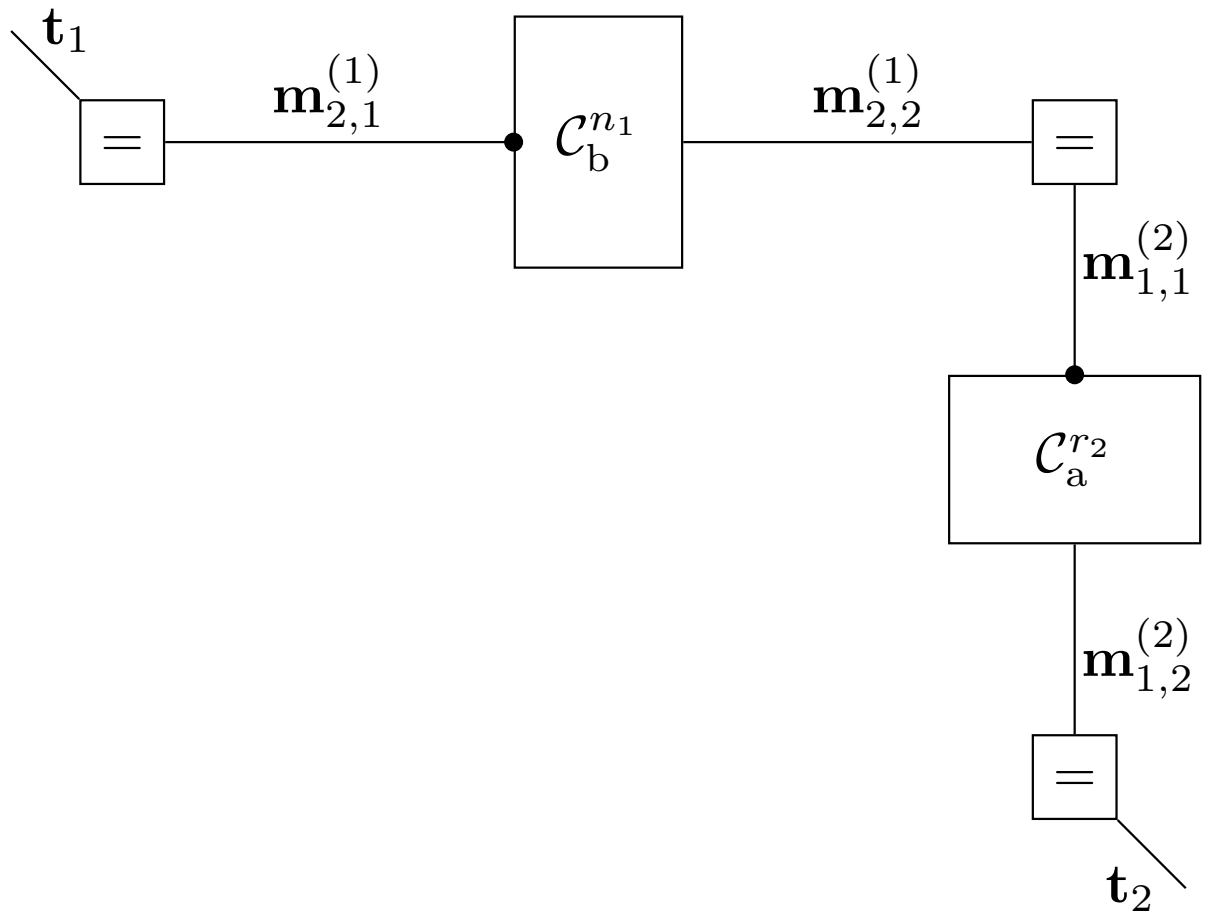
\begin{figure}
  \begin{center}
    \scalebox{1.05}{\begin{tikzpicture}[x=1cm, y= 1cm, node/.style={draw=none},
                          codeH/.style={draw,minimum height = 45pt,minimum width = 30pt},
                          codeW/.style={draw,minimum height = 30pt,minimum width = 45pt},
                          factor/.style={draw,minimum size = 15pt}]
	\node[factor] (E11) at (0,0) {};
	\node at ([yshift=-.75pt]E11) {$=$};
	\node[factor] (E13) at (6,0) {};
	\node at ([yshift=-.75pt]E13) {$=$};
	\node[factor] (E33) at (6,-4) {};
	\node at ([yshift=-.75pt]E33) {$=$};
	
	\node[codeH] (C12) at (3,0) {$\Cb^{n_1}$};
	\node[codeW] (C23) at (6,-2) {$\Ca^{r_2}$};
	
	\draw[-*] (E11) -- (C12);
	\draw (C12) -- (E13);
	
	\draw[-*] (E13) -- (C23);
	\draw (C23) -- (E33);
	
	\path (E11) edge[draw] ([xshift=-.7cm, yshift=.7cm]E11);
	\path (E33) edge[draw] ([xshift=.7cm, yshift=-.7cm]E33);
	
	\node at ([xshift=-10pt, yshift=20pt]E11) {$\mathbf{t}_{1}$};
	\node at ([xshift=10pt, yshift=-20pt]E33) {$\mathbf{t}_{2}$};
	\node at (1.3, 0.3) {$\mathbf{m}_{2,1}^{(1)}$};
	\node at (4.7, 0.3) {$\mathbf{m}_{2,2}^{(1)}$};
	\node at (6.4, -0.75) {$\mathbf{m}_{1,1}^{(2)}$};
	\node at (6.4, -3.25) {$\mathbf{m}_{1,2}^{(2)}$};
\end{tikzpicture}}
  \end{center}
  \caption{An NFG representing the code $\cS$.}
  \label{fig:NFG_for_C_S}
\end{figure}

\subsection{Toric Codes}
\label{sec:LDPC_QECC_Toric:5b}
Similar steps can also be applied to prove the self-orthogonality of the stabilizer label code of a toric code introduced by Kitaev~\cite{kitaev2003fault}.
\begin{definition}[Toric code by Kitaev~\cite{kitaev2003fault}]
Let the binary linear codes $\cC_{X}^{\perp}$ and $\cC_{Z}^{\perp}$ be defined as the codes represented by the NFG (solid, black) and the NFG (dashed, red) in Fig.~\ref{fig:NFG_for_toric_C_X_combined}, respectively,
where the NFGs lie on a torus.
The toric code by Kitaev~\cite{kitaev2003fault} is the CSS stabilizer QECC based on the codes $\cC_{X}$ and $\cC_{Z}$. 
The parameters of the resulting CSS stabilizer QECC are $\llbracket n, 2, d \rrbracket$, where the block length of the code is $n$, the dimension is $2$, and the minimum distance $d$ is proportional to $\sqrt{n}$.
\end{definition}
Now we have a simple proof of the self-orthogonality of the stabilizer label code of the toric code using the NFG representations, i.e., $\cC_{X}^{\perp} \subseteq \cC_{Z}$.
\begin{itemize}
\item The NFG (solid, black) in Fig.~\ref{fig:NFG_for_toric_C_X_combined} is the image representation of $\cC_{X}^{\perp}$ since we can assign values to the equal nodes and collect values on the half edges.
\item Let the code $\cS$ be the code represented by the NFG (dashed, red) in Fig.~\ref{fig:NFG_for_toric_C_X_combined}.
The code $\cS$ is a superset of $\cC_{X}^{\perp}$ since $\cS$ is the code satisfying the redundant checks for $\cC_{X}^{\perp}$.
The NFG (dashed, red) in Fig.~\ref{fig:NFG_for_toric_C_X_combined} is the kernel representation of $\cS$. 
\item When switching from the kernel representation to the image representation, we find the dual of the NFG for $\cS$ is exactly the same NFG for $\cC_{Z}$, i.e., $\cS^{\perp} = \cC_{Z}$, which implies $\cC_{X}^{\perp} \subseteq \cS = \cC_{Z}$.
\end{itemize}
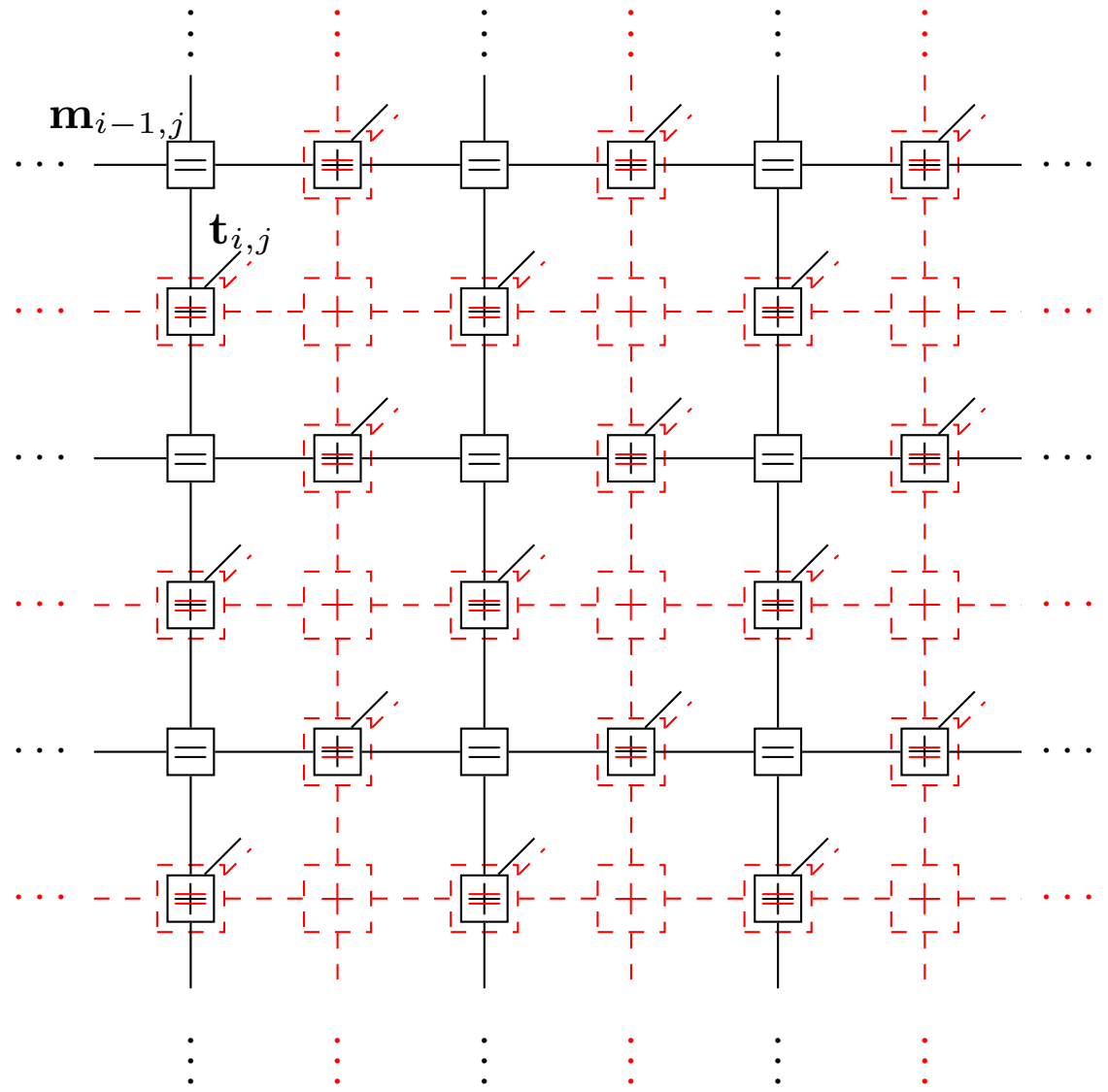
\begin{figure}
  \begin{center}
    \begin{tikzpicture}[x=1cm, y= 1cm, node/.style={draw=none},
                          factor/.style={draw,minimum size = 9pt},
                          dualfactor/.style={draw,minimum size = 13pt, color=red, dashed},]
\foreach \i in {1,3,5}{
	\foreach \j in {2,4,6}{
		\node[factor] (node\i\j) at (\j,-\i) {};
		\node at (node\i\j) {\small{$+$}}; 
		\path ([xshift=-2pt]node\i\j.north east) edge[draw] ([xshift=5pt, yshift=7pt]node\i\j.north east);
		\node[dualfactor] (dualnode\i\j) at (node\i\j) {};
		\node[red] at ([yshift=-.7pt]dualnode\i\j) {\small{$=$}}; 
		\path[red,dashed] ([yshift=-2pt]dualnode\i\j.north east) edge[draw] ([xshift=5pt, yshift=3pt]dualnode\i\j.north east);
}}
\foreach \i in {2,4,6}{
	\foreach \j in {1,3,5}{
		\node[factor] (node\i\j) at (\j,-\i) {};
		\node at (node\i\j) {\small{$+$}};
		\path ([xshift=-2pt]node\i\j.north east) edge[draw] ([xshift=5pt, yshift=7pt]node\i\j.north east);
		\node[dualfactor] (dualnode\i\j) at (node\i\j) {};
		\node[red] at ([yshift=-.7pt]dualnode\i\j) {\small{$=$}};
		\path[red,dashed] ([yshift=-2pt]dualnode\i\j.north east) edge[draw] ([xshift=5pt, yshift=3pt]dualnode\i\j.north east);
}}
\foreach \k in {1,3,5}{  
	\node (node0\k) at (\k,0) {\small{$\vdots$}};
	\node (node7\k) at (\k,-7) {\small{$\vdots$}};
	\node (node\k0) at (0,-\k) {\small{$\cdots$}};
	\node (node\k7) at (7,-\k) {\small{$\cdots$}};
}
\foreach \i in {1,3,5}{
	\foreach \j in {1,3,5}{
		\node[factor] (node\i\j) at (\j,-\i) {};
		\node at ([yshift=-.7pt]node\i\j) {\small{$=$}}; 
}}
\foreach \k in {2,4,6}{ 
	\node (node0\k) at (\k,0) {};
	\node (node7\k) at (\k,-7) {};
	\node (node\k0) at (0,-\k) {};
	\node (node\k7) at (7,-\k) {};
	\node[red] (dualnode0\k) at ([xshift=0cm, yshift=0cm]node0\k) {\small{$\vdots$}};
	\node[red] (dualnode7\k) at ([xshift=0cm, yshift=0cm]node7\k) {\small{$\vdots$}};
	\node[red] (dualnode\k0) at ([xshift=0cm, yshift=0cm]node\k0) {\small{$\cdots$}};
	\node[red] (dualnode\k7) at ([xshift=0cm, yshift=0cm]node\k7) {\small{$\cdots$}};
}
\foreach \i in {2,4,6}{
	\foreach \j in {2,4,6}{
		\node (node\i\j) at (\j,-\i) {};
		\node[dualfactor] (dualnode\i\j) at ([xshift=0cm, yshift=0cm]node\i\j) {};
		\node[red] at (dualnode\i\j) {\small{$+$}}; 
}}
\foreach \i in {1,3,5}{
	\foreach \j in {2,4,6}{
		\pgfmathtruncatemacro\temp{\j+1}; 
		\draw (node\i\j) -- (node\i\temp);
		\pgfmathtruncatemacro\temp{\i+1}; 
		\draw[red, dashed] (dualnode\i\j) -- (dualnode\temp\j);
}}
\foreach \i in {2,4,6}{
	\foreach \j in {1,3,5}{
		\pgfmathtruncatemacro\temp{\i+1}; 
		\draw (node\i\j) -- (node\temp\j);
		\pgfmathtruncatemacro\temp{\j+1}; 
		\draw[red, dashed] (dualnode\i\j) -- (dualnode\i\temp);
}}
\foreach \i in {1,3,5}{
	\foreach \j in {1,3,5}{
		\pgfmathtruncatemacro\temp{\i+1};
		\draw (node\i\j) -- (node\temp\j);
		\pgfmathtruncatemacro\temp{\j+1};
		\draw (node\i\j) -- (node\i\temp);
}}
\foreach \k in {1,3,5}{   
	\draw (node\k0) -- (node\k1);
	\draw (node0\k) -- (node1\k);
}
\foreach \i in {2,4,6}{
	\foreach \j in {2,4,6}{
		\pgfmathtruncatemacro\temp{\i+1};
		\draw[red, dashed] (dualnode\i\j) -- (dualnode\temp\j);
		\pgfmathtruncatemacro\temp{\j+1};
		\draw[red, dashed] (dualnode\i\j) -- (dualnode\i\temp);
}}
\foreach \k in {2,4,6}{   
	\draw[red, dashed] (dualnode\k0) -- (dualnode\k1);
	\draw[red, dashed] (dualnode0\k) -- (dualnode1\k);
}

\node (t_x) at ([xshift=10pt, yshift=15pt]node21) {\small$\mb{t}_{i,j}$};
\node (m_ij) at ([xshift=-14pt, yshift=8pt]node11) {\small$\mb{m}_{i-1,j}$};
\end{tikzpicture}
  \end{center}
  \caption{NFGs appearing in the self-orthogonality proof of toric codes, 
  where the NFG (solid, black) represents $\cC_{X}^{\perp} \defeq \Rm{rowspace}(\vect{H}{X})$ and the NFG (dashed, red) represents $\cC_{S}$ and $\cC_{Z}^{\perp} \defeq \Rm{rowspace}(\vect{H}{Z})$.
  The dots indicate the pair of edges at any two horizontal/vertical ends are identical.}
  \label{fig:NFG_for_toric_C_X_combined}
\end{figure}

\section{Entanglement-Assisted Quantum Codes}
\label{sec:EAQC:5}

As the name suggested, the entanglement-assisted QECCs are constructed with the help of highly entangled quantum qubits pairs which are shared between the transmitter and the receiver.
Given a non-Abelian subgroup $\cS\subseteq \cG_n$ of size $2^{\tilde{m}}$, there exists a set of generators 
$\{\overline{Z}_1, \ldots, \overline{Z}_{s+c}, \overline{X}_1, \ldots, \overline{X}_{c}\}$ for some $s$ and $c$ with $s+2c=\tilde{m}$, 
where they satisfy 
\begin{align}
\label{eq:X_and_Z_condition1}
\overline{X}_j \overline{X}_k &= \overline{X}_k \overline{X}_j\ \forall j,k,\\
\label{eq:X_and_Z_condition2}
\overline{Z}_j \overline{Z}_k &= \overline{Z}_k \overline{Z}_j\ \forall j,k,\\ 
\label{eq:X_and_Z_condition3}
\overline{X}_j \overline{Z}_k &= (-1)^{\delta(j,k)}\overline{Z}_k \overline{X}_j\ \forall j,k.
\end{align}
An $\llbracket n, k\,|\,  c \rrbracket$ ( or $\llbracket n, k, d\,|\, c \rrbracket$ or $\llbracket n, k; m\,|\, c \rrbracket$ when the minimum distance $d$ or the number of memory qubits $m$ for convolutional codes are known) entanglement-assisted QECC can be constructed using the above generators, where $k=n\!-\!s\!-\!c$.

More details about the construction of entanglement-assisted QECCs can be found in~\cite{brun2006correcting} and~\cite{lidar2013quantum}.
Here, we demonstrate the idea using an example from~\cite{brun2006correcting} and~\cite{lidar2013quantum}.

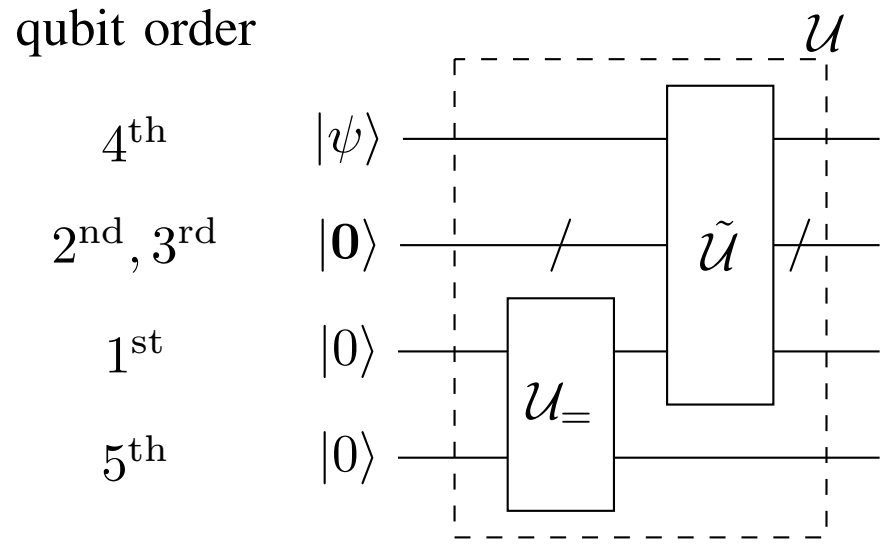
\begin{figure}
  \begin{center}
    \begin{tikzpicture}[x=1cm, y= 1cm, node/.style={draw=none},
                          block/.style={draw,minimum height = 50pt,minimum width = 30pt},
                          block3/.style={draw,minimum height = 60pt, minimum width = 20pt},
                          block2/.style={draw, minimum height = 40pt, minimum width = 20pt},
                          factor/.style={draw,minimum size = 15pt,inner sep = 0pt}
                          ]
     \node         at (-30pt,20pt) {qubit order};
	\node (N00) at (-30pt,0) {$4^{\Rm{th}}$};
	\node (N02) at (-30pt,-20pt) {$2^{\Rm{nd}},3^{\Rm{rd}}$};
	\node (N04) at (-30pt,-40pt) {$1^{\Rm{st}}$}; 
	\node (N06) at (-30pt, -60pt) {$5^{\Rm{th}}$}; 
	
	\node (N4) at   (10pt, 0pt) {$|\psi \rangle$};
	\node (N23) at (10pt, -20pt) {$|\mb{0}\rangle$};
	\node (N1) at   (10pt, -40pt) {$|0\rangle$};
	\node (N5) at   (10pt, -60pt) {$|0\rangle$};
	
	\node[block2] (N30) at (50pt, -50pt) {$\Cal{U}_{=}$};
	\path (N1) edge[draw] (40pt, -40pt);
	\path (N5) edge[draw] (40pt, -60pt);
	
	\node[block3] (N40) at (80pt, -20pt) {$\tilde{\Cal{U}}$};
	\path (N4) edge[draw] (70pt, 0pt);
	\path (N23) edge[draw] (70pt, -20pt);
	\draw (60pt, -40pt) -- (70pt, -40pt);
	
	\node at (50pt, -20pt) {$/$};
	
	\draw (90pt, 0pt) -- (110pt, 0pt);
	\draw (90pt, -20pt) -- (110pt, -20pt);
	\draw (90pt, -40pt) -- (110pt, -40pt);
	\draw (60pt, -60pt) -- (110pt, -60pt);
	
	\node at (95pt, -20pt) {$/$};
	
	\draw[dashed] (30pt, 15pt) rectangle (100pt,-75pt);
	\node at (100pt,20pt) {$\Cal{U}$};
\end{tikzpicture}
  \end{center}
  \caption{Encoding circuit of the entanglement-assisted QECC of Example~\ref{ex:EAQECC1}.}
  \label{fig:EAQECC_circuit_01}
\end{figure}

\begin{example}[A $\llbracket 4, 1\,|\, 1 \rrbracket$ entanglement-assisted QECC $\Cal{C}$]
\label{ex:EAQECC1}
This an example from~\cite[Chapter 7]{brun2006correcting}.
Consider a subgroup $\cS$ of $\cG_n$ generated by a non-commuting set of Pauli operators, i.e., 
\begin{align}
\label{eq:M1}
g'_1 & = \fZ \fX \fZ \fI,\\
\label{eq:M2}
g'_2 & = \fZ \fZ \fI \fZ,\\
\label{eq:M3}
g'_3 & = \fX \fY \fX \fI,\\
\label{eq:M4}
g'_4 & = \fX \fX \fI \fX.
\end{align}
The set of generators 
$\{\overline{Z}_1, \ldots, \overline{Z}_{s+c}, \overline{X}_1, \ldots, \overline{X}_{c}\}$ for some $s+2c=m$
satisfying~\eqref{eq:X_and_Z_condition1}-~\eqref{eq:X_and_Z_condition3} can be found using Algorithm~\ref{algo:Gram-Schmidt_procedure}.

We obtain the following set of generators using Algorithm~\ref{algo:Gram-Schmidt_procedure} with the set of inputs~\eqref{eq:M1}-~\eqref{eq:M4}:
\begin{align}
\label{eq:Z1}
\overline{Z}_1 & = \fZ \fX \fZ \fI,\\
\label{eq:X1}
\overline{X}_1 & = \fZ \fZ \fI \fZ,\\ 
\label{eq:Z2}
\overline{Z}_2 & = \fY \fX \fX \fZ,\\ 
\label{eq:Z3}
\overline{Z}_3 & = \fX \fZ \fZ \fY. 
\end{align}

Fig.~\ref{fig:EAQECC_circuit_01} shows the encoding circuit of the entanglement-assisted QECC $\cC$, where $\tilde{\Cal{U}}$ and $\Cal{U}_{=}$ are two Clifford transformations, which is equivalent to a stabilizer QECC $\cC'$ with encoding circuit defined by Clifford transformation $\Cal{U} = (\tilde{\Cal{U}} \otimes \mb{I}_{2}) (\mb{I}_{8} \otimes \Cal{U}_{=})$.
The Clifford transformation $\Cal{U}_{=}$ transforms the zero state to a bell pair.
The Clifford transformations $\Cal{U}_{=}$ and $\tilde{\Cal{U}}$ satisfy the following equations
\begin{align*}
\Cal{U}_{=} Z_1 \Cal{U}_{=}^{\HP} &= ZZ,\\  
\Cal{U}_{=} Z_2 \Cal{U}_{=}^{\HP} &= XX,\\
\tilde{\Cal{U}} Z_j \tilde{\Cal{U}}^{\HP} &= \overline{Z}_j\ \ \forall j\in\{1,2,3\}, \\
\tilde{\Cal{U}} X_1 \tilde{\Cal{U}}^{\HP} &= \overline{X}_1.
\end{align*}
One possible encoding circuit for $\Cal{U}_{=}$ is a Hadamard gate on the second qubit followed by a CNOT on the first qubit controlled by the second qubit and the corresponding encoding matrix is 
\begin{equation*}
U_{=} \defeq \left[ \begin{array}{cc : cc}
1 & 0 & 0 & 0\\
0 & 0 & 0 & 1\\
0 & 0 & 1 & 1\\
1 & 1 & 0 & 0
\end{array}\right].
\end{equation*}
The qubits are ordered as in Fig.~\ref{fig:EAQECC_circuit_01}, where the $5^{\Rm{th}}$ qubit is the pre-shared ebit, the first three qubits are ancillas, and the $4^{\Rm{th}}$ qubit contains the information to be encoded.
It is easy to check that $\cC'$ has a stabilizer generated by the following generators
\begin{align*}
\overline{Z}_1' &\defeq \Cal{U} Z_1 \Cal{U}^{\HP} = \fZ \fX \fZ \fI | \fZ,\\ 
\overline{Z}_2' &\defeq \Cal{U} Z_2 \Cal{U}^{\HP} = \fY \fX \fX \fZ | \fI,\\
\overline{Z}_3' &\defeq \Cal{U} Z_3 \Cal{U}^{\HP} = \fX \fZ \fZ \fY | \fI,\\ 
\overline{Z}_5' &\defeq \Cal{U} Z_5 \Cal{U}^{\HP} = \fZ \fZ \fI \fZ | \fX. 
\end{align*}
\exampleend
\end{example}

\begin{algorithm}
\caption{Gram-Schmidt Procedure}
\label{algo:Gram-Schmidt_procedure}
\begin{algorithmic}[1]
\REQUIRE $S_0 = \{ g'_1, \ldots, g'_{m} \}\subseteq\cG_n$
\ENSURE $\{\overline{Z}_1, \ldots, \overline{Z}_{s+c}, \overline{X}_1, \ldots, \overline{X}_{c}\}$ for some $s+2c=m$
\STATE Initialize $S_{Z} := [\ ]$, $S_{X} := [\ ]$, and $j:=1$.
\WHILE{$S_0$ is a non-commuting set}
	\STATE Find a pair of anticommuting Pauli operators from $S_0$ and let one be $\overline{Z}_j$ and the other be $\overline{X}_j$
	\STATE $S_0 := S_0 \backslash \{ \overline{Z}_j, \overline{X}_j \}$
	\STATE $S_Z := S_Z \cup \{ \overline{Z}_j \}$
	\STATE $S_X := S_X \cup \{ \overline{X}_j \}$
	\FOR {$g' \in S_0$}
		\IF{$g' \overline{Z}_j = - \overline{Z}_j g'$}
			\STATE $g' := g' \overline{X}_j$
		\ENDIF
		\IF{$g' \overline{X}_j = - \overline{X}_j g'$}
			\STATE $g' := g' \overline{Z}_j$
		\ENDIF
	\ENDFOR
	\STATE $j := j+1$
\ENDWHILE
\STATE $S_Z := S_Z \cup S_0$
\STATE Output $S_Z \cup S_X$
\end{algorithmic}
\end{algorithm}

The NFG in Fig.~\ref{fig:QP4_EAQECC_DE} is a quantum-probability NFG representation of an entanglement-assisted QECC.
The transmitted qubits are affected by errors while the preshared equbits stay unchanged.
\section{Conclusion}
We have developed two complementary normal factor graph representations of stabilizer quantum error-correcting codes. Quantum-code NFGs describe stabilizer and normalizer label codes and use graph duality to establish self-orthogonality and support code construction. Quantum-probability NFGs describe the probabilities associated with encoding, noise, and measurement, with a double-edge formulation that clarifies their relationship to quantum-code NFGs. Applications to convolutional, tail-biting convolutional, and low-density parity-check stabilizer codes illustrate the framework, while the entanglement-assisted extension broadens its scope. An important direction for future work is to develop decoding algorithms that exploit these representations while explicitly accounting for code degeneracy.

\appendices
\section{Proof of Theorem~\ref{thm:selforthogonal_thm}}
\label{appendix:thm1}
\begin{proof}
First, let us show that $\hat{\cB} \subseteq \cB^{\perp}$.
For any $(\hat{\mb{t}}_{1}, \hat{\mb{t}}_{2}) \in \hat{\cB}$, there exists an $\hat{\mb{m}}$ such that $(\hat{\mb{t}}_i, \hat{\mb{m}}) \in \cB_i^{\perp},\ i\in\{1,2\}$.
For any $(\vect{t}{1}, \vect{t}{2}) \in \cB$, there exists an $\mb{m}$ such that $(\vect{t}{i}, \mb{m}) \in \cB_i,\ i\in\{1,2\}$.
Then we have
\begin{align*}
\big\langle (\vect{t}{1}, \vect{t}{2}), (\hat{\mb{t}}_{1}, \hat{\mb{t}}_{2}) \big\rangle_{\syp}
& = \langle \vect{t}{1}, \hat{\mb{t}}_{1} \rangle_{\syp} + \langle \vect{t}{2}, \hat{\mb{t}}_{2} \rangle_{\syp}\\
& \stackrel{(a)}{=} \langle \mb{m}, \hat{\mb{m}} \rangle_{\syp} + \langle \mb{m}, \hat{\mb{m}} \rangle_{\syp}\\
& = 0,
\end{align*}
where step $(a)$ follows from $(\mb{t}_i,\mb{m})\in\codeB_i$ and $(\hat{\mb{t}}_i,\hat{\mb{m}})\in\codeB_i^{\perp}$, which imply $\langle \mb{t}_i,\hat{\mb{t}}_i\rangle_{\syp}  + \langle \mb{m},\hat{\mb{m}}\rangle_{\syp} = 0$ for $i\in\{1, 2\}$.
Therefore, $(\hat{\mb{t}}_{1}, \hat{\mb{t}}_{2}) \in \cB^{\perp}$ and so $\hat{\cB} \subseteq \cB^{\perp}$.

Let us now show that $\cB^{\perp} \subseteq \hat{\cB}$.
Let the binary code $\Cal{M}$ be defined as
\begin{equation*}
\Cal{M} \defeq \Cal{M}_1 + \Cal{M}_2 \defeq \left\{ \vect{m}{1} + \vect{m}{2}\ \middle|\ \vect{m}{i} \in \Cal{M}_i,\ i=1,2 \right\},
\end{equation*}
where for $i\in\{1,2\}$, 
\begin{equation*}
\Cal{M}_i \defeq \left\{ \mb{m}\ \middle|\ \text{there exists }\vect{t}{i} \text{ such that }(\vect{t}{i}, \mb{m}) \in \cB_i \right\}.
\end{equation*}
Suppose the code $\Cal{M}$ is a binary linear code of dimension $k$ and length $2n$ with a generator matrix $G$.
Let $R$ be a right inverse of $G$, i.e., $G R = \mb{I}_{k}$.
Let $W$ be defined as $W \defeq R^{\msf{T}} \mb{\Lambda}_n$.
Denote the $i^{\Rm{th}}$ rows of $G$ and $W$ as $\vect{g}{i}$ and $\vect{w}{i}$, respectively.
Since $G \mb{\Lambda}_n W^{\msf{T}} = G \mb{\Lambda}_n \mb{\Lambda}_n R = GR = \mb{I}_{k}$, we have
\begin{equation}
\label{eq:g_w_inner_product}
\langle \vect{g}{i}, \vect{w}{j} \rangle_{\syp} = [i=j]\ \forall i,j.
\end{equation}
For any $\hat{\mb{t}} \defeq (\hat{\mb{t}}_{1}, \hat{\mb{t}}_{2}) \in \cB^{\perp}$, we want to show that there exists an $\hat{\mb{m}}$ such that $(\hat{\mb{t}}_{i}, \hat{\mb{m}}) \in \cB_i^{\perp}$ for $i\in\{1,2\}$, i.e., $\hat{\mb{t}} \in \hat{\cB}$.
We observe that for $(\vect{t}{i}, \vect{m}{i}) \in \cB_i$, $i\in\{1,2\}$, the value of $\langle \vect{t}{i}, \hat{\mb{t}}_{i} \rangle_{\syp}$ only depends on $\vect{m}{i}$ and $\hat{\mb{t}}$ and then we define $h_{\hat{\mb{t}}}$ in order to have an explicit form of $\hat{\mb{m}}$.
Let the function $h_{\hat{\mb{t}}}: \Cal{M} \rightarrow \GF{2}$ be defined as
\begin{equation}
\label{eq:h_def}
h_{\hat{\mb{t}}}(\mb{m}) \defeq \langle \vect{t}{1}, \hat{\mb{t}}_{1} \rangle_{\syp} + \langle \vect{t}{2}, \hat{\mb{t}}_{2} \rangle_{\syp},
\end{equation}
where $\mb{m} = \vect{m}{1} + \vect{m}{2}$, $\vect{m}{i}\in\Cal{M}_i$, and $(\vect{t}{i}, \vect{m}{i}) \in \cB_i$, $i\in\{1,2\}$.
The function $h_{\hat{\mb{t}}}$ is well-defined for the following reason.
For any $\mb{m} = \vect{m}{1} + \vect{m}{2} = \vect{m}{1}' + \vect{m}{2}'$, $\vect{m}{i}, \vect{m}{i}' \in\Cal{M}_i$, $(\vect{t}{i}, \vect{m}{i}), (\vect{t}{i}', \vect{m}{i}') \in \cB_i$, $i\in\{1,2\}$, 
we have 
\begin{align*}
& \left(\langle\vect{t}{1}, \hat{\mb{t}}_{1} \rangle_{\syp} + \langle \vect{t}{2}, \hat{\mb{t}}_{2} \rangle_{\syp}\right) - \left(\langle\vect{t}{1}', \hat{\mb{t}}_{1} \rangle_{\syp} + \langle \vect{t}{2}', \hat{\mb{t}}_{2} \rangle_{\syp}\right) \\
&= \langle \vect{t}{1} - \vect{t}{1}', \hat{\mb{t}}_{1}\rangle_{\syp} + \langle \vect{t}{2} - \vect{t}{2}', \hat{\mb{t}}_{2}\rangle_{\syp} \\
&\stackrel{(a)}{=} 0,
\end{align*}
where step $(a)$ follows from the fact that 
if we let $\triangle \mb{m} \defeq \vect{m}{1} - \vect{m}{1}' = \vect{m}{2}' - \vect{m}{2} = \vect{m}{2} - \vect{m}{2}'$,
then we have $(\vect{t}{i} - \vect{t}{i}', \triangle \mb{m}) \in \cB_i$ for $i=1,2$, which implies 
$(\vect{t}{1} - \vect{t}{1}', \vect{t}{2} - \vect{t}{2}') \in \cB$.
It is also easy to check that the function $h_{\hat{\mb{t}}}$ is linear and that 
$h_{\hat{\mb{t}}}(\vect{m}{i}) = \langle \vect{t}{i}, \hat{\mb{t}}_{i} \rangle_{\syp}$ 
where $(\vect{t}{i}, \vect{m}{i}) \in \cB_i$, $i=1,2$,
by choosing $(\vect{t}{j}, \vect{m}{j}) = (\mb{0}, \mb{0}) \in \cB_{j}$, $j \in \{1,2\}\backslash\{i\}$.

Now we choose $\hat{\mb{m}}$ to be $\hat{\mb{m}} \defeq \sum_{i=1}^{k} h_{\hat{\mb{t}}}(\vect{g}{i}) \vect{w}{i}$.
For any $\mb{m}\in\Cal{M}_i$, $i=1,2$, we have $\mb{m}\in\Cal{M}_i \subseteq \Cal{M}$ and there exist $c_1, \ldots, c_k$ such that $\mb{m} = \sum_{j=1}^{k} c_j \vect{g}{j}$.
Then we have 
\begin{align*}
\langle \mb{m}, \hat{\mb{m}} \rangle_{\syp} 
& = \Big\langle \sum_{j=1}^{k} c_j \vect{g}{j}, \sum_{j'=1}^{k} h_{\hat{\mb{t}}}(\vect{g}{j'}) \vect{w}{j'} \Big\rangle_{\syp}\\
& \stackrel{(a)}{=} \sum_{j=1}^{k} \sum_{j'=1}^{k} c_j h_{\hat{\mb{t}}}(\vect{g}{j'}) \big\langle \vect{g}{j},  \vect{w}{j'} \big\rangle_{\syp}\\
& \stackrel{(b)}{=} \sum_{j=1}^{k}  c_j h_{\hat{\mb{t}}}(\vect{g}{j}) \\
& \stackrel{(c)}{=} h_{\hat{\mb{t}} }
\left(\sum_{j=1}^{k} c_j\vect{g}{j}\right) \\
& = h_{\hat{\mb{t}}}(\mb{m})\\
& = \langle \vect{t}{i}, \hat{\mb{t}}_{i} \rangle_{\syp},
\end{align*}
where step $(a)$ follows from the fact that the symplectic inner product is bilinear,
step $(b)$ follows from~\eqref{eq:g_w_inner_product}, and 
step $(c)$ follows from the linearity of $h_{\hat{\mb{t}}}$.
Therefore, we have $(\hat{\mb{t}}_{i}, \hat{\mb{m}}) \in \cB_i^{\perp}$ for $i\in\{1,2\}$ and hence $\hat{\mb{t}} \in \hat{\cB}$.
\end{proof}
\section{Proof of Theorem~\ref{thm:TBC_QECC_code_design_NFG_single_g}}
\label{appendix:thm2}
\begin{proof}
First, we can show that $\cB_{\Rm{e}}$ is self-orthogonal under the trace inner product.
It is sufficient to prove the rows of the generator matrix of $\cB_{\Rm{e}}$ are orthogonal to each other.
For any $i,j \in \set{I}_{\nu+1}$, we have
\begin{align*}
&\langle [\vect{s}{j},\ \vect{g}{j},\ \vect{s}{j+1}], [\vect{s}{i},\ \vect{g}{i},\ \vect{s}{i+1}] \rangle_{\tr}\\
&=\ \langle \vect{s}{j}, \vect{s}{i}\rangle_{\tr} + \langle \vect{g}{j}, \vect{g}{i}\rangle_{\tr} +  \langle \vect{s}{j+1}, \vect{s}{i+1}\rangle_{\tr}\\
\label{eq:gamma_g_relation}
&\stackrel{(a)}{=}\ \gamma_{j,i} + \langle \vect{g}{j}, \vect{g}{i}\rangle_{\tr} + \gamma_{j+1,i+1}\\
&\stackrel{(b)}{=}\ 2 \cdot \gamma_{j+1,i+1}\\
&=\ 0\ (\text{mod }2),
\end{align*}
where step $(a)$ follows from~\eqref{eq:gamma_state_relation} and 
step $(b)$ follows from the definition of $\gamma_{j,i}$.

Then, let us show that $\codeB = \codeB_{\graph{G}}$.
\begin{itemize}
\item We first prove that $\codeB \subseteq \codeB_{\graph{G}}$.
Consider the fact that both $\codeB$ and $\codeB_{\graph{G}}$ are additive code over $\GF{4}$.
It suffices to show that all the generators of $\codeB$ are in $\codeB_{\graph{G}}$, i.e., $\vg(p) \in \codeB_{\graph{G}}\ \forall p \in \mbb{Z}_{N}$.
Similar to the way we define $\vg$ and $\vg(p)$ in~\eqref{eq:g_TBC} and~\eqref{eq:g_p_TBC}, we define $\mb{s}$ in the $n$-symbol form and $\mb{s}(p)$ which is the $n$-symbol cyclic left shift of $\mb{s}$ as
\begin{align}
\mb{s} & \defeq [\vect{s}{0},\ \vect{s}{1},\ \ldots,\ \vect{s}{N-1}],\\
\mb{s}(p) & \defeq [\vect{s}{p},\ \vect{s}{p+1},\ \ldots,\ \vect{s}{p+N-1}],
\end{align}
respectively, where the indices are taken modulo $N$.
For any $i \in \mbb{Z}_{N}$, we have 
$[\vect{s}{p+i},\ \vect{g}{p+i},\ \vect{s}{p+i+1}] \in \codeB_{i}$ since this vector is in $\codeBe$ and hence 
$f(\vg(p), \mb{s}(p)) = 1$ which implies $\vg(p) \in \codeB_{\graph{G}}$.
\item We then prove that $\codeB \supseteq \codeB_{\graph{G}}$.
For any $\mb{t} \in \codeB_{\graph{G}}$,
we have $f(\mb{t}, \mb{m})=1$ for some 
$\mb{t} = [\vect{t}{0},\ \vect{t}{1},\ \ldots,\ \vect{t}{N-1}]$ and 
$\mb{m} = [\vect{m}{0},\ \vect{m}{1},\ \ldots,\ \vect{m}{N-1}]$,
where $\vect{t}{\ell} \in \GF{4}^{n}$ and $\vect{m}{\ell} \in \GF{4}^{\nu}$ for all $\ell \in\mbb{Z}_{N}$.
More specifically, we have 
\begin{equation}
[\vect{m}{\ell}^{\Rm{L}},\ \vect{t}{\ell},\ \vect{m}{\ell+1}^{\Rm{R}}] \in \cB_{\ell}\ \forall \ell \in \mbb{Z}_{N}, 
\end{equation}
where $\vect{m}{\ell}^{\Rm{L}} \defeq \vect{m}{\ell}$ and $\vect{m}{\ell}^{\Rm{R}} \defeq \vect{m}{\ell}$ for all $\ell \in \mbb{Z}_{N}$.
Suppose there are some coefficients $\alpha_{i,\ell} \in \GF{2}$, $i\in \mbb{Z}_{N}$, $\ell\in \mbb{Z}_{\nu+1}$ such that 
\begin{equation}
[\vect{m}{i}^{\Rm{L}},\ \vect{t}{i},\ \vect{m}{i+1}^{\Rm{R}}] 
= \sum_{\ell=0}^{\nu} \alpha_{i,\ell} [\vect{s}{\ell},\ \vect{g}{\ell},\ \vect{s}{\ell+1}]\ \forall i\in\mbb{Z}_{N}.
\end{equation}
For any $i\in\mbb{Z}_{N}$, consider the value of $\vect{m}{i}$ obtained from the left and the right, i.e.,
\begin{align}
\label{eq:m_i_L}
&\vect{m}{i} = \vect{m}{i}^{\Rm{L}} 
= \sum_{\ell=0}^{\nu} \alpha_{i,\ell}\vect{s}{\ell}
= \sum_{\ell=0}^{\nu-1} \alpha_{i,\ell+1}\vect{s}{\ell+1}\\
\label{eq:m_i_R}
&\vect{m}{i} = \vect{m}{i}^{\Rm{R}} 
= \sum_{\ell=0}^{\nu} \alpha_{i-1,\ell}\vect{s}{\ell+1}
= \sum_{\ell=0}^{\nu-1} \alpha_{i-1,\ell}\vect{s}{\ell+1}
\end{align}
As mentioned in Remark~\ref{rm:Alg1}, the set $\tilde{\set{S}}$ is linearly independent over $\GF{2}$.
By comparing~\eqref{eq:m_i_L} and~\eqref{eq:m_i_R}, we can obtain 
\begin{equation}
\label{eq:alpha_i}
\alpha_{i,\ell+1} = \alpha_{i-1,\ell}\ \forall i\in\mbb{Z}_{N}, \ell\in\set{I}_{\nu}.
\end{equation}
Hence 
\begin{align}
\vect{t}{i} &= \sum_{\ell=0}^{N-1} \alpha_{i,\ell} \vect{g}{\ell} 
\stackrel{(a)}{=} \sum_{\ell=0}^{N-1} \alpha_{i-\ell,0} \vect{g}{\ell}\ \forall i\in\mbb{Z}_{N},
\end{align}
where step $(a)$ follows from~\eqref{eq:alpha_i}.
Hence
\begin{align}
\mb{t} &= \sum_{\ell=0}^{N-1} \alpha_{i-\ell,0} \bvg(-\ell) \in \codeB,
\end{align}
from which we can show that $\codeB \supseteq \codeB_{\graph{G}}$.
\end{itemize}
Finally, according to Algorithm~\ref{alg:TBC_QECC_single_generator}, we have the state vectors $\vect{s}{\ell}\in\GF{4}^{\nu}$, $\ell\in\set{I}_{\nu+1}^{+}$.
Therefore, we need $\nu$ memory qubits for these state vectors.
\end{proof}

\bibliographystyle{IEEEtran}
\bibliography{ref}
\end{document}